\documentclass[aps,reprint,superscriptaddress,longbibliography,bibnotes]{revtex4-2}

\usepackage{graphicx}
\usepackage{dcolumn}
\usepackage{bm, amssymb, color}
\usepackage{hyperref}
\usepackage{natbib}
\usepackage[mathscr]{euscript}
\usepackage{booktabs,tabulary}
\usepackage{multirow}
\usepackage{amsmath,soul}
\usepackage[export]{adjustbox}

\usepackage[caption=false]{subfig}
\usepackage{array}
\usepackage{cases}

\usepackage{float}

\newcommand{\fn}[2]{\mathinner{#1\mathopen{\left(#2\right)}}}

\newcommand{\spD}[1]{\fn{\tilde{\chi}_{_V}}{#1}}

\usepackage[colorinlistoftodos,shadow,textwidth=18mm]{todonotes}

\newcolumntype{L}[1]{>{\raggedright\let\newline\\\arraybackslash\hspace{0pt}}m{#1}}
\newcolumntype{C}[1]{>{\centering\let\newline\\\arraybackslash\hspace{0pt}}m{#1}}
\newcolumntype{R}[1]{>{\raggedleft\let\newline\\\arraybackslash\hspace{0pt}}m{#1}}
\setstcolor{green}

\graphicspath{{./figs/}}

\begin{document}
\title{Characterization of void space, large-scale structure, and transport properties of maximally random jammed packings of superballs}

\author{Charles Emmett Maher}
\affiliation{Department of Chemistry, Princeton University, Princeton, New Jersey 08544, USA}
\author{Frank H. Stillinger}
\affiliation{Department of Chemistry, Princeton University, Princeton, New Jersey 08544, USA}
\author{Salvatore Torquato}%
\email{torquato@princeton.edu}
\homepage{http://chemlabs.princeton.edu}
\affiliation{Department of Chemistry, Princeton University, Princeton, New Jersey 08544, USA}
\affiliation{Department of Physics, Princeton University, Princeton, New Jersey 08544, USA}
\affiliation{Princeton Institute for the Science and Technology of Materials, Princeton University, Princeton, New Jersey 08544, USA}
\affiliation{Program in Applied and Computational Mathematics, Princeton University, Princeton, New Jersey 08544, USA}

\date{\today}
\begin{abstract}
Dense, disordered packings of particles are useful models of low-temperature amorphous phases of matter, biological systems, granular media, and colloidal systems.
The study of dense packings of nonspherical particles enables one to ascertain how rotational degrees of freedom affect packing behavior.
Here, we study superballs, a large family of deformations of the sphere, defined in three dimensions by $|x_1|^{2p}+|x_2|^{2p}+|x_3|^{2p}\leq1$, where $p\in(0,\infty)$ is a \textit{deformation parameter} indicating to what extent the shape deviates from a sphere.
As $p$ increases from the sphere point ($p=1$), the superball tends to a cuboidal shape and approaches a cube in the $p\rightarrow\infty$ limit.
As $p\rightarrow 0.5$, it approaches an octahedron, becomes a concave body with octahedral symmetry for $p < 0.5$, and approaches a three-dimensional cross in the limit $p\rightarrow0$.
Previous characterization of superball packings has shown that they have a maximally random jammed (MRJ) state, whose properties (e.g., packing fraction $\phi$, average contact number $\bar{Z}$) vary nonanalytically as $p$ diverges from unity.
Here, we use an event-driven molecular dynamics algorithm to produce MRJ superball packings with $0.85 \leq p \leq 1.50$. 
To supplement the previous work on such packings, we characterize their large-scale structure by examining the behaviors of their structure factors $S(Q)$ and spectral densities $\spD{Q}$, as the wave number $Q$ tends to zero, and find that these packings are effectively hyperuniform for all values of $p$ examined.
We show that the mean width $\bar{w}$ is a useful length scale to make distances dimensionless in order to compare systematically superballs of different shape.
Moreover, we compute the complementary cumulative pore-size distribution $F(\delta)$ and find that the pore sizes tend to decrease as $|1-p|$ increases.
From $F(\delta)$, we estimate how the fluid permeability, mean survival time, and principal diffusion relaxation time vary as a function of $p$.
Additionally, we compute the diffusion ``spreadability" $\mathcal{S}(t)$ [Torquato, Phys. Rev. E, \textbf{104}, 054102, (2021)] of these packings and find that the long-time power-law scaling indicates these packings are hyperuniform with a small-$Q$ power law scaling of the spectral density $\spD{Q}\sim Q^{\alpha}$ with an exponent $\alpha$ that ranges from 0.64 at the sphere point to 0.32 at $p = 1.50$, and decreases as $|1-p|$ increases.
Each of the structural characteristics computed here exhibits an extremum at the sphere point and varies nonanalytically as $p$ departs the sphere point.
We find the nonanalytic behavior in $\phi$ on either side of the sphere point is nearly linear, and determine that the rattler fraction $\phi_R$ decreases rapidly as $|1-p|$ increases.
Our results can be used to help inform the design of colloidal or granular materials with targeted densities and transport properties.

\end{abstract}

\maketitle
\section{Introduction}\label{sec:Intro}
A particle packing is a collection of nonoverlapping bodies in $d$-dimensional Euclidean space $\mathbb{R}^d$.
The packing fraction $\phi$ is the fraction of $\mathbb{R}^d$ covered by these bodies.
Dense packings have been used to model physical phenomena in a wide variety of contexts including condensed and soft matter physics \cite{hughel_1965, zallen_2007, RHM_Text, chaikin_lubensky_2000}, materials science \cite{RHM_Text, mehta_1994}, and biology \cite{LIANG_bio1, Gevertx_bio3, Purohit_bio2}, among many others (see Refs. [\citenum{Torquato_JamHardPart}] and [\citenum{Torquato_PackingPersp}]).
A \textit{jammed} packing is one in which each particle is contacted by its neighbors such that mechanical stability of a particular type is conferred to the packing \cite{Torquato_JamHardPart}. 
Such systems grant insight into the structure and bulk properties of crystals, glasses, and granular media \cite{Torquato_JamHardPart, Parisi_Mean}. 

Jammed packings can be organized into three mathematically precise categories based on the nature of the mechanical stability conferred, which in order of increasing stability are as follows \cite{Torquato_JamHardPart, Torqauto_Multiplicity}:
(1) \textit{Local jamming}: no individual particle can be moved while holding all other particles fixed.
(2) \textit{Collective jamming}: the packing is locally jammed, and no collective motion of a finite subset of particles is possible. 
(3) \textit{Strict jamming}: the packing is collectively jammed and all volume-nonincreasing deformations are disallowed by the impenetrability constraint.

A jammed state of particular interest is the \textit{maximally random jammed} (MRJ) state.
Such packings are the most disordered configuration (as measured by a set of scalar order metrics) subject to a particular jamming category, and can be viewed as prototypical glasses because of their maximal disorder and mechanical rigidity \cite{Torquato_JamHardPart, Torquato_RCPBad}. 
Moreover, these packings are \textit{hyperuniform}, meaning their infinite-wavelength density fluctuations are anomalously suppressed compared to those in typical disordered systems \cite{Torquato_HUDef, Donev_Unexpected, Torquato_HURev}.
Numerical simulations in $\mathbb{R}^3$ have produced, to a very good approximation, monodisperse MRJ packings of several particle shapes including spheres \cite{Torquato_RCPBad}, ellipsoids \cite{Donev_MRJelip}, superballs \cite{Jiao_MRJballs}, the Platonic solids \cite{Jiao_MRJplatonic}, and truncated tetrahedra \cite{Chen_MRJtruntet}.
While MRJ sphere packings have been characterized extensively \cite{Klatt_MRJI, Klatt_MRJII, Klatt_MRJIII, Donev_Unexpected, Donev_g2, Atkinson_Rattler}, other shapes are less well understood due to their relative mathematical complexity. 

MRJ packings of spheres in $\mathbb{R}^3$ are \textit{isostatic} \cite{Donev_g2, Ohern_Iso}, meaning that the total number of interparticle contacts (constraints) is equal to the number of degrees of freedom in the system and that all of the constraints are linearly independent.
In such packings, it is thus implied that the average number of contacts per particle $\bar{Z}$ is equal to twice the number of degrees of freedom per particle (i.e., $\bar{Z}=2f$) in the thermodynamic limit, which has been verified to a high numerical accuracy \cite{Donev_g2, Ohern_Iso}.
Additionally, analyses approximating the nonaffine elastic response of disordered solids have shown the isostaticity of jammed sphere packings (see, e.g., Ref. \citenum{Zaccone_nonaff}).
While, e.g., spheres, polyhedra \cite{Jiao_MRJplatonic, Torquato_ASCNat}, and lenses \cite{Cinacchi_LensMRJ} have isostatic MRJ packings, this is not a general signature of the MRJ state.
In particular, certain aspherical particles with smooth boundaries, e.g., ellipsoids \cite{Donev_MRJelip}, superellipsoids \cite{Delaney_Superell}, and superballs \cite{Jiao_MRJballs} have \textit{hypostatic} MRJ packings, meaning that $\bar{Z} < 2f$.
Using second- and higher-order jamming analyses, Donev \textit{et al.} \cite{Donev_EllipFunct} have rigorously shown that if the curvature of nonspherical particles at their contact points is considered, then hypostatic packings of nonspherical particles can indeed be jammed.

In practice, disordered jammed packings of monodisperse spheres and nearly-spherical particles produced via simulation or experiment contain a small concentration of \textit{rattlers} ($\lesssim\!\!3\%$ of particles \cite{Torquato_JamHardPart,Torquato_PackingPersp}), which are unjammed particles locally imprisoned by their jammed neighbors \cite{Torquato_JamHardPart, Torquato_RCPBad}.
Jamming precludes the existence of rattlers \cite{Torquato_JamHardPart, Torqauto_Multiplicity}.
Nevertheless, it is the significant majority of particles that confers rigidity to the packing, and in any case, the rattlers could be removed (in computer simulations) without disrupting the remaining jammed particles \cite{Torquato_JamHardPart,Torquato_PackingPersp}.
The \textit{rattler fraction} $\phi_R$ is greatest in packings of spherically symmetric particles ($\sim$2.5\% in $\mathbb{R}^3$), decreases in packings of particles with rotational degrees of freedom \cite{Maher_Kin} (vanishing in the case of sufficiently aspherical particles \cite{Jiao_MRJballs, Donev_EllipFunct}), and increases in lower spatial dimensions ($\sim$3.5\% in $\mathbb{R}^2$) \cite{Atkinson_2DMono}.
Atkinson \textit{et al.} \cite{Atkinson_Slowdown} have shown that removing rattlers from MRJ packings results in a nonhyperuniform packing, meaning the subset of jammed particles alone is not hyperuniform.

Torquato and Stillinger \cite{Torquato_HUDef} suggested that certain defect-free, strictly jammed packings of identical spheres are hyperuniform.
Specifically, they conjectured that any strictly jammed, saturated, infinite packing of identical spheres is hyperuniform.
A saturated packing is one in which there is insufficient space to add another particle of the same type to the packing.
The conjecture excludes packings that contain rattlers because they are by definition not strictly jammed (see above).
Nonetheless, such packings are effectively hyperuniform \cite{Donev_Unexpected, Zachary_QLRLet, Bert_MRJNum, Week_MRJnum, Drey_MRJnum, Zachary_QLRII, Zachary_QLRI}.
The Torquato-Stillinger conjecture is supported by recent theoretical considerations involving free-volume theory \cite{Torquato_FVT} and numerical investigations \cite{Rissone_TSC} involving large-scale correlations in three-dimensional sphere packings at jamming.
One expects this conjecture to extend to certain defect-free strictly jammed packings of nonspherical particles.
Zachary \textit{et. al} \cite{Zachary_QLRI, Zachary_QLRII, Zachary_QLRLet} have shown that two-dimensional MRJ packings of polydisperse and/or nonspherical particles are effectively hyperuniform with respect to the spectral density, the proper spectral measure for packings of polydisperse or noncircular particles.
Here, we use the spectral density to show that MRJ packings of nonspherical particles in three dimensions are effectively hyperuniform.

It is fundamentally and practically important to understand the packing behavior of nonspherical particles \cite{ABREU_Spherocyl, Williams_SpheroCyl, Donev_MRJelip, Donev_Rect, Yatsenko_Rod, Torquato_Org, Torquato_ASCNat, Jiao_MRJplatonic, Maher_Kin}.
In particular, this allows us to better understand real granular media and low-temperature states of matter. 
Supramolecular chemistry of organic compounds, which attain a wide range of symmetry groups \cite{kitaigorodsky_orgcryst}, can also be modeled using particles with the same symmetry.

\begin{figure}[t]
    \centering
    \includegraphics[width = 0.45\textwidth]{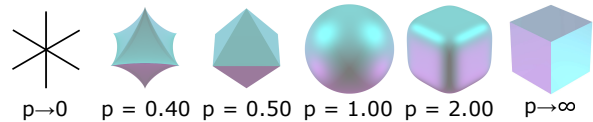}
    \caption{Superballs with different values of the deformation parameter $p$.}
    \label{fig:SB_assortl}
\end{figure}

Studies of the dense packings of superballs in two and three dimensions were introduced by Jiao, Stillinger, and Torquato \cite{Jiao_MRJballs, Jiao_OptSD, Jaio_OptSB} to ascertain the effect of deforming from a sphere and the resulting rotational degrees of freedom.
Superballs in $\mathbb{R}^d$ are a family of centrally symmetric shapes with $d$ equal semiaxes, defined by the inequality
\begin{equation}\label{eq:SB}
|x_1|^{2p}+|x_2|^{2p}+\cdots+|x_d|^{2p}\leq1,    
\end{equation}
where $x_i (i = 1,\dots,d)$ are Cartesian coordinates and $p \geq 0$ is the \textit{deformation parameter} which indicates the extent to which the shape is deformed away from the $d$-dimensional sphere ($p=1)$.
Henceforth, the term superball will refer to the three-dimensional object, while \textit{superdisk} will refer to the two-dimensional object.
As $p$ increases from unity superballs attain cubic symmetry, becoming a cube in the limit of $p \rightarrow\infty$, and as $p$ decreases attain octahedral symmetry (see Fig. \ref{fig:SB_assortl}) becoming an octahedron at $p = 0.5$, concave at $p < 0.5$, and a three-dimensional cross in the $p\rightarrow0$ limit.

A superball breaks the rotational symmetry of a sphere differently from an ellipsoid \cite{Jiao_MRJballs}.
In the direct vicinity of the sphere point, the superball attains either cubic ($p > 1$) or octahedral ($p <1$) symmetry, while ellipsoids are simply affine transformations of the sphere.
Far from the sphere point, the asphericity \cite{Torquato_ASCNat, Jiao_MRJplatonic}, defined as the ratio of the radii of the circumsphere and insphere of a nonspherical particle, can increase without limit as the aspect ratio $a$ grows for ellipsoids, while it is always bounded and close to unity for convex superballs (i.e., $p \geq 0.5)$.
These unique geometric properties of superballs result in rich phase behavior \cite{Batten_SBPhase, Ni_SBphase, Odriozola_Phase}, including a multitude of crystalline phases, plastic crystalline phases, and complex melting transitions.
Moreover, novel inorganic synthetic methods allow for the production of superball-shaped colloids \cite{Rossi_SBcolloid, Zhang_SBass, meijer_SBsolidsolid}.
Thus, the results presented herein can be used to inform the design of disordered colloidal materials with desired properties using such particles.

Previous theoretical studies of superball and superdisk packings include the determination of their densest known packings for all values of $p$, which are Bravais lattices with symmetries consistent with those native to the superball \cite{Jaio_OptSB}, as well as their phase behavior (see above).
Moreover, Jiao \textit{et al.} have studied the properties of MRJ superball packings \cite{Jiao_MRJballs}.
In particular, they found that $\phi$ and $\bar{Z}$ increase rapidly and nonanalytically as the superballs become more aspherical (i.e., as $|1-p|$ increases), while $\phi_R$ decreases, which is attributed to the symmetry breaking as $|1-p|$ increases from 0.
This increase in $\phi$ is a result of the superball shape becoming more efficient at filling space as $p$ deviates from unity.
Ellipsoids, however, have a critical aspect ratio $a^*$ \cite{Donev_MRJelip} at which $\phi$ is maximized and thereafter begins to decrease due to very elongated or plate-like ellipsoids having strong exclusion-volume effects in orientationally disordered packings (see, e.g., Ref. [\citenum{Williams_spherocyl}]) which causes the MRJ packing fraction to decrease.
A comparison of the ellipsoid and superball MRJ packing fractions as a function of the scaled mean width (see Secs. \ref{Sec:MW}, \ref{Sec:MWR}) is given in Fig. \ref{fig:phicomp}.
These observations are consistent with the notion that packings of particles with rotational degrees of freedom tend to have larger $\phi$ and smaller $\phi_R$.
The nonanalyticity around $p=1$ is also observed in optimal superball and superdisk packings \cite{Jaio_OptSB,Jiao_OptSD}.
By contrast, the densest known ellipsoid packings have a smooth increase in $\phi$ as $a$ deviates from unity \cite{Donev_OptEllip}.
They also show that MRJ superball packings are \textit{highly} hypostatic, meaning $\bar{Z}$ is much less than $2f$, which requires non-trivially correlated local arrangements of particles.
These so-called ``non-generic" local configurations, in which a particle has fewer contacts than average, are counter-intuitively not rare.
This property of superballs stands in contrast to ellipsoid packings, which are \textit{slightly} hypostatic, meaning $\bar{Z}$ is only slightly less than $2f$.

Similar nonanalytic behavior is also observed in MRJ packings of bidisperse superdisks in which there are equimolar amounts of superdisks with a size ratio of 1.4 \cite{Jiao_MRJballs}.
To our knowledge, however, monodisperse MRJ superdisk packings for $p\neq1$ have not been observed.
Monodisperse disks in $\mathbb{R}^2$ lack \textit{geometrical frustration}, i.e., the densest local packing arrangement is compatible with the global densest packing arrangement (the triangular lattice).
As a result, typical packing algorithms generate polycrystalline disk packings with a probability of nearly unity \cite{Atkinson_2DMono}.
Thus, the definition of \textit{random close packing} (RCP), which identifies the most probable packings as the most disordered, misleadingly identifies these polycrystalline packings as the RCP state, a dubious proposition for the MRJ state \cite{Donev_MRJDisk, Torquato_JamHardPart}.
Atkinson \textit{et al.} \cite{Atkinson_2DMono} produced MRJ-like disk packings using the Torquato-Jiao sequential linear programming algorithm \cite{TJ_Algo}, which are isostatic and qualitatively distinct from commonly observed polycrystalline packings.

Herein, we aim to build on the foundational study of MRJ superballs by Jiao \textit{et al.} \cite{Jiao_MRJballs}.
We use the Donev-Torquato-Stillinger (DTS) algorithm \cite{DTS_AlgI, DTS_AlgII}, which employs event-driven molecular dynamics to produce dense packings of centrally symmetric objects (more details in Sec. \ref{sec:DTS}), to produce large MRJ superball packings with $0.85 \leq p \leq 1.50$.
We show that these packings are effectively hyperuniform and compute the \textit{structure factor} $S(Q)$ and \textit{spectral density} $\spD{Q}$ and examine the small-$Q$ power-law scaling given by:
\begin{equation}\label{eq:powerlaw}
    \spD{Q}\sim Q^{\alpha},
\end{equation}
where $Q$ is the wave number and $\alpha>0$ is the hyperuniformity scaling exponent, which are defined in Sec. \ref{sec:FT}.
The structure factor and spectral density (mathematically defined in Sec. \ref{sec:FT}) are related to the Fourier transforms of the pair correlation function $g_2(r)$ and autocovariance of the phase indicator function $\chi_{_V}(r)$, respectively, and can be obtained via scattering experiments \cite{RHM_Text, Debye_Scattering}. 
Using $\spD{Q}$, we compute the \textit{spreadability} $\mathcal{S}(t)$ and find the long-time scaling indicates these packings are hyperuniform with $\alpha\in(0.32,0.64)$ which decreases as $|1-p|$ increases \cite{Torqauto_Spread}.
We additionally compute the complementary cumulative pore-size distribution $F(\delta)$, and find that pore sizes tend to be smaller as $|1-p|$ increases.
We also note that each packing is saturated. 
From $F(\delta)$, we are able to estimate the fluid permeability $k$, mean survival time $\tau$, and principal diffusion relaxation time $T_1$.
We observe that $k$, $\tau$, and $T_1$ have maxima at $p = 1$ and decrease as $|1-p|$ increases.
By producing MRJ packings with $|1-p|$ closer to zero than in previous work, we determine that $\phi$ increases nearly linearly for sufficiently small values of $|1-p|$.
We also characterize $\phi_R$ as a function of $p$, which decreases rapidly and nonanalytically as $p$ diverges from unity.
We expect these results to be useful in the design of disordered colloidal materials with desired properties.

The rest of the paper is organized as follows. 
Section \ref{sec:Backg} contains the pertinent background and mathematical definitions. 
Section \ref{sec:Algo} describes the methods used to produce and characterize the MRJ superball packings.
In Sec. \ref{sec:StructChar} we present results from the structural characterization of said packings and in Sec. \ref{sec:Props} we present their transport properties.
We then offer conclusions and potential future studies in Sec. \ref{sec:Conc}.

\section{Background and Definitions}\label{sec:Backg}
\subsection{Structure factor and spectral density}\label{sec:FT}
A system comprising point particles in $\mathbb{R}^d$ can be completely characterized by a set of $n$-particle probability density functions $\rho_n(\mathbf{r}_1,\dots,\mathbf{r}_n) \;\forall\;n \geq 1$, which are proportional to the probability of finding $n$ particles at the positions $\mathbf{r}_1,\dots,\mathbf{r}_n$.
For statistically homogeneous systems, $\rho_1(\mathbf{r}_1) = \rho$, where $\rho$ is the number density, and $\rho_2(\mathbf{r}_1,\mathbf{r}_2) = \rho^2g_2(\mathbf{r})$, where $\mathbf{r} = \mathbf{r}_2 - \mathbf{r}_1$ and $g_2(\mathbf{r})$ is the pair correlation function.
If the system is also statistically isotropic, $g_2(\mathbf{r}) = g_2(r)$, where $r = \|\mathbf{r}\|$.
The ensemble-averaged structure factor $S(\mathbf{Q})$ is defined as
\begin{equation}\label{eq:SK_hdefn}
    S(\mathbf{Q}) = 1+ \rho\tilde{h}(\mathbf{Q}),
\end{equation}
where $\tilde{h}(\mathbf{Q})$ is the Fourier transform of the total correlation function $h(\mathbf{r}) = g_2(\mathbf{r})-1$, and $\mathbf{Q}$ is a wave vector.

For single periodic configurations with $N$ point particles with positions $\mathbf{r}^N= (\mathbf{r}_1,\dots,\mathbf{r}_N)$ within a fundamental cell $F$ of a lattice $\Lambda$, the scattering intensity $\mathbb{S}(\mathbf{Q})$ is given by
\begin{equation}\label{eq:ScattInt}
    \mathbb{S}(\mathbf{Q}) = \frac{\left|\sum_{i = 1}^N\textrm{exp}(-i\mathbf{Q}\cdot\mathbf{r}_i)\right|^2}{N}.
\end{equation}
In the thermodynamic limit, an ensemble of $N$-particle configurations in $F$ is related to $S(\mathbf{Q})$ by
\begin{equation}\label{eq:SkScatt}
    \lim_{N, V_F\rightarrow\infty}\langle\mathbb{S}(\mathbf{Q})\rangle = (2\pi)^d\rho\delta(\mathbf{Q})+S(\mathbf{Q}),
\end{equation}
where $V_F$ is the volume of the fundamental cell and $\delta$ is the Dirac delta function \cite{Torquato_HURev}.
For finite $N$ simulations under periodic boundary conditions, Eq. (\ref{eq:ScattInt}) is used to compute $S(\mathbf{Q})$ directly by averaging over configurations. 
Here, we compute the angular-averaged $S(Q)$ by applying Eq. (\ref{eq:ScattInt}) to the superball centroids.

Packings can be interpreted as two-phase heterogeneous media, where the matrix phase $\mathcal{V}_1$ is the void (pore) space between the particles, and the particle phase $\mathcal{V}_2$ is the space occupied by the particles, such that $\mathcal{V}_1\cup\mathcal{V}_2=V\subset\mathbb{R}^d$ \cite{Cinacchi_LensMRJ}. 
The packing microstructure can be fully characterized by a countably infinite set of $n$-point probability functions $S_n^{(i)}$, defined by \cite{RHM_Text}
\begin{equation*}\label{eq:npt_probab}
     S_n^{(i)}(\mathbf{x}_1,\dots,\mathbf{x}_n)=\left\langle\prod^n_{j=1}\mathcal{I}^{(i)}(\mathbf{x}_n)\right\rangle,
\end{equation*}
where $\mathcal{I}^{(i)}$ is the indicator function for phase $i$:
\begin{equation}
    \mathcal{I}^{(i)}(\mathbf{x}) =
    \begin{cases}
    1, & \mathbf{x} \in \mathcal{V}_i\\
    0, & \textrm{else}.
    \end{cases}
\end{equation}
The function $S_n^{(i)}$ gives the probability of finding $n$ points at positions $\mathbf{x}_1,\dots,\mathbf{x}_n$ in phase $i$.
In what follows, we drop the superscript $i$, and restrict our discussion to the particle phase $\mathcal{V}_2$.

For statistically homogeneous media, $S_n(\mathbf{x}_1,\dots,\mathbf{x}_n)$ is translationally invariant and, in particular, the one-point correlation function is independent of position and equal to the packing fraction
\begin{equation}
    S_1(\mathbf{x})=\phi,
\end{equation}
while the two-point correlation function $S_2(\mathbf{r})$ depends on the displacement vector $\mathbf{r}\equiv\mathbf{x}_2-\mathbf{x}_1$.
The corresponding two-point autocovariance function $\chi_{_V}(\mathbf{r})$ \cite{RHM_Text,StochaticGeo_Text, Quint_Autoco} is obtained by subtracting the long-range behavior from $S_2(\mathbf{r})$:
\begin{equation}
    \chi_{_V}(\mathbf{r})=S_2(\mathbf{r})-\phi^2 
\end{equation}
The nonnegative \textit{spectral density} $\spD{\mathbf{Q}}$ is defined as the Fourier transform of $\chi_{_V}(\mathbf{r})$ \cite{RHM_Text}, i.e.,
\begin{equation}
    \spD{\mathbf{Q}} = \int_{\mathbb{R}^d}\chi_{_V}(\mathbf{r})e^{-i\mathbf{Q}\cdot\mathbf{r}}d\mathbf{r}.
\end{equation}
For a monodisperse packing of particles $\mathbf{\Omega}$ with arbitrary shape, it is known that \cite{Torquato_SD1, RHM_Text, Torquato_DisorderHUHet}
\begin{equation}
    \spD{\mathbf{Q}}=\rho|\tilde{m}(\mathbf{Q};\mathbf{R})|^2S(\mathbf{Q}),
\end{equation}
where $\mathbf{R}$ denotes the geometrical parameters of the particle shape and $\tilde{m}(\mathbf{Q};\mathbf{R})$ is the Fourier transform of the particle indicator function (form factor) defined as
\begin{equation}
    m(\mathbf{r};\mathbf{R}) =
    \begin{cases}
    1, & \mathbf{r} \textrm{ is in }\mathbf{R}\\
    0, & \textrm{otherwise,}
    \end{cases}
\end{equation}
where $\mathbf{r}$ is a vector measured with respect to the particle centroid.

For single finite configurations of $N$ identical hard particles under periodic boundary conditions, $\spD{\mathbf{Q}}$ can be expressed as \cite{Zachary_QLRI}
\begin{equation}
\begin{split}\label{eq:spD}
    \spD{\mathbf{Q}} = &\frac{\left|\sum_{j=1}^N\textrm{exp}(-i\mathbf{Q}\cdot\mathbf{r}_j)\tilde{m}(\mathbf{Q};\mathbf{R}_j)\right|^2}{V_F}\\
    &(\mathbf{Q}\neq0),    
\end{split}
\end{equation}
where $\{\mathbf{r}_\mathit{j}\}$ denotes the set of particle centroids and $\mathbf{R}_j$ denotes the $j^{th}$ particle.
Moreover, for such systems $\phi$ is given by
\begin{equation}
    \phi = \frac{Nv_1}{V_F},
\end{equation}
where $v_1$ is the volume of a single particle.

To our knowledge, $\tilde{m}(\textbf{Q};\textbf{R})$ has not been computed for superballs.
To circumvent this issue, we create a cubic voxelization of the packings using an efficient method (described in Sec \ref{sec:vox}) and apply Eq. (\ref{eq:spD}) to the result, requiring only $\tilde{m}(\mathbf{Q}; \mathbf{R})$ for the cube. 
We then compute the angular-averaged $\spD{Q}$ by applying Eq. (\ref{eq:spD}) to the voxelized packings.
Additionally, approximations of $\tilde{m}(\mathbf{Q}; \mathbf{R})$ for the superballs used in this work, computed using this voxelization procedure, are given in the Supplemental Material \cite{Supp}.

In this paper, we consider both the angular-averaged structure factor of the superball centroids and the angular-averaged spectral density.
The spectral density takes into account the structure and orientation of the particle volume, information, which is lost when only considering $S(\mathbf{Q})$.
Superballs (for $p\neq1$) are nonspherical, thus requiring the use of  $\spD{\mathbf{Q}}$ to fully characterize the density fluctuations in their packings.
A more detailed discussion of the importance of computing the spectral density is given in Sec. \ref{sec:specresults} and Refs. [\citenum{Zachary_QLRI, Zachary_QLRII, Zachary_QLRLet}].

A system has hyperuniform local-number-density fluctuations if as $Q\rightarrow0,\; S(Q)\rightarrow0$, while it has hyperuniform local-volume-fraction fluctuations if as $Q\rightarrow0,\;\spD{Q}\rightarrow0$ \cite{Torquato_HURev}.
Spectral densities that approach the origin with the power-law form given in Eq. (\ref{eq:powerlaw}) can be divided into three different classes based on their hyperuniformity scaling exponent $\alpha$ \cite{Torquato_DisorderHUHet}:
\begin{equation}\label{eq:classes}
    \spD{Q}\sim Q^\alpha
    \begin{cases}
    \alpha > 1,& \textrm{Class I}\\
    \alpha = 1,& \textrm{Class II}\\
    \alpha < 1,& \textrm{Class III}.
    \end{cases}
\end{equation}
Classes I and III are the strongest and weakest forms of hyperuniformity, respectively.
Such classes apply analogously to structure factors that approach the origin with a power-law form \cite{Torquato_HUDef, Zachary2009, Zachary_Skclasses}.
Moreover, a system is deemed to be ``effectively hyperuniform" (i.e., that a system is, for all intents and purposes, hyperuniform) when the hyperuniformity index $H$, defined as \cite{Hchi_cutoff},
\begin{equation}\label{eq:HX}
    H = \frac{\tilde{\chi}_{_V}(0)}{\spD{Q_{peak}}},
\end{equation}
where $\spD{Q_{peak}}$ is the largest peak of the spectral density, is less than about $10^{-2}$ \cite{Hchi_cutoff}.
Equation (\ref{eq:HX}) is exact in the infinite system limit, but a numerical extrapolation is required for numerical simulations due to finite system sizes.

\subsection{Pore-size distribution and transport properties}\label{sec:PSD}
\subsubsection{Pore-size distribution}
We characterize the void space in the superball packings by the distribution of their pore sizes $\delta$, i.e., the maximum radius of a spherical pore that can be assigned to a random point in the void space such that the pore lies entirely in the void space.
The probability density function $P(\delta)$ of the pore sizes, also known as the ``pore-size distribution" \cite{PRAGER_PSD, RHM_Text}, is normalized $\int_0^{\infty}P(\delta)d\delta=1$ and has units of inverse length.
For a randomly selected point in the void space, $P(\delta)d\delta$ is the probability that the shortest distance to the nearest void-particle interface is between $\delta$ and $\delta+d\delta$.

Equivalently, one can use the \textit{complementary cumulative pore-size distribution function} $F(\delta)$ defined as
\begin{equation}\label{eq:fdel}
    F(\delta)=\int^{\infty}_{\delta}P(r)dr,
\end{equation}
which can be interpreted as the fraction of void space that can admit a pore with a radius greater than $\delta$.
By definition, $F(0)=1$, $F(\infty)=0$, and $F(\delta)$ is unitless.
With $F(\delta)$, we can compute the \textit{mean pore size} $\langle\delta\rangle$ and the \textit{second moment} $\langle\delta^2\rangle$ of $P(\delta)$ using \cite{RHM_Text}:
\begin{equation}
    \langle\delta\rangle=\int_0^{\infty}F(\delta)d\delta,
\end{equation}
\begin{equation}
    \langle\delta^2\rangle=2\int_0^{\infty}F(\delta)\delta d\delta.
\end{equation}
These two quantities can be interpreted as characteristic length scales of the matrix phase and used to compute the transport properties of heterogeneous media \cite{RHM_Text}.

\subsubsection{Fluid permeability}
Darcy's law, which describes the slow flow of an incompressible viscous fluid through a porous medium, defines the \textit{fluid permeability} $k$, and can be rigorously derived via homogenization theory \cite{rubinstein_torquato_1989}.
The quantity $k$ has dimensions of (length)$^2$ and can be interpreted as an effective pore channel area of the dynamically connected part of the pore space \cite{RHM_Text}. 
Using the solutions of the unsteady Stokes equations for the fluid velocity vector field, Avellaneda and Torquato \cite{Avellaneda_fluid} derived the following relationship between $k$, the formation factor $\mathcal{F}$ of the porous medium, and a length scale $\mathcal{L}$ that is determined by the eigenvalues of the Stokes operator:
\begin{equation}\label{eq:kappa}
    k = \frac{\mathcal{L}^2}{\mathcal{F}},
\end{equation}
where $\mathcal{L}$ is a certain weighted sum of the viscous relaxation times $\Theta_n$ (i.e., inversely proportional to the eigenvalues of the Stokes operator), and $\mathcal{F}=\sigma_1/\sigma_e$ where $\sigma_e$ is the effective conductivity of a porous medium with a conducting fluid of conductivity $\sigma_1$ and a perfectly insulating solid phase.
Qualitatively, $\mathcal{F}$ quantifies the ``windiness" of the entire void space and is a monotonically decreasing function of the porosity \cite{TORQUATO_water}. 
Note, $\mathcal{L}$ in Eq. (\ref{eq:kappa}) absorbs a factor of 8 compared to the definition given in Ref. [\citenum{rubinstein_torquato_1989}], specifically, $\mathcal{L} = L/8$.

The theoretical prediction of $k$ is a difficult problem because it is nontrivial to estimate $\mathcal{L}$.
Recently, Torquato \cite{TORQUATO_water} proposed that, for well-connected pore spaces, $\mathcal{L}^2$ can be approximated by $\langle\delta^2\rangle$,
\begin{equation}\label{eq:kapprox}
    k \approx \frac{\langle\delta^2\rangle}{\mathcal{F}}
\end{equation}
which was verified by Torquato \cite{TORQUATO_water} for BCC, equilibrium, and stealthy sphere packings and by Klatt \textit{et al.} \cite{Klatt_Pore} for a number of other ordered and disordered sphere systems.
Additionally, to approximate $\mathcal{F}$, we use a tight lower bound derived by Torquato \cite{Torquato_F} for any porous medium in $\mathbb{R}^3$ that accounts for up to four-point information.
To an excellent approximation, the four-point parameter vanishes for a class of ordered and disordered dispersions of particles, thus yielding the accurate estimate
\begin{equation}\label{eq:F}
    \mathcal{F}\approx\frac{2+\phi-(1-\phi)\zeta_2}{(1-\phi)(2-\zeta_2)},
\end{equation}
where $\zeta_2\in(0,1]$ is a \textit{three-point microstructural parameter}, which is a weighted integral involving $S_i$ for $i = 1, 2, 3$.
This approximation is in excellent agreement with simulations of a variety of ordered and disordered sphere dispersions \cite{Torquato_F, Kim_F, ROBINSON_F, GILLMAN_FA, gillman_FB, Nguyen_F}.
In the $\zeta_2=0$ case, Eq. (\ref{eq:F}) reduces to the well-known Hashin-Shtrikman lower bound on $\mathcal{F}$ \cite{RHM_Text, HASHIN_F}.

\subsubsection{Mean survival time and principal diffusion relaxation time}
Another set of material descriptors concern the diffusion of a species through a pore space with a diffusion coefficient $\mathcal{D}$ that reacts at the pore-solid interface with a reaction rate $\kappa$.
The \textit{diffusion controlled-limit} is reached as $\kappa\rightarrow\infty$, while taking $\kappa\rightarrow 0$ corresponds to a perfectly reflective interface.
One related quantity of interest is the mean survival time $\tau$, or the average lifetime of the diffusing species before reacting with the interface.
Additionally, the principal relaxation time $T_1$ is associated with the time-dependent decay of an initially uniform concentration field of the diffusing particles \cite{RHM_Text}, and is also pertinent to the description of viscous flow in porous media \cite{Torquato_T1}.

Using variational principles, Torquato and Avellaneda \cite{Torquato_T1} derived the following upper bounds on $\tau$ and $T_1$ in terms of lower-order moments of the pore-size probability density function
\begin{equation}\label{eq:tau}
    \tau \leq \frac{\langle\delta\rangle^2}{\mathcal{D}}+\frac{(1-\phi)}{\kappa s},
\end{equation}
\begin{equation}\label{eq:T1}
    T_1 \leq \frac{\langle\delta^2\rangle}{\mathcal{D}} + \frac{3(1-\phi)\langle\delta\rangle^2}{4\kappa s\langle\delta^2\rangle},
\end{equation}
where $s$ is the specific surface of the medium.
In this paper, we only consider the diffusion-controlled limit.

\subsubsection{Spreadability}\label{sec:spread_def}
Recent work \cite{Torqauto_Spread} has revealed that the time-dependent \textit{spreadability} is a powerful new dynamic-based figure of merit to probe and classify the spectrum of possible microstructures of two-phase media across length scales.
Consider the time-dependent problem describing the mass transfer of a solute between two phases and assume that the solute is initially only present in one phase, specifically the particle phase, and both phases have the same $\mathcal{D}$.
The fraction of total solute present in the void space as a function of time $\mathcal{S}(t)$, is termed the \textit{spreadability} because it is a measure of the spreadability of diffusion information as a function of time.
Qualitatively, given two different two-phase systems at some time $t$, the one with a larger value of $\mathcal{S}(t)$ spreads diffusion information more rapidly.
Recently, Torqauto has shown that the \textit{excess spreadability} $\mathcal{S}(\infty)-\mathcal{S}(t)$ can be expressed in Fourier space in any dimension $d$ as \cite{Torqauto_Spread}: 
\begin{equation}\label{eq:FT_spread}
    \mathcal{S}(\infty)-\mathcal{S}(t) = \frac{d\omega_d}{(2\pi)^d \phi}\int^{\infty}_0 Q^{d-1}\spD{Q}\textrm{exp}[-Q^2\mathcal{D}t]dk,
\end{equation}
where $\omega_d$ is the volume of a $d$-dimensional unit sphere:
\begin{equation}
    \omega_d=\frac{\pi^{d/2}}{\Gamma(1+d/2)}.
\end{equation}

Consider the particular case of two-phase media with $\spD{Q}$ that obeys a power-law scaling in the $Q\rightarrow0$ limit:
\begin{equation}
    \lim_{Q\rightarrow 0} \spD{Q} = B|Q\ell|^{\alpha},
\end{equation}
where $B$ is a positive dimensionless constant, $\ell$ represents some characteristic microscopic length scale, and $\alpha\in(-d,\infty)$.
Such media can then be classified by their value of $\alpha$, in particular if $\alpha > 0$, the medium is hyperuniform; if $\alpha = 0$, it is a typical nonhyperuniform disordered medium; and if $\alpha < 0$, it is \textit{anti}hyperuniform, meaning $\spD{Q}$ diverges at the origin.
The long-time behavior of the excess spreadability for this class of media can be written as \cite{Torqauto_Spread},

\begin{equation}\label{eq:PLF}
    \mathcal{S}(\infty)-\mathcal{S}(t) \sim 1/t^{(d+\alpha)/2}
\end{equation}
Thus, hyperuniform two-phase media have a decay rate faster than $1/t^{d/2}$ at large $t$.
Here, we use Eq. (\ref{eq:FT_spread}) to compute $\mathcal{S}(\infty)-\mathcal{S}(t)$ for the superball packings.

\subsection{Mean width}\label{Sec:MW}
We will show that the mean width $\bar{w}$ is a useful means to make distances dimensionless when comparing superballs with different values of $p$.
The mean width $\bar{w}$ is a Minkowski functional with dimensions of length \cite{StochaticGeo_Text}.
Consider a convex $d$-dimensional body trapped between two parallel $(d-1)$-dimensional hyperplanes.
The ``width" of the convex body $w(\mathbf{n})$ in the direction $\mathbf{n}$ is the distance between the closest pair of parallel hyperplanes that do not intersect the body.
The average of $w(\mathbf{n})$ such that $\mathbf{n}$ is uniformly distributed over the unit (hyper)sphere in $\mathbb{R}^d$ is $\bar{w}$ \cite{Torquato_Perc1, StochaticGeo_Text}.
\begin{figure}[t]
    \centering
    \includegraphics[width = 0.45\textwidth]{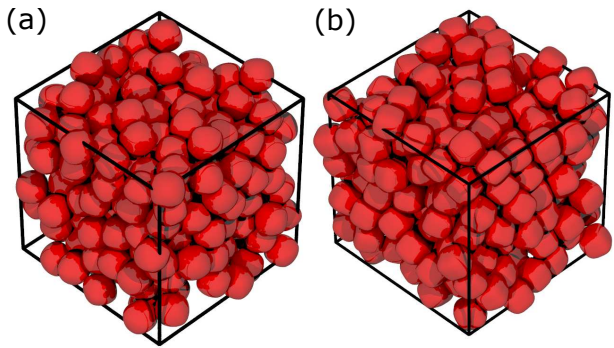}
    \caption{Illustrative configurations of MRJ packings of superballs with N = 250 for two different values of the deformation parameter: (a) $p$ = 0.85, (b) $p$ = 1.5.}
    \label{fig:ex_pack}
\end{figure}

\medskip
\medskip
\medskip

\section{Algorithmic and Experimental Details}\label{sec:Algo}
\subsection{Donev-Torquato-Stillinger algorithm}\label{sec:DTS}
We use an event-driven molecular dynamics simulation developed by Donev, Torquato, and Stillinger \cite{DTS_AlgI, DTS_AlgII} (henceforth referred to as the DTS algorithm) to generate MRJ packings of superballs.
This algorithm generalizes the Lubachevsky-Stillinger (LS) sphere packing algorithm \cite{LS_ALG} to accommodate other centrally symmetric convex particles, e.g, ellipsoids and superballs.
Initial conditions are produced by randomly distributing randomly oriented particles, without overlap, in a cubic periodic simulation box (fundamental cell).
In this paper, we choose $\phi_{initial} = 0.2$.
Particles are then assigned random translational and rotational velocities, and their motions are followed as they collide elastically and expand uniformly with an expansion rate $\gamma$.
Eventually, a jammed state with a diverging collision rate is reached, as is a local maximum in $\phi$.
To enforce randomness in our packings, we use an initially large expansion rate [$\gamma\in(0.05,0.005)$ up to a dimensionless pressure $P=10^6$] to avoid following the equilibrium branch of the phase diagram, which leads to crystallization. 
As the jamming point is approached, we decrease the expansion rate ($\gamma\sim0.001$) to ensure a truly jammed packing with a well-defined contact network is produced.
Previous work on spheres \cite{Donev_g2}, ellipsoids \cite{Donev_MRJelip}, and superballs \cite{Jiao_MRJballs} indicates that this is a reliable method to produce MRJ-like packings.
Illustrative examples of the packings produced are shown in Fig. \ref{fig:ex_pack}.
We find that using initial expansion rates larger than $\gamma=0.05$ causes the initial expansion step to terminate far from the jamming point.
This allows the configuration much more space to rearrange during the second, slower expansion step which tends to result in configurations with larger $\phi$ and $\bar{Z}$, which indicates they are not representative of the MRJ state.

In this paper, we study superballs with $p\in[0.85,1.5]$.
For $p < 0.85$ the resulting superballs are polyhedronlike, which causes numerical instability in the algorithm.
While previous work \cite{Jiao_MRJballs} uses values of $p$ up to 3, we find that our larger system sizes ($N = 5000$ compared to $N = 1000$) are unable to produce high-quality MRJ packings in a feasible amount of time.
For packings with $N=5000 , 1000$ the terminal pressure is $P=10^{14}$.
The results for $\phi$, $\bar{Z}$, and $\phi_R$ for $p\neq0.975,1,1.025, 1.05$ are averaged over 50 $N=5000$ configurations, and over 10 $N = 5000$ configurations for $p=0.975, 1.025, 1.05$.
For packings of spheres with $N > 1000$, the DTS algorithm is known to take an impractical amount of time to produce high-quality contact networks \cite{Donev_Unexpected}, so for the $\phi, \bar{Z},$ and $\phi_R$ of spheres we use 10 $N = 1000$ configurations.
Particles are considered to be in contact if they are less than a distance of $10^{-10}D$ apart, where $D$ is the length of one of the superball's major axes.
For $F(\delta)$, 1000000 pore sizes are computed in each of 10 $N = 5000$ configurations.
The results for $S(Q)$ and $\spD{Q}$ are averaged over 50 $N=5000$ configurations.

\subsection{Voxelization}\label{sec:vox}
Here, we present a procedure by which we voxelize a packing of superballs in a cubic fundamental cell, allowing us to easily compute certain properties of these packings, such as $\spD{Q}$ and $F(\delta)$.
To do so, we leverage the superellipsoid shape function given in Refs. [\citenum{donev_thesis, Donev_EllipFunct}].
The following assumes a monodisperse superball packing in $\mathbb{R}^3$, but an analogous procedure can be carried out for polydisperse packings, packings in $\mathbb{R}^2$, or packings of superellipsoids [so long as value of $p$ for each coordinate is equal, (cf. Eq. (\ref{eq:SB})], with appropriate modifications.

First, we divide a cube with the same dimensions as the simulation box into $M\times M\times M$ cubic voxels, where increasing $M$ results in a higher-resolution packing.
For each superball in the packing, we then carry out the following steps.
Let
\begin{equation}
    \mathbf{O}^{-1}=
    \begin{pmatrix}
    1/(R\epsilon) & 0 & 0 \\
    0 & 1/(R\epsilon) & 0 \\
    0 & 0 & 1/(R\epsilon) \\
    \end{pmatrix}
\end{equation}
be a matrix that describes a \textit{sphere} that has a radius $R$ equal to the major semiaxes of the superball and $\epsilon$ is a small parameter that scales the superballs such that $\phi$ of the voxelization better matches the true value of $\phi$ \cite{Coker_Digit}.
Then, using the location of the centroid in the simulation box, we find the corresponding voxel, which contains it, and check all voxels within a cubic neighborhood around this voxel such that the entire superball is contained within the neighborhood.
To determine if a particular voxel should be filled or not, we evaluate
\begin{equation}
    \xi(\mathbf{r}) = g\left[\tilde{\xi}(\mathbf{\tilde{r}})\right]-1
\end{equation}
where $\mathbf{\tilde{r}} = \mathbf{O}^{-1}\mathbf{Q}(\mathbf{r}-\mathbf{r}_0)$ is the relative position rotated and scaled according to the orientation and shape of the superball, where $\mathbf{Q}$ is a rotational matrix describing the orientation of the superball, $\mathbf{r}$ is the location of the voxel, and $\mathbf{r}_0$ is the location of the centroid of the superball.
For single exponent superballs, we have
\begin{equation}
    g(x) = x^{1/p}
\end{equation}
and
\begin{equation}
    \tilde{\xi}(\tilde{\mathbf{r}}) = \mathbf{e}^Tf(\tilde{\mathbf{r}})
\end{equation}
where $\mathbf{e}$ is $(1,1,1)$, and $f(x) = |x|^{2p}$.
If $\xi(\mathbf{r}) < 0$ then the superball overlaps the voxel and we set its value to 1 (filled), else we set its value to 0 (empty). 

To compute $\spD{Q}$, we apply Eq. (\ref{eq:spD}) to the result of the above procedure, which is effectively a packing of cubes.
For each of the $N = 5000$ packings used to compute $\spD{Q}$, we produce a voxelization with a resolution of $300\times300\times300$ voxels.
To compute $F(\delta)$, we use a procedure very similar to the one given in Ref. \citenum{Pore_Alg}.
In brief, we choose a random point in the void space of the packing, determine the voxel it lies in, and compute the distance to the nearest filled voxel.
To do this quickly, we precompute a list of vectors $\nu = (n,n,n)$, where $n\in\mathbb{N}_0$, in order of increasing magnitude, then check the voxels by iterating through this list and adding the vector $\nu$ to the index of the starting voxel until a filled voxel is found.
For each of the $N = 5000$ packings used to compute $F(\delta)$, we produce a voxelization with a resolution of $500\times500\times500$ voxels.

\begin{figure}[t!]
    \centering
    \includegraphics[width=0.45\textwidth]{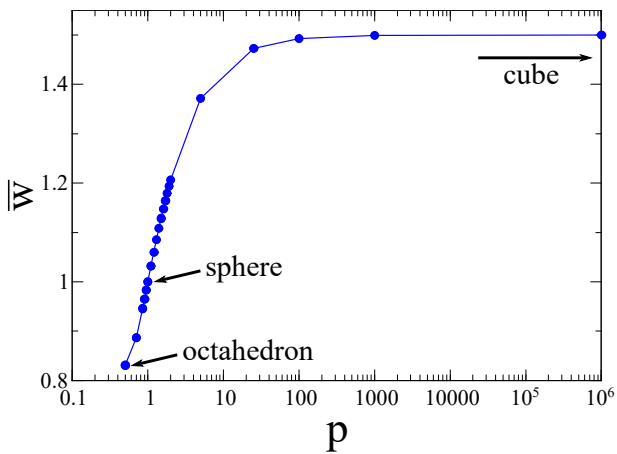}
    \caption{Mean width $\bar{w}$ [defined by Eq. (\ref{eq:w})] as a function of deformation parameter $p$ of a superball with unit-length axes, with attention called to the octahedron ($p = 0.5$), sphere ($p = 1.0$), and cube ($p \rightarrow \infty$) cases.}
    \label{fig:SB_MW}
\end{figure}

\subsection{Mean width details} \label{Sec:MWR}
To account for the differences in particle geometry as $p$ changes, we scale distances by the mean width $\bar{w}$.
To our knowledge, there is no known analytical formula for $\bar{w}$ of the superball.
Thus, we create a polygonal mesh to approximate the surface of the superball and use \cite{santalo}
\begin{equation}\label{eq:w}
    \bar{w} = \frac{1}{4\pi}\sum_il_i\theta_i,
\end{equation}
where $l_i$ is the length of the $i$th edge of the mesh, and $\theta_i$ is the angle between the normals of the faces, which meet at the $i$th edge, which exactly computes $\bar{w}$ for any convex polyhedron.
Figure \ref{fig:SB_MW} shows $\bar{w}$ for superballs over a wide range of $p$ using this approximation.
To demonstrate the effectiveness of this approach, we note that $\bar{w}$ can be computed exactly for octahedra, spheres, and cubes, which given a unit diameter are $\frac{3}{\pi\sqrt{2}}\textrm{arccos}(1/3)$, 1, and 3/2, respectively.
Using the approximation above, we find that the octahedron and sphere cases ($p=0.5, 1.0$) agree up to 4 decimal places, and that the nearly cubic superball ($p = 10^6$) value agrees up to 6 decimal places, all of which are slightly smaller than the expected value.
We note that when $Q$ is scaled by $\bar{w}$ the principal peaks of both $S(Q)$ and $\spD{Q}$ become very closely clustered (see Sec. \ref{sec:specresults}), indicating that this is a reasonable choice of scale.
Scaling by other, seemingly sensible, length scales, such as the major axis length of the superballs results in a larger distribution of principal peak positions, and as such are not the proper choice (see Supplemental Material \cite{Supp}).

\begin{figure}[t!]
    \centering
    \includegraphics[width = 0.45\textwidth]{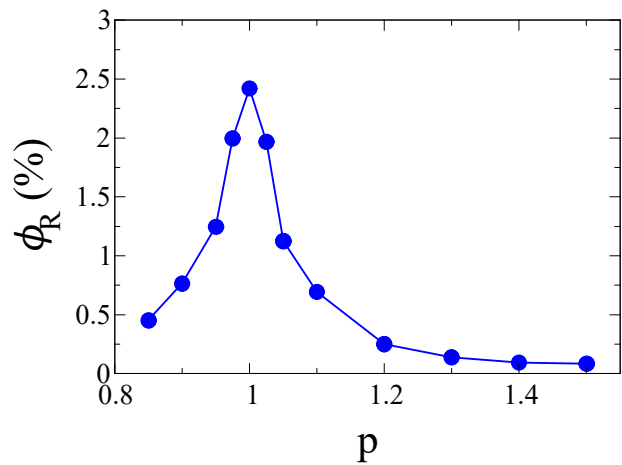}
    \caption{The percentage rattler fraction $\phi_R$ of MRJ superball packings as a function of deformation parameter $p$.}
\label{fig:ratfrac}
\end{figure}

\section{Structural Characteristics of MRJ Superball Packings}\label{sec:StructChar}
\subsection{Rattler fraction}
We carefully examine $\phi_R$ in superball packings, the results of which are shown in Fig. \ref{fig:ratfrac}.
Previously, it was stated that the rattler fraction decreases as $|1-p|$ increases, and nearly vanishes for $p > 2.75$ \cite{Jiao_MRJballs}.
Such a decrease in $\phi_R$ was also observed in ellipsoid packings as $a$ was increased \cite{Donev_MRJelip}.
The present findings are consistent with that notion.
We find that sphere packings have the largest $\phi_R$, which decreases rapidly and monotonically as $|1-p|$ increases.
We expect that $\phi_R$ will vanish in the $p\rightarrow\infty$ limit.
Thus, for the majority of values of $p$, the dense disordered superball packings cannot be regarded to be idealized MRJ states (for reasons noted in Introduction), but can be regarded to be good approximations of MRJ packings given the small concentration of rattlers.
The introduction of rotational degrees of freedom results in a decrease in $\phi_R$ due to the increased number of contacts required for jamming, as well as an increased difficulty in forming the isotropic cages needed to house rattlers \cite{Jiao_MRJballs}.

\begin{figure}[t!]
    \centering
    \includegraphics[width = 0.45\textwidth]{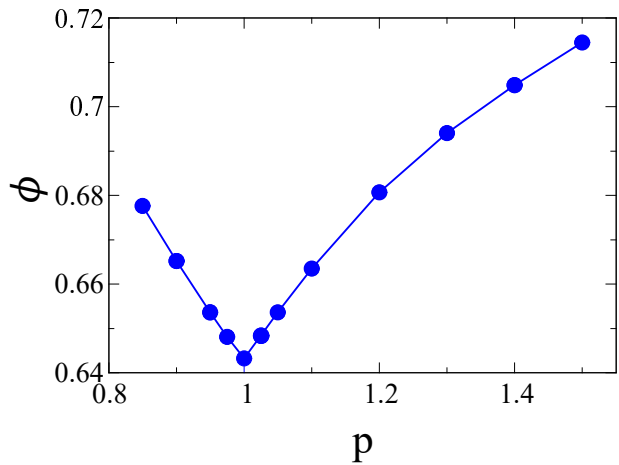}
    \caption{The packing fraction $\phi$ of MRJ superball packings as a function of deformation parameter $p$.}
    \label{fig:packfrac}
\end{figure}

\subsection{Packing fraction}
Figure \ref{fig:packfrac} shows $\phi$ for MRJ superball packings as a function of $p$.
The sharp, nonanalytic increase we observe as $|1-p|$ increases is due chiefly to the breaking of the spherical symmetry of the particle.
Non-spherical particles can more efficiently cover space by orienting themselves such that their protuberances occupy open spaces that spheres would be unable to.
In previous work \cite{Jiao_MRJballs}, the closest values of $p$ to the sphere point studied were $p=0.95$ and 1.10.
To better characterize the cusp at the sphere point, we produce packings with $p = 0.975, 1.025,$ and 1.05, and use a linear regression to fit this data.
We find that the slope for $p < 1$ is -0.2084 (coefficient of determination $R^2=0.9984$) and 0.207 ($R^2 = 1$) for $p > 1$, showing that the increase in $\phi$ as $|1-p|$ increases is very nearly linear.
The qualitative behavior of $\phi$ observed here is otherwise consistent with previous findings \cite{Jiao_MRJballs}.

\begin{figure}[t!]
    \centering
    \includegraphics[width = 0.45\textwidth]{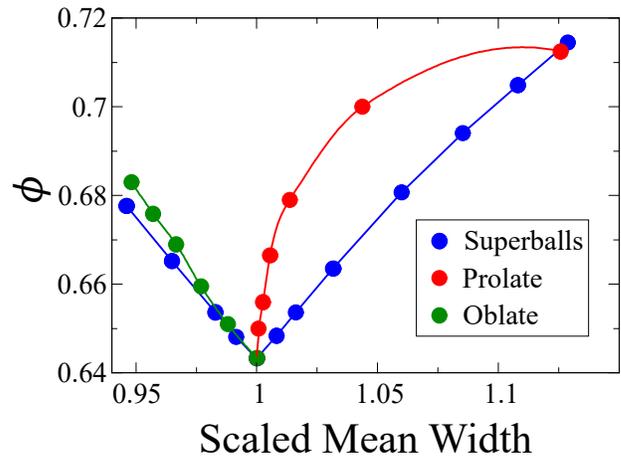}
    \caption{Packing fraction $\phi$ for superballs, oblate, and prolate spheroids \cite{Donev_EllipFunct} as a function of their scaled mean widths. Superball mean widths are scaled by the length of their major axes, while the spheroid mean widths are scaled by the length of their two equivalent major axes. The ratios of the semiaxes for the prolate spheroids are 1:1:$a$, and 1:$a$:$a$ for oblate spheroids, where $a$ is the aspect ratio.}
    \label{fig:phicomp}
\end{figure}

In Fig. \ref{fig:phicomp} we compare $\phi$ for MRJ packings of superballs for $0.85 \leq p \leq 1.50$ and prolate and oblate spheroids as a function of the scaled mean width (see Ref. \citenum{Torquato_Perc1} for spheroid mean width formulas).
Notably, for superballs, $\phi$ varies linearly (slope of -0.6402, $R^2 > 0.99$ for $p \leq 1$, slope of 0.5579, $R^2>0.99$ for $p \geq 1$) over the entire range of mean widths considered.
The packing fraction of MRJ oblate spheroid packings also increases roughly linearly, and more rapidly than superballs, as the scaled mean width decreases.
By contrast, $\phi$ for MRJ prolate spheroid packings increases quickly and then begins to plateau as the scaled mean width increases because of the increased effect of the anisotropic exclusion volume of such deformed spheroids \cite{Donev_MRJelip}.
For the range of scaled mean widths considered in Fig. \ref{fig:phicomp}, the spheroids have larger $\phi$ because they require more contacts per particle than superballs to achieve mechanical stability, which requires a denser packing of particles \cite{Donev_MRJelip}.

\begin{figure}[t!]
    \centering
    \includegraphics[width = 0.45\textwidth]{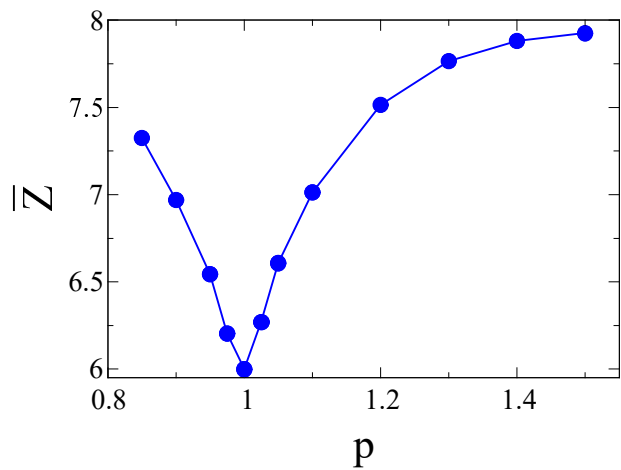}
    \caption{The average contact number $\bar{Z}$ of MRJ superball packings as a function of deformation parameter $p$.}
\label{fig:Z_av}
\end{figure}

\begin{figure}[t!]
    \centering
    \includegraphics[width = 0.41\textwidth]{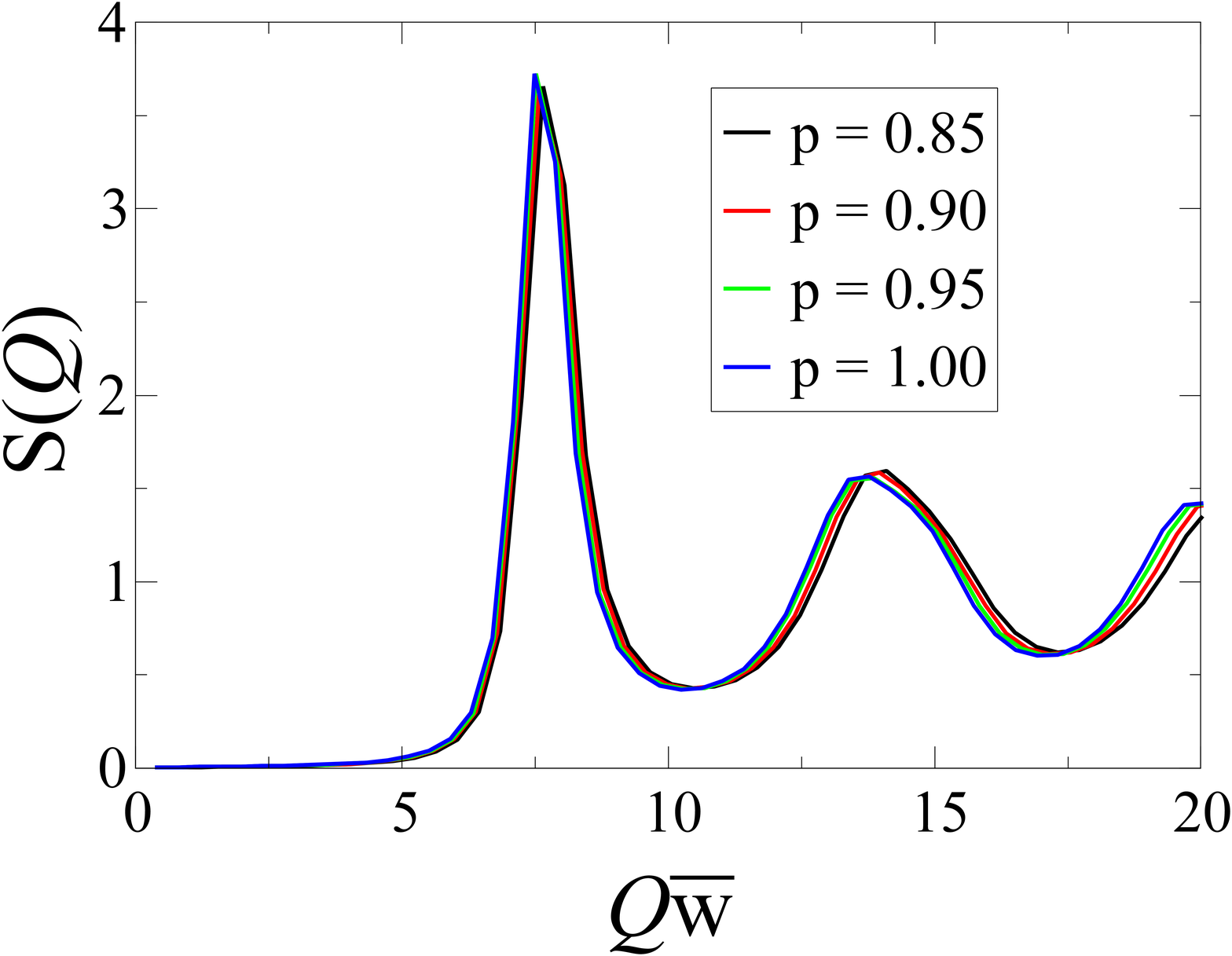}
    \caption{The structure factor $S(Q)$ [defined by Eq. (\ref{eq:ScattInt})] as a function of wave number $Q$ scaled by the mean width $\bar{w}$ for values of the deformation parameter $p \leq 1$. }
\label{fig:Sk_less1}
\end{figure}
\begin{figure}[t!]
    \centering
    \includegraphics[width = 0.41\textwidth]{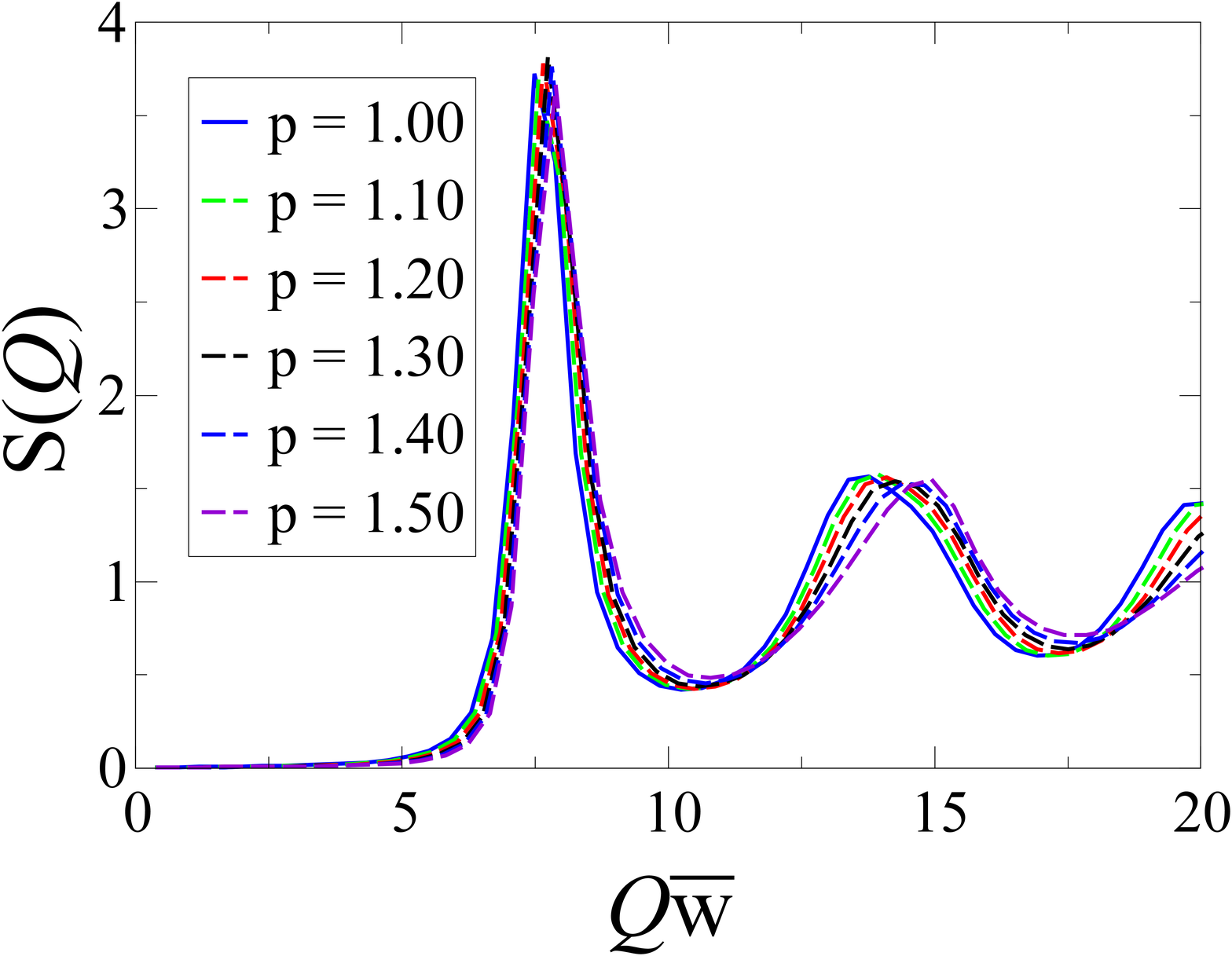}
    \caption{The structure factor $S(Q)$ [defined by Eq. (\ref{eq:ScattInt})] as a function of wave number $Q$ scaled by the mean width $\bar{w}$ for values of the deformation parameter $p \geq 1$. }
\label{fig:Sk_great1}
\end{figure}

\subsection{Average contact number}
Much like with $\phi$ and $\phi_R$, $\bar{Z}$ also exhibits the characteristic ``cusp" at $p=1$, but unlike $\phi$, does not have linear growth in its vicinity.
Figure \ref{fig:Z_av} shows $\bar{Z}$ as a function of $p$ in the MRJ superball packings.
Note that rattlers are ignored when computing $\bar{Z}$.
The values of $\bar{Z}$ increase sharply for small values of $|1-p|$, then begin to plateau for large values of $p$.
This sharp increase occurs because additional contacts are required to constrain the new rotational degrees of freedom that arise when the spherical symmetry is broken.
For $p = 1$ the packings are \textit{exactly} isostatic, as expected \cite{Donev_g2}, while the remainder of the packings are \textit{highly} hypostatic (specifically, $\bar{Z}<12$).
In particular, we find that $\bar{Z}\approx7.9$ (cf. Ref. [\citenum{Jiao_MRJballs}]) for sufficiently large $p$.
By contrast, for large $a$, prolate spheroids plateau at $\bar{Z}\approx9.9$, while other, more asymmetric, ellipsoids plateau at $\bar{Z}\approx11.75$ \cite{Donev_MRJelip}.
The qualitative behavior of $\bar{Z}$ is consistent with previous findings \cite{Jiao_MRJballs}.
Notably, when $\phi_R$ begins to plateau at $p\approx1.4$, so too does $\bar{Z}$.

\begin{figure}[t!]
    \centering
    \includegraphics[width = 0.47\textwidth]{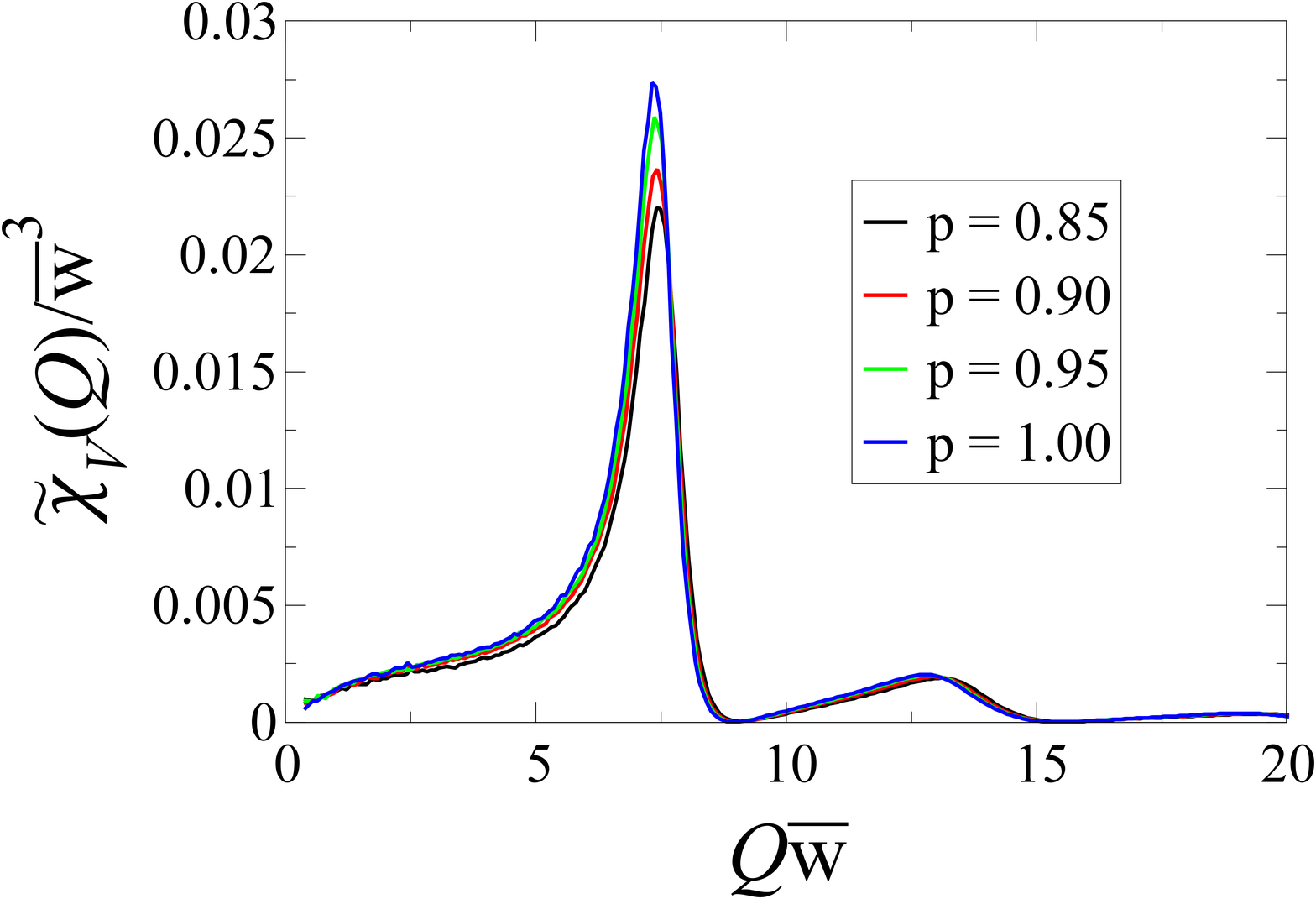}
    \caption{The scaled spectral density $\spD{Q}/\bar{w}^3$ [defined by Eq. (\ref{eq:spD})] as a function of wave number $Q$ scaled by the mean width $\bar{w}$ for values of the deformation parameter $p \leq 1$. }
\label{fig:Xk_less1}
\end{figure}
\begin{figure}[t!]
    \centering
    \includegraphics[width = 0.47\textwidth]{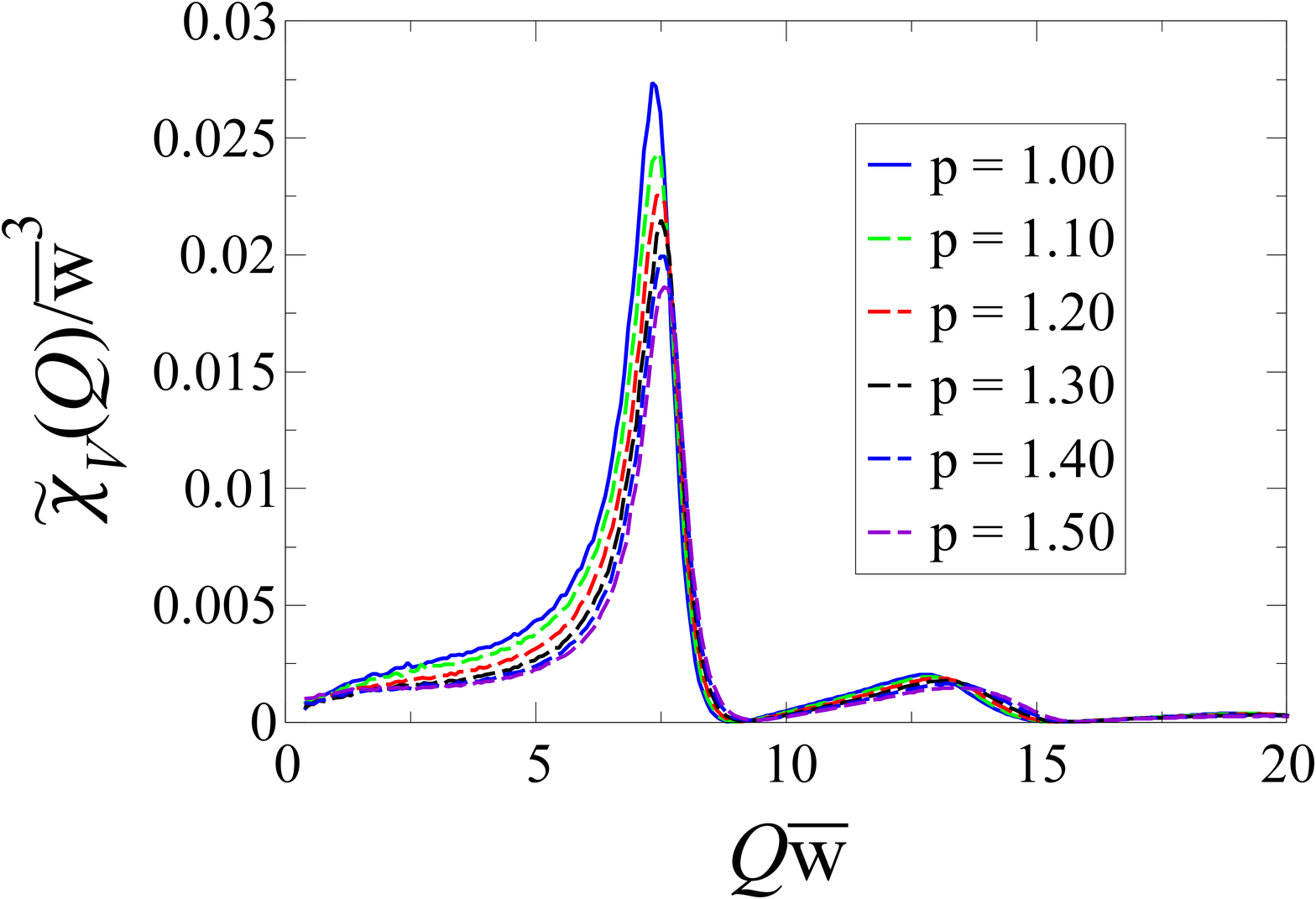}
    \caption{The scaled spectral density $\spD{Q}/\bar{w}^3$ [defined by Eq. (\ref{eq:spD})] as a function of wave number $Q$ scaled by the mean width $\bar{w}$ for values of the deformation parameter $p \geq 1$. }
\label{fig:Xk_great1}
\end{figure}

\subsection{Effective hyperuniformity of MRJ superball packings}\label{sec:specresults}

Zachary \textit{et al.} have shown that the spectral density $\spD{Q}$ needs to be used to fully characterize the density fluctuations for packings of polydisperse and/or nonspherical particles \cite{Zachary_QLRLet, Zachary_QLRI, Zachary_QLRII}.
The structure and orientation of the particle volume are accounted for in $\spD{Q}$, while $S(Q)$ only considers the particle positions.
Thus, because superballs lack spherical symmetry (for $p\neq1$), we must consider $\spD{Q}$.
To examine the shape effects of the slightly aspherical superballs (i.e., those with $p$ very close to unity), we also consider $S(Q)$.
These characterizations allow us to ascertain the degree of hyperuniformity in these superball packings.
Notably, our paper appropriately utilizes the spectral density $\spD{Q}$ to ascertain whether an MRJ packing of nonspherical particles is effectively hyperuniform for the first time in $\mathbb{R}^3$.
Previous studies considering the hyperuniformity of MRJ packings of nonspherical particles in $\mathbb{R}^3$, (e.g., Ref. [\citenum{Jiao_MRJplatonic}]) consider only $S(Q)$.

Figures \ref{fig:Sk_less1} and \ref{fig:Sk_great1} show $S(Q)$ for the centroids of MRJ superball packings with $p \leq 1$ and $p \geq 1$, respectively.
By scaling $Q$ by $\bar{w}$, we find that the principal peaks all collapse to nearly the same position.
Moreover, as $|1-p|$ increases, peak positions are pushed to larger values of $Q\bar{w}$.
Second and subsequent peak heights also begin to see attenuation, which increases in magnitude as $|p-1|$ increases.

\begin{table}[t!]
\caption{The fit parameters $a_0$ and $a_1$ obtained by fitting Eq. (\ref{eq:fit}) to  the small-$Q$ values of $\spD{Q}$ ($Q\bar{w}\lesssim1.4$) with a fixed hyperuniformity scaling exponent $\alpha$ computed via the the excess spreadability [cf. Eq. (\ref{eq:FT_spread})], and the corresponding value of the hyperuniformity index $H$ for each value of the deformation parameter $p$ considered.}
\label{tab:fit}
\begin{tabular}{|l|l|l|l|l|}
\hline
$p$    & $a_0\times10^{-4}$ & $a_1\times10^{-3}$ & $\alpha$ & $H\times10^{-3}$ \\ \hline
0.85 & 1.74             & 1.16             & 0.540  & 7.94           \\ \hline
0.90 & 1.18             & 1.24             & 0.618  & 5.00           \\ \hline
0.95 & 1.48             & 1.24             & 0.624  & 5.74           \\ \hline
1.00 & 0.26             & 0.14             & 0.640  & 0.96           \\ \hline
1.10 & 2.00             & 1.11             & 0.600  & 8.22           \\ \hline
1.20 & 0.97             & 1.08             & 0.540  & 4.29           \\ \hline
1.30 & 1.68             & 0.87             & 0.500  & 7.81           \\ \hline
1.40 & 0.99             & 1.01             & 0.380  & 4.98           \\ \hline
1.50 & 2.10             & 1.06             & 0.320  & 11.30          \\ \hline
\end{tabular}
\end{table}
\begin{figure}[t!]
    \centering
    \includegraphics[width = 0.45\textwidth]{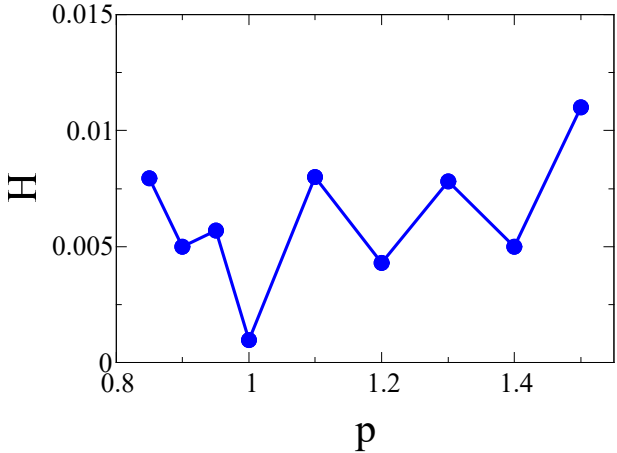}
    \caption{The $H$ index [defined by Eq. (\ref{eq:HX})] as a function of the deformation parameter $p$ computed from the spectral density $\spD{Q}$.}
\label{fig:H_Xk}
\end{figure}

While the centroids in MRJ sphere packings are known to be hyperuniform \cite{Donev_Unexpected}, this is not generally true of packings of particles with anisotropy or polydispersity \cite{Zachary_QLRI, Zachary_QLRII, Zachary_QLRLet}.
Due to the striking similarities between $S(Q)$ for spheres and all other superballs considered here, it is reasonable to conclude that superballs behave like \textit{effective spheres} inscribed within the superballs.
This sphere-like behavior occurs because there is sufficient orientational disorder in the superball packings.
This distribution of orientations averages out local inhomogeneities in the spatial distribution of particle centroids caused by the particle anisotropy on large scales.

Figures \ref{fig:Xk_less1} and \ref{fig:Xk_great1} show $\spD{Q}$ for MRJ superball packings with $p \leq 1$ and $p \geq 1$, respectively.
As above, scaling $Q$ by $\bar{w}$ results in the principal peaks clustering tightly.
Likewise, minor peaks are shifted to larger values of $Q\bar{w}$ and have their heights attenuated as $|1-p|$ increases.
The significant attenuation of the principal peak heights in $\spD{Q}$ as $|1-p|$ increases compared to those in $S(Q)$ is a result of $\tilde{m}(Q;\mathbf{R})$ [cf. Eq. (\ref{eq:spD})].

\begin{figure}[b!]
    \centering
    \includegraphics[width = 0.45\textwidth]{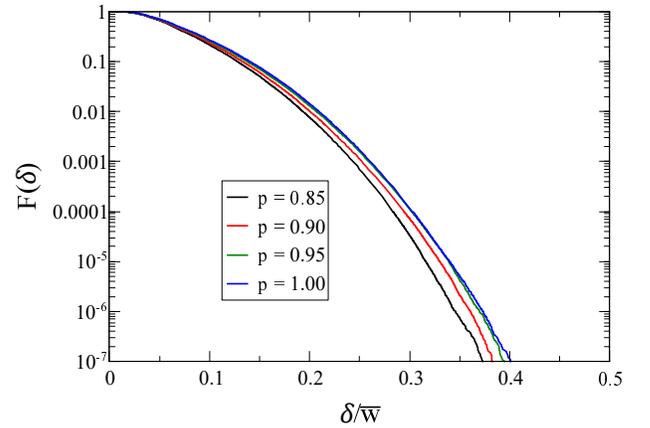}
    \caption{The complementary cumulative pore-size distribution function $F(\delta)$ [defined by Eq. (\ref{eq:fdel})] as a function of pore radius $\delta$ scaled by the mean width $\bar{w}$ for values of the deformation parameter $p \leq 1$.}
\label{fig:PSD_smallp}
\end{figure}
\begin{figure}[b!]
    \centering
    \includegraphics[width = 0.45\textwidth]{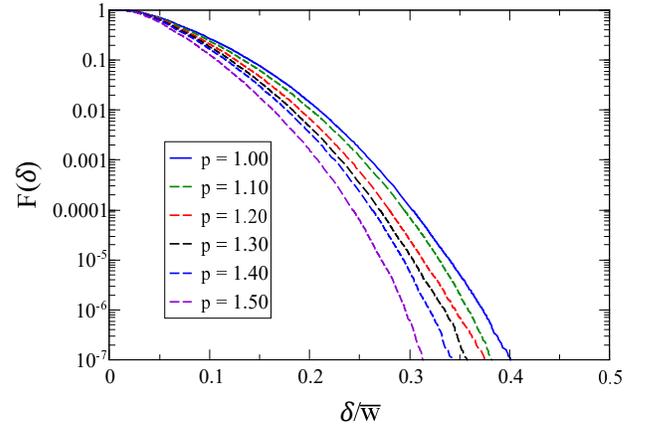}
    \caption{The complementary cumulative pore-size distribution function $F(\delta)$ [defined by Eq. (\ref{eq:fdel})] as a function of pore radius $\delta$ scaled by the mean width $\bar{w}$ for values of the deformation parameter $p \geq 1$.}
\label{fig:PSD_largep}
\end{figure}

To estimate $\spD{0}$ to obtain $H$ defined by Eq. (\ref{eq:HX}) we must fit the small-$Q$ region of $\spD{Q}$ (i.e., $Q\bar{w} \lesssim 1.4$) with
\begin{equation}\label{eq:fit}
    \spD{Q} = a_0 + a_1Q^{\alpha},
\end{equation}
where $a_0 = \spD{0}$ and $a_1$ are fit parameters and $\alpha$ is the hyperuniformity scaling exponent.
As a result of the voxelization procedure and small system size, the small-$Q$ values of the spectral density are noisy.
Thus, a direct numerical fit using Eq. (\ref{eq:fit}) with the triplet $a_0, a_1$, and $\alpha$ as free parameters has a strong dependence on the range of $Q$ values to which the fit is applied, resulting in a large range of values for both $a_0$ and $\alpha$.
For example, for $p = 1.50$, by varying the fit window between $Q\bar{w}\in(0.415, 1.16)$ and $Q\bar{w}\in(0.415,2.07)$ we find $\alpha\in(0.244,0.711)$ and $a_0\in(3.82\times10^{-8},6.94\times10^{-4})$.
Computing the excess spreadability (see Sec. \ref{sec:spread_Results} for additional details) is a robust and accurate way to find $\alpha$.
To reduce the variability in our results for $H$ when fitting $\spD{Q}$ for a given value of $p$, we fix $\alpha$ in Eq. (\ref{eq:fit}) to be the value computed via the excess spreadability.
Over the same range of fits above for $p=1.50$, $a_0\in(2.10\times10^{-4},3.64\times10^{-4})$ when $\alpha$ is fixed to be the value found via the excess spreadability.
Table \ref{tab:fit} contains the fit parameters $a_0$ and $a_1$, the fixed value of $\alpha$ from the excess spreadability calculation used in the fit, and the corresponding value of $H$ for each $p$ value examined.
Figure \ref{fig:H_Xk} shows $H$ as a function of $p$ computed by using Eq. (\ref{eq:HX}).
We find the values of $H$ are on the order of, or less than, the effective hyperuniformity threshold (10$^{-2}$), thus we consider the packings to be effectively hyperuniform, as one would expect for MRJ packings of anisotropic particles.

\section{Effective Properties of MRJ Superball Packings}\label{sec:Props}
\subsection{Pore-size distribution function and transport properties}
Figures \ref{fig:PSD_smallp} and \ref{fig:PSD_largep} show $F(\delta)$ for MRJ superball packings with $p \leq 1$ and $p \geq 1$, respectively.
The maximum value of $\delta$ in these packings is bounded and less than $\bar{w}/2$, indicating that each packing is saturated.
Again, the $p=1$ point represents an extreme value; the MRJ packings of such superballs have the largest pore sizes, reflective of sphere packings having the lowest $\phi$ (largest volume of void space).
As $p$ diverges from unity (with a concomitant increase in $\phi$), we find that the pore sizes tend to become smaller.

\begin{figure}[t!]
    \centering
    \includegraphics[width = 0.48\textwidth]{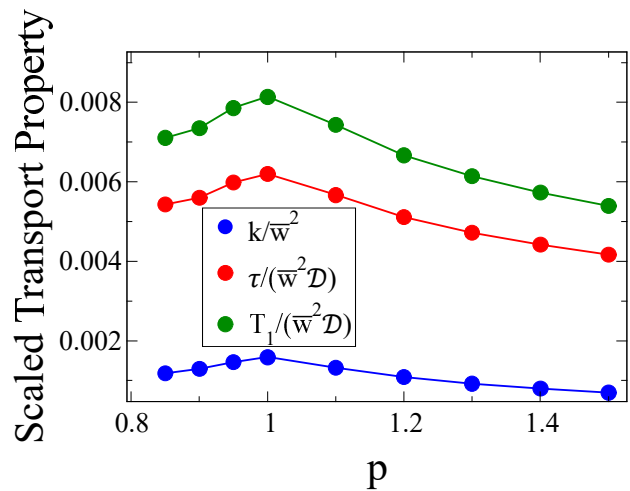}
    \caption{The fluid permeability $k$ [Eq. (\ref{eq:kappa})] scaled by the mean width squared $\bar{w}^2$, mean survival time $\tau$ [Eq. (\ref{eq:tau})] scaled by the diffusion coefficient $\mathcal{D}$ and $\bar{w}^2$, and principal diffusion relaxation time $T_1$ [Eq. (\ref{eq:T1})] scaled by $\mathcal{D}$ and $\bar{w}^2$ of the MRJ superball packings as a function of the deformation parameter $p$.}
    \label{fig:transprop}
\end{figure}

Using the equations in Sec. \ref{sec:PSD}, we approximate $k$, and compute upper bounds on $\tau$, and $T_1$ in the diffusion-controlled limit (see Fig. \ref{fig:transprop}) for a range of $p$ values.
To approximate $\mathcal{F}$ in Eq. (\ref{eq:kapprox}), we use the following formula for $\zeta_2$ based on an interpolation analysis on data involving packings of spheres and packings of cubes from Ref. [\citenum{RHM_Text}]:
\begin{equation}
    \zeta_2=0.21068\phi+|p-1|(0.11882+0.772236\phi).
\end{equation}
We find that our predictions of $\tau$ and $T_1$ for MRJ sphere packings are slightly larger than (but of the same order as) previous results, which we attribute to the voxelization procedure slightly overestimating the pore sizes, while $k$ falls within the previously computed bounds \cite{Klatt_MRJIII}.
The transport properties have a maximum at $p = 1$ (where $\phi$ is at a minimum), and decrease as $|1-p|$ increases, which is consistent with sphere packings having larger pore sizes than spherically asymmetric superballs.

\begin{figure}[t!]
    \centering
    \includegraphics[width = 0.45\textwidth]{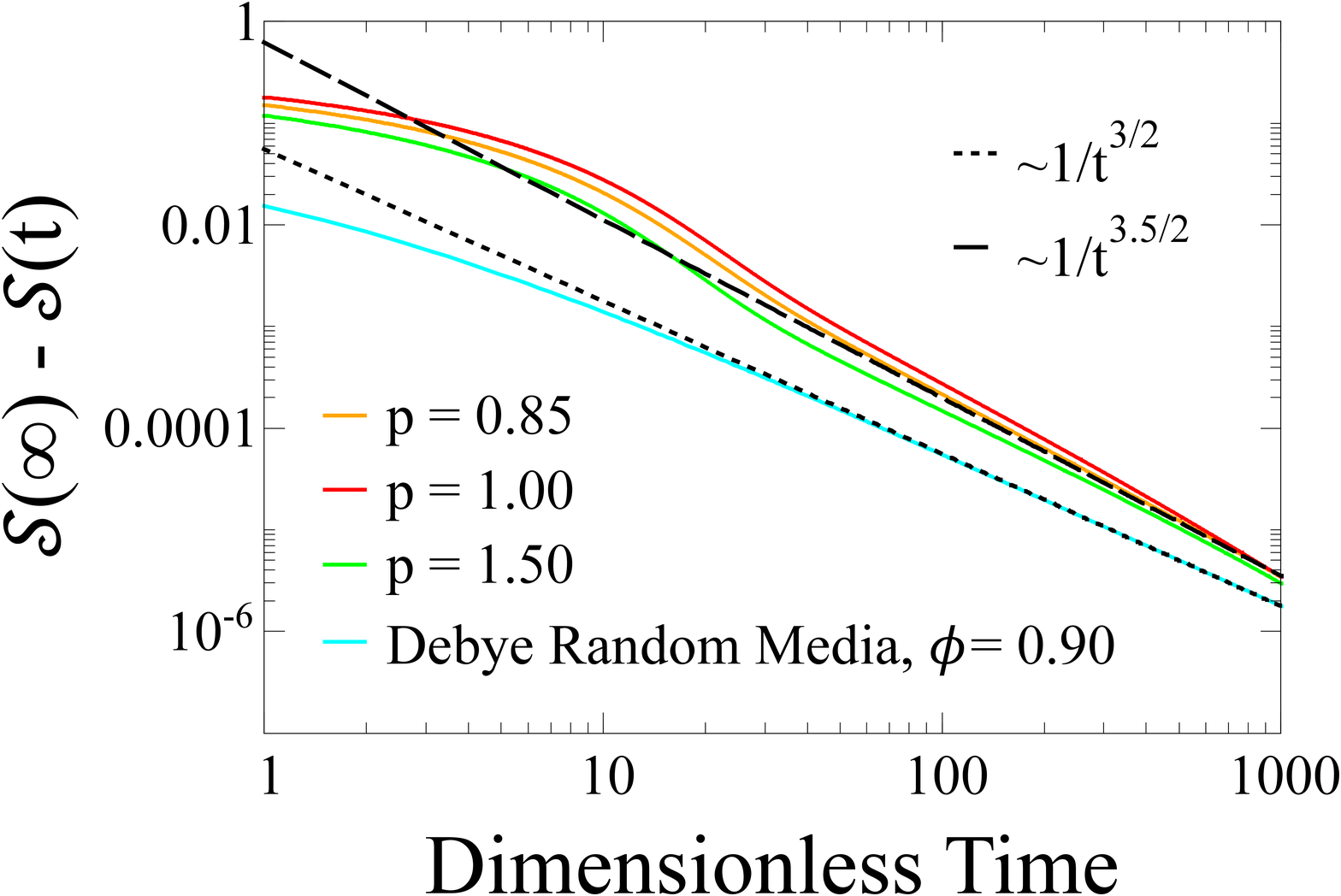}
    \caption{The excess spreadability $\mathcal{S}(\infty)-\mathcal{S}(t)$ [defined by Eq. (\ref{eq:FT_spread})] as a function of dimensionless time for the MRJ superballs packings, as well as for nonhyperuniform Debye random media, for comparison. The dimensionless times are $t\mathcal{D}/\hat{a}^2$ for Debye random media, where $\hat{a}$ is the length scale given in Eq. (\ref{eq:deb}), and $t\mathcal{D}/\bar{w}^2$ for superball packings. The dashed lines are eye guides to show the long-time scaling behavior.}
    \label{fig:spread}
\end{figure}

\subsection{Spreadability}\label{sec:spread_Results}
In three dimensions, if the long-time excess spreadability of a medium decays more quickly than $t^{-3/2}$, then the medium is hyperuniform [cf. Eq. (\ref{eq:FT_spread})] \cite{Torqauto_Spread}.
To contrast, consider a typical disordered nonhyperuniform medium such as Debye random media \cite{Debye_Scattering}, whose spectral density in three dimensions is given by \cite{Torqauto_Spread}
\begin{equation}\label{eq:deb}
   \spD{k}=\frac{\phi(1-\phi)\pi a^3}{(1+(k\hat{a})^2)^2},
\end{equation}
where $\hat{a}$ is the characteristic length scale of the medium, with long-time excess spreadability scaling of exactly $t^{-3/2}$.
Figure \ref{fig:spread} shows the excess spreadability for MRJ superball packings with $p = 0.85,1.00,1.50$, and Debye random media for comparison.
These packings, for all values of $p$ considered, have long-time scaling exponents between -1.66 and -1.82.
The spreadability decay tends to become slower as $|1-p|$ increases, indicating packings of less spherical superballs are more weakly hyperuniform.

\begin{figure}[t!]
    \centering
    \includegraphics[width = 0.42\textwidth]{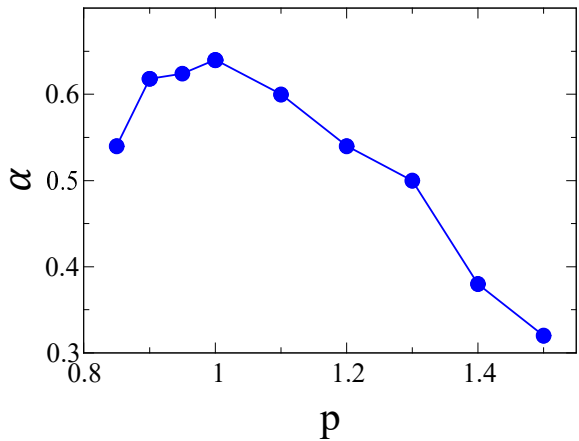}
    \caption{The hyperuniformity scaling exponent $\alpha$ determined via fitting the long-time behavior of Eq. (\ref{eq:FT_spread}) to Eq. (\ref{eq:PLF}) as a function of $p$.}
    \label{fig:alpha}
\end{figure}

The corresponding values of $\alpha$ based on these scaling exponents are given in Fig. \ref{fig:alpha}.
We find the MRJ superball packings belong to Class III [see Eq. (\ref{eq:classes})] for all values of $p$ considered.
In the special case of the sphere ($p=1$), a recent numerical study of randomly close packed spheres \cite{Wilken_ClassIII} also reports that such packings are of Class III, but with a value of $\alpha=0.24$, extracted from the structure factor, that is substantially smaller than for our MRJ sphere packings.
It is not surprising that the spreadability offers a robust and accurate method to compute $\alpha$ that is less susceptible to pointwise variations in $\spD{Q}$ itself, since the spreadability can be regarded to be a Gaussian smoothing of the spectral density as derived in Ref. [\citenum{Torqauto_Spread}].
Due to finite system sizes, the small-$Q$ region of $\spD{Q}$ tends to have low resolution and is prone to being noisy, which can lead to large variability in the resulting numerical fits.
Computing the excess spreadability does not rely on these potentially dubious extrapolations, and as such is more robust against noise than other measures of $\alpha$.
One can compare these results to those from Ref. [\citenum{Torqauto_Spread}], which demonstrates that these packings are ``less hyperuniform" than the stealthy hyperuniform systems considered therein, which is consistent with the classification of these hyperuniform systems in previous work (see, e.g. Table 1 in Ref. [\citenum{Torquato_HURev}]).

\section{Conclusions}\label{sec:Conc}

In this paper, we examined superballs, a family of centrally symmetric shapes defined by Eq. (\ref{eq:SB}), that take on both cube-like ($p>1$) and octahedron-like shapes ($p < 1$).
Using the DTS algorithm, we produced large MRJ packings of superballs with $p\in[0.85,1.50]$.
We also generated voxelized versions of these packings using an efficient method to aid in the computation of $\spD{Q}$ and $F({\delta})$.
To characterize these packings, we computed $\phi$, $\bar{Z}$, and $\phi_R$.
We also calculated the mean width $\bar{w}$, and find that it is a useful length scale to make distances dimensionless to compare superballs of different shape.
In particular, $\bar{w}$ is a better choice of scale than other, seemingly sensible, choices of scale, like the length of the major axes of the superball.
In addition, we determined $S(Q)$ of the superball centroids and $\spD{Q}$ of the voxelized packings and examined the small-$Q$ behavior to characterize the large-scale properties.
We also computed $F(\delta)$ and the transport properties $k$, $\tau$, and $T_1$, as well as the excess spreadability $\mathcal{S}(\infty)-\mathcal{S}(t)$.
Novel experimental techniques have allowed for the synthesis of colloidal particles with superball-like shapes \cite{Rossi_SBcolloid, Zhang_SBass, meijer_SBsolidsolid}. 
Thus, careful characterization of simulated packings can be helpful in the design of real colloidal materials fabricated via these methods. 

To build on previous work in this area \cite{Jiao_MRJballs}, we more closely characterized the nonanalytic ``cusp" in $\phi$ at $p = 1$ and find that $\phi$ increases nearly linearly on either side of the sphere point as $|1-p|$ increases for sufficiently small $|1-p|$.
The notion that $p=1$ is an extreme point persists in each of the subsequent characterizations of the packings, although the linear scaling does not.
We additionally determined $\phi_R$  as a function of $p$ and find that it rapidly decreases as $|1-p|$ increases.
Notably, as $\bar{Z}$ begins to plateau at $p\approx1.4$, so too does $\phi_R$.
Because $\phi_R$ monotonically decreases, we expect it to vanish in the $p\rightarrow\infty$ limit.

In the present work, we have used $\spD{Q}$ to assess the hyperuniformity of packings of nonspherical particles in $\mathbb{R}^3$.
We found that the MRJ superball packings are effectively hyperuniform with respect to $\spD{Q},$ which is the appropriate spectral measure for nonspherical particles.
Further, due to the striking similarities between $S(Q)$ of spheres and those of the nonspherical superballs, we conclude that superballs behave like \textit{effective spheres} inscribed within the superballs.

Moreover, we found that the pore sizes in MRJ superball packings tend to become smaller away from the sphere point (i.e., as $\phi$ increases).
The maximum pore sizes in these packings are also bounded and less than the semiaxes of the superballs, indicating that the packings are saturated.
The fluid permeability $k$, mean survival time $\tau$, and principal diffusion relaxation time $T_1$ all have a maximum at $p = 1$ and decrease as $|1-p|$ increases. 
Additionally, the long-time excess spreadability indicates that these packings are hyperuniform with $\alpha\in(0.32,0.68)$ that decreases as $|1-p|$ increases.
Use of the spreadability to compute $\alpha$ is robust to noise, and as such a reliable way to compute $\alpha$ for numerically- or experimentally-generated two-phase media.

Due to stability issues in the DTS algorithm, we are presently unable to simulate the behavior of superballs with $p > 3.0$ (cube limit) or $p < 0.85$ (octahedral limit, concave superballs). 
While the the MRJ state of octahedra has been studied using other methods (see Refs. [\citenum{Jiao_MRJplatonic, Torquato_ASCNat}]), the MRJ state of cubes is still undiscovered.
Further, very little is known about the behavior of hard concave superballs, e.g., only predictions for the densest packings of concave superballs are known \cite{Jaio_OptSB}, and disordered packings have only been examined via RSA \cite{Concave_SB}.
It is also of great interest to more carefully examine the structure of the contact network beyond $\bar{Z}$ and its relationship with the particle shape.
For example, similar to the analysis carried out in Ref. [\citenum{O'Hern_hypos}], one can examine the relationship between $p$ and the distribution of contact angles (or $\alpha$, in the case of spheroids), which could help determine why MRJ superball packings are more hypostatic than MRJ spheroid packings.

~

\section*{Acknowledgements}
The authors are grateful to Yang Jiao, Murray Skolnick, and Michael Klatt for helpful discussions.
This work was supported by the Air Force Office of Scientific Research Program on Mechanics of Multifunctional Materials and Microsystems under Grant No. FA9550-18-1-0514. 
\appendix





\begin{thebibliography}{95}%
\makeatletter
\providecommand \@ifxundefined [1]{%
 \@ifx{#1\undefined}
}%
\providecommand \@ifnum [1]{%
 \ifnum #1\expandafter \@firstoftwo
 \else \expandafter \@secondoftwo
 \fi
}%
\providecommand \@ifx [1]{%
 \ifx #1\expandafter \@firstoftwo
 \else \expandafter \@secondoftwo
 \fi
}%
\providecommand \natexlab [1]{#1}%
\providecommand \enquote  [1]{``#1''}%
\providecommand \bibnamefont  [1]{#1}%
\providecommand \bibfnamefont [1]{#1}%
\providecommand \citenamefont [1]{#1}%
\providecommand \href@noop [0]{\@secondoftwo}%
\providecommand \href [0]{\begingroup \@sanitize@url \@href}%
\providecommand \@href[1]{\@@startlink{#1}\@@href}%
\providecommand \@@href[1]{\endgroup#1\@@endlink}%
\providecommand \@sanitize@url [0]{\catcode `\\12\catcode `\$12\catcode
  `\&12\catcode `\#12\catcode `\^12\catcode `\_12\catcode `\%12\relax}%
\providecommand \@@startlink[1]{}%
\providecommand \@@endlink[0]{}%
\providecommand \url  [0]{\begingroup\@sanitize@url \@url }%
\providecommand \@url [1]{\endgroup\@href {#1}{\urlprefix }}%
\providecommand \urlprefix  [0]{URL }%
\providecommand \Eprint [0]{\href }%
\providecommand \doibase [0]{https://doi.org/}%
\providecommand \selectlanguage [0]{\@gobble}%
\providecommand \bibinfo  [0]{\@secondoftwo}%
\providecommand \bibfield  [0]{\@secondoftwo}%
\providecommand \translation [1]{[#1]}%
\providecommand \BibitemOpen [0]{}%
\providecommand \bibitemStop [0]{}%
\providecommand \bibitemNoStop [0]{.\EOS\space}%
\providecommand \EOS [0]{\spacefactor3000\relax}%
\providecommand \BibitemShut  [1]{\csname bibitem#1\endcsname}%
\let\auto@bib@innerbib\@empty
\bibitem [{\citenamefont {Bernal}(1965)}]{hughel_1965}%
  \BibitemOpen
  \bibfield  {author} {\bibinfo {author} {\bibfnamefont {J.}~\bibnamefont
  {Bernal}},\ }\href@noop {} {\emph {\bibinfo {title} {Liquids: {S}tructure,
  {P}roperties, {S}olid {I}nteractions}}}\ (\bibinfo  {publisher} {Elsevier
  Pub. Co.},\ \bibinfo {year} {1965})\BibitemShut {NoStop}%
\bibitem [{\citenamefont {Zallen}(2007)}]{zallen_2007}%
  \BibitemOpen
  \bibfield  {author} {\bibinfo {author} {\bibfnamefont {R.}~\bibnamefont
  {Zallen}},\ }\href@noop {} {\emph {\bibinfo {title} {The {P}hysics of
  {A}morphous {S}olids}}}\ (\bibinfo  {publisher} {Wiley-VCH},\ \bibinfo {year}
  {2007})\ pp.\ \bibinfo {pages} {25--50}\BibitemShut {NoStop}%
\bibitem [{\citenamefont {Torquato}(2002)}]{RHM_Text}%
  \BibitemOpen
  \bibfield  {author} {\bibinfo {author} {\bibfnamefont {S.}~\bibnamefont
  {Torquato}},\ }\href@noop {} {\emph {\bibinfo {title} {Random {H}eterogeneous
  {M}aterials: {M}icrostructure and {M}acroscopic {P}roperties}}}\ (\bibinfo
  {publisher} {Springer, New York, NY},\ \bibinfo {year} {2002})\BibitemShut
  {NoStop}%
\bibitem [{\citenamefont {Chaikin}\ and\ \citenamefont
  {Lubensky}(2000)}]{chaikin_lubensky_2000}%
  \BibitemOpen
  \bibfield  {author} {\bibinfo {author} {\bibfnamefont {P.~M.}\ \bibnamefont
  {Chaikin}}\ and\ \bibinfo {author} {\bibfnamefont {T.~C.}\ \bibnamefont
  {Lubensky}},\ }\href@noop {} {\emph {\bibinfo {title} {Principles of
  {C}ondensed {M}atter {P}hysics}}}\ (\bibinfo  {publisher} {Cambridge
  University Press},\ \bibinfo {year} {2000})\BibitemShut {NoStop}%
\bibitem [{\citenamefont {Mehta}(1994)}]{mehta_1994}%
  \BibitemOpen
  \bibfield  {author} {\bibinfo {author} {\bibfnamefont {A.}~\bibnamefont
  {Mehta}},\ }\href@noop {} {\emph {\bibinfo {title} {Granular {M}atter an
  {I}nterdisciplinary {A}pproach}}}\ (\bibinfo  {publisher} {Springer-Verlag},\
  \bibinfo {year} {1994})\BibitemShut {NoStop}%
\bibitem [{\citenamefont {Liang}\ and\ \citenamefont
  {Dill}(2001)}]{LIANG_bio1}%
  \BibitemOpen
  \bibfield  {author} {\bibinfo {author} {\bibfnamefont {J.}~\bibnamefont
  {Liang}}\ and\ \bibinfo {author} {\bibfnamefont {K.~A.}\ \bibnamefont
  {Dill}},\ }\bibfield  {title} {\bibinfo {title} {Are proteins well-packed?},\
  }\href@noop {} {\bibfield  {journal} {\bibinfo  {journal} {Biophys. J.}\
  }\textbf {\bibinfo {volume} {81}},\ \bibinfo {pages} {751 } (\bibinfo {year}
  {2001})}\BibitemShut {NoStop}%
\bibitem [{\citenamefont {Gevertz}\ and\ \citenamefont
  {Torquato}(2008)}]{Gevertx_bio3}%
  \BibitemOpen
  \bibfield  {author} {\bibinfo {author} {\bibfnamefont {J.~L.}\ \bibnamefont
  {Gevertz}}\ and\ \bibinfo {author} {\bibfnamefont {S.}~\bibnamefont
  {Torquato}},\ }\bibfield  {title} {\bibinfo {title} {A novel three-phase
  model of brain tissue microstructure},\ }\href
  {https://doi.org/10.1371/journal.pcbi.1000152} {\bibfield  {journal}
  {\bibinfo  {journal} {PLoS Comput. Biol.}\ }\textbf {\bibinfo {volume} {4}},\
  \bibinfo {pages} {1} (\bibinfo {year} {2008})}\BibitemShut {NoStop}%
\bibitem [{\citenamefont {Purohit}\ \emph {et~al.}(2003)\citenamefont
  {Purohit}, \citenamefont {Kondev},\ and\ \citenamefont
  {Phillips}}]{Purohit_bio2}%
  \BibitemOpen
  \bibfield  {author} {\bibinfo {author} {\bibfnamefont {P.~K.}\ \bibnamefont
  {Purohit}}, \bibinfo {author} {\bibfnamefont {J.}~\bibnamefont {Kondev}},\
  and\ \bibinfo {author} {\bibfnamefont {R.}~\bibnamefont {Phillips}},\
  }\bibfield  {title} {\bibinfo {title} {Mechanics of {DNA} packaging in
  viruses},\ }\href {https://doi.org/10.1073/pnas.0737893100} {\bibfield
  {journal} {\bibinfo  {journal} {Proc. Natl. Acad. Sci. U. S. A.}\ }\textbf
  {\bibinfo {volume} {100}},\ \bibinfo {pages} {3173} (\bibinfo {year}
  {2003})}\BibitemShut {NoStop}%
\bibitem [{\citenamefont {Torquato}\ and\ \citenamefont
  {Stillinger}(2010)}]{Torquato_JamHardPart}%
  \BibitemOpen
  \bibfield  {author} {\bibinfo {author} {\bibfnamefont {S.}~\bibnamefont
  {Torquato}}\ and\ \bibinfo {author} {\bibfnamefont {F.~H.}\ \bibnamefont
  {Stillinger}},\ }\bibfield  {title} {\bibinfo {title} {Jammed hard-particle
  packings: From {K}epler to {B}ernal and beyond},\ }\href
  {https://doi.org/10.1103/RevModPhys.82.2633} {\bibfield  {journal} {\bibinfo
  {journal} {Rev. Mod. Phys.}\ }\textbf {\bibinfo {volume} {82}},\ \bibinfo
  {pages} {2633} (\bibinfo {year} {2010})}\BibitemShut {NoStop}%
\bibitem [{\citenamefont
  {Torquato}(2018{\natexlab{a}})}]{Torquato_PackingPersp}%
  \BibitemOpen
  \bibfield  {author} {\bibinfo {author} {\bibfnamefont {S.}~\bibnamefont
  {Torquato}},\ }\bibfield  {title} {\bibinfo {title} {Perspective: Basic
  understanding of condensed phases of matter via packing models},\ }\href
  {https://doi.org/10.1063/1.5036657} {\bibfield  {journal} {\bibinfo
  {journal} {J. Chem. Phys.}\ }\textbf {\bibinfo {volume} {149}},\ \bibinfo
  {pages} {020901} (\bibinfo {year} {2018}{\natexlab{a}})}\BibitemShut
  {NoStop}%
\bibitem [{\citenamefont {Parisi}\ and\ \citenamefont
  {Zamponi}(2010)}]{Parisi_Mean}%
  \BibitemOpen
  \bibfield  {author} {\bibinfo {author} {\bibfnamefont {G.}~\bibnamefont
  {Parisi}}\ and\ \bibinfo {author} {\bibfnamefont {F.}~\bibnamefont
  {Zamponi}},\ }\bibfield  {title} {\bibinfo {title} {Mean-field theory of hard
  sphere glasses and jamming},\ }\href
  {https://doi.org/10.1103/RevModPhys.82.789} {\bibfield  {journal} {\bibinfo
  {journal} {Rev. Mod. Phys.}\ }\textbf {\bibinfo {volume} {82}},\ \bibinfo
  {pages} {789} (\bibinfo {year} {2010})}\BibitemShut {NoStop}%
\bibitem [{\citenamefont {Torquato}\ and\ \citenamefont
  {Stillinger}(2001)}]{Torqauto_Multiplicity}%
  \BibitemOpen
  \bibfield  {author} {\bibinfo {author} {\bibfnamefont {S.}~\bibnamefont
  {Torquato}}\ and\ \bibinfo {author} {\bibfnamefont {F.~H.}\ \bibnamefont
  {Stillinger}},\ }\bibfield  {title} {\bibinfo {title} {Multiplicity of
  generation, selection, and classification procedures for jammed hard-particle
  packings},\ }\href@noop {} {\bibfield  {journal} {\bibinfo  {journal} {J.
  Phys. Chem. B}\ }\textbf {\bibinfo {volume} {105}},\ \bibinfo {pages} {11849}
  (\bibinfo {year} {2001})}\BibitemShut {NoStop}%
\bibitem [{\citenamefont {Torquato}\ \emph {et~al.}(2000)\citenamefont
  {Torquato}, \citenamefont {Truskett},\ and\ \citenamefont
  {Debenedetti}}]{Torquato_RCPBad}%
  \BibitemOpen
  \bibfield  {author} {\bibinfo {author} {\bibfnamefont {S.}~\bibnamefont
  {Torquato}}, \bibinfo {author} {\bibfnamefont {T.~M.}\ \bibnamefont
  {Truskett}},\ and\ \bibinfo {author} {\bibfnamefont {P.~G.}\ \bibnamefont
  {Debenedetti}},\ }\bibfield  {title} {\bibinfo {title} {Is random close
  packing of spheres well defined?},\ }\href
  {https://doi.org/10.1103/PhysRevLett.84.2064} {\bibfield  {journal} {\bibinfo
   {journal} {Phys. Rev. Lett.}\ }\textbf {\bibinfo {volume} {84}},\ \bibinfo
  {pages} {2064} (\bibinfo {year} {2000})}\BibitemShut {NoStop}%
\bibitem [{\citenamefont {Torquato}\ and\ \citenamefont
  {Stillinger}(2003)}]{Torquato_HUDef}%
  \BibitemOpen
  \bibfield  {author} {\bibinfo {author} {\bibfnamefont {S.}~\bibnamefont
  {Torquato}}\ and\ \bibinfo {author} {\bibfnamefont {F.~H.}\ \bibnamefont
  {Stillinger}},\ }\bibfield  {title} {\bibinfo {title} {Local density
  fluctuations, hyperuniformity, and order metrics},\ }\href
  {https://doi.org/10.1103/PhysRevE.68.041113} {\bibfield  {journal} {\bibinfo
  {journal} {Phys. Rev. E}\ }\textbf {\bibinfo {volume} {68}},\ \bibinfo
  {pages} {041113} (\bibinfo {year} {2003})}\BibitemShut {NoStop}%
\bibitem [{\citenamefont {Donev}\ \emph
  {et~al.}(2005{\natexlab{a}})\citenamefont {Donev}, \citenamefont
  {Stillinger},\ and\ \citenamefont {Torquato}}]{Donev_Unexpected}%
  \BibitemOpen
  \bibfield  {author} {\bibinfo {author} {\bibfnamefont {A.}~\bibnamefont
  {Donev}}, \bibinfo {author} {\bibfnamefont {F.~H.}\ \bibnamefont
  {Stillinger}},\ and\ \bibinfo {author} {\bibfnamefont {S.}~\bibnamefont
  {Torquato}},\ }\bibfield  {title} {\bibinfo {title} {Unexpected density
  fluctuations in jammed disordered sphere packings},\ }\href
  {https://doi.org/10.1103/PhysRevLett.95.090604} {\bibfield  {journal}
  {\bibinfo  {journal} {Phys. Rev. Lett.}\ }\textbf {\bibinfo {volume} {95}},\
  \bibinfo {pages} {090604} (\bibinfo {year} {2005}{\natexlab{a}})}\BibitemShut
  {NoStop}%
\bibitem [{\citenamefont {Torquato}(2018{\natexlab{b}})}]{Torquato_HURev}%
  \BibitemOpen
  \bibfield  {author} {\bibinfo {author} {\bibfnamefont {S.}~\bibnamefont
  {Torquato}},\ }\bibfield  {title} {\bibinfo {title} {Hyperuniform states of
  matter},\ }\href@noop {} {\bibfield  {journal} {\bibinfo  {journal} {Phys.
  Rep.}\ }\textbf {\bibinfo {volume} {745}},\ \bibinfo {pages} {1 } (\bibinfo
  {year} {2018}{\natexlab{b}})},\ \bibinfo {note} {hyperuniform States of
  Matter}\BibitemShut {NoStop}%
\bibitem [{\citenamefont {Donev}\ \emph
  {et~al.}(2004{\natexlab{a}})\citenamefont {Donev}, \citenamefont {Cisse},
  \citenamefont {Sachs}, \citenamefont {Variano}, \citenamefont {Stillinger},
  \citenamefont {Connelly}, \citenamefont {Torquato},\ and\ \citenamefont
  {Chaikin}}]{Donev_MRJelip}%
  \BibitemOpen
  \bibfield  {author} {\bibinfo {author} {\bibfnamefont {A.}~\bibnamefont
  {Donev}}, \bibinfo {author} {\bibfnamefont {I.}~\bibnamefont {Cisse}},
  \bibinfo {author} {\bibfnamefont {D.}~\bibnamefont {Sachs}}, \bibinfo
  {author} {\bibfnamefont {E.~A.}\ \bibnamefont {Variano}}, \bibinfo {author}
  {\bibfnamefont {F.~H.}\ \bibnamefont {Stillinger}}, \bibinfo {author}
  {\bibfnamefont {R.}~\bibnamefont {Connelly}}, \bibinfo {author}
  {\bibfnamefont {S.}~\bibnamefont {Torquato}},\ and\ \bibinfo {author}
  {\bibfnamefont {P.~M.}\ \bibnamefont {Chaikin}},\ }\bibfield  {title}
  {\bibinfo {title} {Improving the density of jammed disordered packings using
  ellipsoids},\ }\href {https://doi.org/10.1126/science.1093010} {\bibfield
  {journal} {\bibinfo  {journal} {Science}\ }\textbf {\bibinfo {volume}
  {303}},\ \bibinfo {pages} {990} (\bibinfo {year}
  {2004}{\natexlab{a}})}\BibitemShut {NoStop}%
\bibitem [{\citenamefont {Jiao}\ \emph {et~al.}(2010)\citenamefont {Jiao},
  \citenamefont {Stillinger},\ and\ \citenamefont {Torquato}}]{Jiao_MRJballs}%
  \BibitemOpen
  \bibfield  {author} {\bibinfo {author} {\bibfnamefont {Y.}~\bibnamefont
  {Jiao}}, \bibinfo {author} {\bibfnamefont {F.~H.}\ \bibnamefont
  {Stillinger}},\ and\ \bibinfo {author} {\bibfnamefont {S.}~\bibnamefont
  {Torquato}},\ }\bibfield  {title} {\bibinfo {title} {Distinctive features
  arising in maximally random jammed packings of superballs},\ }\href
  {https://doi.org/10.1103/PhysRevE.81.041304} {\bibfield  {journal} {\bibinfo
  {journal} {Phys. Rev. E}\ }\textbf {\bibinfo {volume} {81}},\ \bibinfo
  {pages} {041304} (\bibinfo {year} {2010})}\BibitemShut {NoStop}%
\bibitem [{\citenamefont {Jiao}\ and\ \citenamefont
  {Torquato}(2011)}]{Jiao_MRJplatonic}%
  \BibitemOpen
  \bibfield  {author} {\bibinfo {author} {\bibfnamefont {Y.}~\bibnamefont
  {Jiao}}\ and\ \bibinfo {author} {\bibfnamefont {S.}~\bibnamefont
  {Torquato}},\ }\bibfield  {title} {\bibinfo {title} {Maximally random jammed
  packings of platonic solids: Hyperuniform long-range correlations and
  isostaticity},\ }\href {https://doi.org/10.1103/PhysRevE.84.041309}
  {\bibfield  {journal} {\bibinfo  {journal} {Phys. Rev. E}\ }\textbf {\bibinfo
  {volume} {84}},\ \bibinfo {pages} {041309} (\bibinfo {year}
  {2011})}\BibitemShut {NoStop}%
\bibitem [{\citenamefont {Chen}\ \emph {et~al.}(2014)\citenamefont {Chen},
  \citenamefont {Jiao},\ and\ \citenamefont {Torquato}}]{Chen_MRJtruntet}%
  \BibitemOpen
  \bibfield  {author} {\bibinfo {author} {\bibfnamefont {D.}~\bibnamefont
  {Chen}}, \bibinfo {author} {\bibfnamefont {Y.}~\bibnamefont {Jiao}},\ and\
  \bibinfo {author} {\bibfnamefont {S.}~\bibnamefont {Torquato}},\ }\bibfield
  {title} {\bibinfo {title} {Equilibrium phase behavior and maximally random
  jammed state of truncated tetrahedra},\ }\href
  {https://doi.org/10.1021/jp5010133} {\bibfield  {journal} {\bibinfo
  {journal} {J. Phys. Chem. B}\ }\textbf {\bibinfo {volume} {118}},\ \bibinfo
  {pages} {7981} (\bibinfo {year} {2014})}\BibitemShut {NoStop}%
\bibitem [{\citenamefont {Klatt}\ and\ \citenamefont
  {Torquato}(2014)}]{Klatt_MRJI}%
  \BibitemOpen
  \bibfield  {author} {\bibinfo {author} {\bibfnamefont {M.}~\bibnamefont
  {Klatt}}\ and\ \bibinfo {author} {\bibfnamefont {S.}~\bibnamefont
  {Torquato}},\ }\bibfield  {title} {\bibinfo {title} {Characterization of
  maximally random jammed sphere packings: {V}oronoi correlation functions},\
  }\href {https://doi.org/10.1103/PhysRevE.90.052120} {\bibfield  {journal}
  {\bibinfo  {journal} {Phys. Rev. E}\ }\textbf {\bibinfo {volume} {90}},\
  \bibinfo {pages} {052120} (\bibinfo {year} {2014})}\BibitemShut {NoStop}%
\bibitem [{\citenamefont {Klatt}\ and\ \citenamefont
  {Torquato}(2016)}]{Klatt_MRJII}%
  \BibitemOpen
  \bibfield  {author} {\bibinfo {author} {\bibfnamefont {M.}~\bibnamefont
  {Klatt}}\ and\ \bibinfo {author} {\bibfnamefont {S.}~\bibnamefont
  {Torquato}},\ }\bibfield  {title} {\bibinfo {title} {Characterization of
  maximally random jammed sphere packings. {II}. {C}orrelation functions and
  density fluctuations},\ }\href {https://doi.org/10.1103/PhysRevE.94.022152}
  {\bibfield  {journal} {\bibinfo  {journal} {Phys. Rev. E}\ }\textbf {\bibinfo
  {volume} {94}},\ \bibinfo {pages} {022152} (\bibinfo {year}
  {2016})}\BibitemShut {NoStop}%
\bibitem [{\citenamefont {Klatt}\ and\ \citenamefont
  {Torquato}(2018)}]{Klatt_MRJIII}%
  \BibitemOpen
  \bibfield  {author} {\bibinfo {author} {\bibfnamefont {M.}~\bibnamefont
  {Klatt}}\ and\ \bibinfo {author} {\bibfnamefont {S.}~\bibnamefont
  {Torquato}},\ }\bibfield  {title} {\bibinfo {title} {Characterization of
  maximally random jammed sphere packings. {III}. {T}ransport and
  electromagnetic properties via correlation functions},\ }\href
  {https://doi.org/10.1103/PhysRevE.97.012118} {\bibfield  {journal} {\bibinfo
  {journal} {Phys. Rev. E}\ }\textbf {\bibinfo {volume} {97}},\ \bibinfo
  {pages} {012118} (\bibinfo {year} {2018})}\BibitemShut {NoStop}%
\bibitem [{\citenamefont {Donev}\ \emph
  {et~al.}(2005{\natexlab{b}})\citenamefont {Donev}, \citenamefont {Torquato},\
  and\ \citenamefont {Stillinger}}]{Donev_g2}%
  \BibitemOpen
  \bibfield  {author} {\bibinfo {author} {\bibfnamefont {A.}~\bibnamefont
  {Donev}}, \bibinfo {author} {\bibfnamefont {S.}~\bibnamefont {Torquato}},\
  and\ \bibinfo {author} {\bibfnamefont {F.~H.}\ \bibnamefont {Stillinger}},\
  }\bibfield  {title} {\bibinfo {title} {Pair correlation function
  characteristics of nearly jammed disordered and ordered hard-sphere
  packings},\ }\href {https://doi.org/10.1103/PhysRevE.71.011105} {\bibfield
  {journal} {\bibinfo  {journal} {Phys. Rev. E}\ }\textbf {\bibinfo {volume}
  {71}},\ \bibinfo {pages} {011105} (\bibinfo {year}
  {2005}{\natexlab{b}})}\BibitemShut {NoStop}%
\bibitem [{\citenamefont {Atkinson}\ \emph {et~al.}(2013)\citenamefont
  {Atkinson}, \citenamefont {Stillinger},\ and\ \citenamefont
  {Torquato}}]{Atkinson_Rattler}%
  \BibitemOpen
  \bibfield  {author} {\bibinfo {author} {\bibfnamefont {S.}~\bibnamefont
  {Atkinson}}, \bibinfo {author} {\bibfnamefont {F.~H.}\ \bibnamefont
  {Stillinger}},\ and\ \bibinfo {author} {\bibfnamefont {S.}~\bibnamefont
  {Torquato}},\ }\bibfield  {title} {\bibinfo {title} {Detailed
  characterization of rattlers in exactly isostatic, strictly jammed sphere
  packings},\ }\href {https://doi.org/10.1103/PhysRevE.88.062208} {\bibfield
  {journal} {\bibinfo  {journal} {Phys. Rev. E}\ }\textbf {\bibinfo {volume}
  {88}},\ \bibinfo {pages} {062208} (\bibinfo {year} {2013})}\BibitemShut
  {NoStop}%
\bibitem [{\citenamefont {O'Hern}\ \emph {et~al.}(2003)\citenamefont {O'Hern}
  \emph {et~al.}}]{Ohern_Iso}%
  \BibitemOpen
  \bibfield  {author} {\bibinfo {author} {\bibfnamefont {C.~S.}\ \bibnamefont
  {O'Hern}},\ \bibinfo {author} {\bibfnamefont {L.~E.}\ \bibnamefont
  {Silbert}},\ \bibinfo {author} {\bibfnamefont {A.~J.}\ \bibnamefont
  {Liu}},\ and\ \bibinfo {author} {\bibfnamefont {S.~R.}\ \bibnamefont
  {Nagel}},\ } \bibfield  {title} {\bibinfo {title} {Jamming at
  zero temperature and zero applied stress: The epitome of disorder},\ }\href
  {https://doi.org/10.1103/PhysRevE.68.011306} {\bibfield  {journal} {\bibinfo
  {journal} {Phys. Rev. E}\ }\textbf {\bibinfo {volume} {68}},\ \bibinfo
  {pages} {011306} (\bibinfo {year} {2003})}\BibitemShut {NoStop}%
\bibitem [{\citenamefont {Zaccone}\ and\ \citenamefont
  {Scossa-Romano}(2011)}]{Zaccone_nonaff}%
  \BibitemOpen
  \bibfield  {author} {\bibinfo {author} {\bibfnamefont {A.}~\bibnamefont
  {Zaccone}}\ and\ \bibinfo {author} {\bibfnamefont {E.}~\bibnamefont
  {Scossa-Romano}},\ }\bibfield  {title} {\bibinfo {title} {Approximate
  analytical description of the nonaffine response of amorphous solids},\
  }\href@noop {} {\bibfield  {journal} {\bibinfo  {journal} {Phys. Rev. B}\
  }\textbf {\bibinfo {volume} {83}},\ \bibinfo {pages} {184205} (\bibinfo
  {year} {2011})}\BibitemShut {NoStop}%
\bibitem [{\citenamefont {Torquato}\ and\ \citenamefont
  {Jiao}(2009)}]{Torquato_ASCNat}%
  \BibitemOpen
  \bibfield  {author} {\bibinfo {author} {\bibfnamefont {S.}~\bibnamefont
  {Torquato}}\ and\ \bibinfo {author} {\bibfnamefont {Y.}~\bibnamefont
  {Jiao}},\ }\bibfield  {title} {\bibinfo {title} {Dense packings of the
  {P}latonic and {A}rchimedean solids},\ }\href
  {https://doi.org/10.1038/nature08239} {\bibfield  {journal} {\bibinfo
  {journal} {Nature}\ }\textbf {\bibinfo {volume} {460}},\ \bibinfo {pages}
  {876} (\bibinfo {year} {2009})}\BibitemShut {NoStop}%
\bibitem [{\citenamefont {Cinacchi}\ and\ \citenamefont
  {Torquato}(2019)}]{Cinacchi_LensMRJ}%
  \BibitemOpen
  \bibfield  {author} {\bibinfo {author} {\bibfnamefont {G.}~\bibnamefont
  {Cinacchi}}\ and\ \bibinfo {author} {\bibfnamefont {S.}~\bibnamefont
  {Torquato}},\ }\bibfield  {title} {\bibinfo {title} {Hard convex lens-shaped
  particles: Characterization of dense disordered packings},\ }\href
  {https://doi.org/10.1103/PhysRevE.100.062902} {\bibfield  {journal} {\bibinfo
   {journal} {Phys. Rev. E}\ }\textbf {\bibinfo {volume} {100}},\ \bibinfo
  {pages} {062902} (\bibinfo {year} {2019})}\BibitemShut {NoStop}%
\bibitem [{\citenamefont {Delaney}\ and\ \citenamefont
  {Cleary}(2010)}]{Delaney_Superell}%
  \BibitemOpen
  \bibfield  {author} {\bibinfo {author} {\bibfnamefont {G.~W.}\ \bibnamefont
  {Delaney}}\ and\ \bibinfo {author} {\bibfnamefont {P.~W.}\ \bibnamefont
  {Cleary}},\ }\bibfield  {title} {\bibinfo {title} {The packing properties of
  superellipsoids},\ }\href {https://doi.org/10.1209/0295-5075/89/34002}
  {\bibfield  {journal} {\bibinfo  {journal} {{EPL}}\ }\textbf {\bibinfo
  {volume} {89}},\ \bibinfo {pages} {34002} (\bibinfo {year}
  {2010})}\BibitemShut {NoStop}%
\bibitem [{\citenamefont {Donev}\ \emph {et~al.}(2007)\citenamefont {Donev},
  \citenamefont {Connelly}, \citenamefont {Stillinger},\ and\ \citenamefont
  {Torquato}}]{Donev_EllipFunct}%
  \BibitemOpen
  \bibfield  {author} {\bibinfo {author} {\bibfnamefont {A.}~\bibnamefont
  {Donev}}, \bibinfo {author} {\bibfnamefont {R.}~\bibnamefont {Connelly}},
  \bibinfo {author} {\bibfnamefont {F.~H.}\ \bibnamefont {Stillinger}},\ and\
  \bibinfo {author} {\bibfnamefont {S.}~\bibnamefont {Torquato}},\ }\bibfield
  {title} {\bibinfo {title} {Underconstrained jammed packings of nonspherical
  hard particles: Ellipses and ellipsoids},\ }\href@noop {} {\bibfield
  {journal} {\bibinfo  {journal} {Phys. Rev. E}\ }\textbf {\bibinfo {volume}
  {75}},\ \bibinfo {pages} {051304} (\bibinfo {year} {2007})}\BibitemShut
  {NoStop}%
\bibitem [{\citenamefont {Maher}\ \emph {et~al.}(2021)\citenamefont {Maher},
  \citenamefont {Stillinger},\ and\ \citenamefont {Torquato}}]{Maher_Kin}%
  \BibitemOpen
  \bibfield  {author} {\bibinfo {author} {\bibfnamefont {C.~E.}\ \bibnamefont
  {Maher}}, \bibinfo {author} {\bibfnamefont {F.~H.}\ \bibnamefont
  {Stillinger}},\ and\ \bibinfo {author} {\bibfnamefont {S.}~\bibnamefont
  {Torquato}},\ }\bibfield  {title} {\bibinfo {title} {Kinetic frustration
  effects on dense two-dimensional packings of convex particles and their
  structural characteristics},\ }\href
  {https://doi.org/10.1021/acs.jpcb.1c00497} {\bibfield  {journal} {\bibinfo
  {journal} {J. Phys. Chem. B}\ }\textbf {\bibinfo {volume} {125}},\ \bibinfo
  {pages} {2450} (\bibinfo {year} {2021})}\BibitemShut {NoStop}%
\bibitem [{\citenamefont {Atkinson}\ \emph {et~al.}(2014)\citenamefont
  {Atkinson}, \citenamefont {Stillinger},\ and\ \citenamefont
  {Torquato}}]{Atkinson_2DMono}%
  \BibitemOpen
  \bibfield  {author} {\bibinfo {author} {\bibfnamefont {S.}~\bibnamefont
  {Atkinson}}, \bibinfo {author} {\bibfnamefont {F.~H.}\ \bibnamefont
  {Stillinger}},\ and\ \bibinfo {author} {\bibfnamefont {S.}~\bibnamefont
  {Torquato}},\ }\bibfield  {title} {\bibinfo {title} {Existence of isostatic,
  maximally random jammed monodisperse hard-disk packings},\ }\href
  {https://doi.org/10.1073/pnas.1408371112} {\bibfield  {journal} {\bibinfo
  {journal} {Proc. Natl. Acad. Sci. U. S. A.}\ }\textbf {\bibinfo {volume}
  {111}},\ \bibinfo {pages} {18436} (\bibinfo {year} {2014})}\BibitemShut
  {NoStop}%
\bibitem [{\citenamefont {Atkinson}\ \emph {et~al.}(2016)\citenamefont
  {Atkinson}, \citenamefont {Zhang}, \citenamefont {Hopkins},\ and\
  \citenamefont {Torquato}}]{Atkinson_Slowdown}%
  \BibitemOpen
  \bibfield  {author} {\bibinfo {author} {\bibfnamefont {S.}~\bibnamefont
  {Atkinson}}, \bibinfo {author} {\bibfnamefont {G.}~\bibnamefont {Zhang}},
  \bibinfo {author} {\bibfnamefont {A.~B.}\ \bibnamefont {Hopkins}},\ and\
  \bibinfo {author} {\bibfnamefont {S.}~\bibnamefont {Torquato}},\ }\bibfield
  {title} {\bibinfo {title} {Critical slowing down and hyperuniformity on
  approach to jamming},\ }\href {https://doi.org/10.1103/PhysRevE.94.012902}
  {\bibfield  {journal} {\bibinfo  {journal} {Phys. Rev. E}\ }\textbf {\bibinfo
  {volume} {94}},\ \bibinfo {pages} {012902} (\bibinfo {year}
  {2016})}\BibitemShut {NoStop}%
\bibitem [{\citenamefont {Zachary}\ \emph
  {et~al.}(2011{\natexlab{a}})\citenamefont {Zachary}, \citenamefont {Jiao},\
  and\ \citenamefont {Torquato}}]{Zachary_QLRLet}%
  \BibitemOpen
  \bibfield  {author} {\bibinfo {author} {\bibfnamefont {C.~E.}\ \bibnamefont
  {Zachary}}, \bibinfo {author} {\bibfnamefont {Y.}~\bibnamefont {Jiao}},\ and\
  \bibinfo {author} {\bibfnamefont {S.}~\bibnamefont {Torquato}},\ }\bibfield
  {title} {\bibinfo {title} {Hyperuniform long-range correlations are a
  signature of disordered jammed hard-particle packings},\ }\href
  {https://doi.org/10.1103/PhysRevLett.106.178001} {\bibfield  {journal}
  {\bibinfo  {journal} {Phys. Rev. Lett.}\ }\textbf {\bibinfo {volume} {106}},\
  \bibinfo {pages} {178001} (\bibinfo {year} {2011}{\natexlab{a}})}\BibitemShut
  {NoStop}%
\bibitem [{\citenamefont {Berthier}\ \emph {et~al.}(2011)\citenamefont
  {Berthier}, \citenamefont {Chaudhuri}, \citenamefont {Coulais}, \citenamefont
  {Dauchot},\ and\ \citenamefont {Sollich}}]{Bert_MRJNum}%
  \BibitemOpen
  \bibfield  {author} {\bibinfo {author} {\bibfnamefont {L.}~\bibnamefont
  {Berthier}}, \bibinfo {author} {\bibfnamefont {P.}~\bibnamefont {Chaudhuri}},
  \bibinfo {author} {\bibfnamefont {C.}~\bibnamefont {Coulais}}, \bibinfo
  {author} {\bibfnamefont {O.}~\bibnamefont {Dauchot}},\ and\ \bibinfo {author}
  {\bibfnamefont {P.}~\bibnamefont {Sollich}},\ }\bibfield  {title} {\bibinfo
  {title} {Suppressed compressibility at large scale in jammed packings of
  size-disperse spheres},\ }\href@noop {} {\bibfield  {journal} {\bibinfo
  {journal} {Phys. Rev. Lett.}\ }\textbf {\bibinfo {volume} {106}},\ \bibinfo
  {pages} {120601} (\bibinfo {year} {2011})}\BibitemShut {NoStop}%
\bibitem [{\citenamefont {Kurita}\ and\ \citenamefont
  {Weeks}(2011)}]{Week_MRJnum}%
  \BibitemOpen
  \bibfield  {author} {\bibinfo {author} {\bibfnamefont {R.}~\bibnamefont
  {Kurita}}\ and\ \bibinfo {author} {\bibfnamefont {E.~R.}\ \bibnamefont
  {Weeks}},\ }\bibfield  {title} {\bibinfo {title} {Incompressibility of
  polydisperse random-close-packed colloidal particles},\ }\href@noop {}
  {\bibfield  {journal} {\bibinfo  {journal} {Phys. Rev. E}\ }\textbf {\bibinfo
  {volume} {84}},\ \bibinfo {pages} {030401} (\bibinfo {year}
  {2011})}\BibitemShut {NoStop}%
\bibitem [{\citenamefont {Dreyfus}\ \emph {et~al.}(2015)\citenamefont
  {Dreyfus}, \citenamefont {Xu}, \citenamefont {Still}, \citenamefont {Hough},
  \citenamefont {Yodh},\ and\ \citenamefont {Torquato}}]{Drey_MRJnum}%
  \BibitemOpen
  \bibfield  {author} {\bibinfo {author} {\bibfnamefont {R.}~\bibnamefont
  {Dreyfus}}, \bibinfo {author} {\bibfnamefont {Y.}~\bibnamefont {Xu}},
  \bibinfo {author} {\bibfnamefont {T.}~\bibnamefont {Still}}, \bibinfo
  {author} {\bibfnamefont {L.~A.}\ \bibnamefont {Hough}}, \bibinfo {author}
  {\bibfnamefont {A.~G.}\ \bibnamefont {Yodh}},\ and\ \bibinfo {author}
  {\bibfnamefont {S.}~\bibnamefont {Torquato}},\ }\bibfield  {title} {\bibinfo
  {title} {Diagnosing hyperuniformity in two-dimensional, disordered, jammed
  packings of soft spheres},\ }\href@noop {} {\bibfield  {journal} {\bibinfo
  {journal} {Phys. Rev. E}\ }\textbf {\bibinfo {volume} {91}},\ \bibinfo
  {pages} {012302} (\bibinfo {year} {2015})}\BibitemShut {NoStop}%
\bibitem [{\citenamefont {Zachary}\ \emph
  {et~al.}(2011{\natexlab{b}})\citenamefont {Zachary}, \citenamefont {Jiao},\
  and\ \citenamefont {Torquato}}]{Zachary_QLRII}%
  \BibitemOpen
  \bibfield  {author} {\bibinfo {author} {\bibfnamefont {C.~E.}\ \bibnamefont
  {Zachary}}, \bibinfo {author} {\bibfnamefont {Y.}~\bibnamefont {Jiao}},\ and\
  \bibinfo {author} {\bibfnamefont {S.}~\bibnamefont {Torquato}},\ }\bibfield
  {title} {\bibinfo {title} {Hyperuniformity, quasi-long-range correlations,
  and void-space constraints in maximally random jammed particle packings.
  {II}. anisotropy in particle shape},\ }\href
  {https://doi.org/10.1103/PhysRevE.83.051309} {\bibfield  {journal} {\bibinfo
  {journal} {Phys. Rev. E}\ }\textbf {\bibinfo {volume} {83}},\ \bibinfo
  {pages} {051309} (\bibinfo {year} {2011}{\natexlab{b}})}\BibitemShut
  {NoStop}%
\bibitem [{\citenamefont {Zachary}\ \emph
  {et~al.}(2011{\natexlab{c}})\citenamefont {Zachary}, \citenamefont {Jiao},\
  and\ \citenamefont {Torquato}}]{Zachary_QLRI}%
  \BibitemOpen
  \bibfield  {author} {\bibinfo {author} {\bibfnamefont {C.~E.}\ \bibnamefont
  {Zachary}}, \bibinfo {author} {\bibfnamefont {Y.}~\bibnamefont {Jiao}},\ and\
  \bibinfo {author} {\bibfnamefont {S.}~\bibnamefont {Torquato}},\ }\bibfield
  {title} {\bibinfo {title} {Hyperuniformity, quasi-long-range correlations,
  and void-space constraints in maximally random jammed particle packings. {I}.
  polydisperse spheres},\ }\href {https://doi.org/10.1103/PhysRevE.83.051308}
  {\bibfield  {journal} {\bibinfo  {journal} {Phys. Rev. E}\ }\textbf {\bibinfo
  {volume} {83}},\ \bibinfo {pages} {051308} (\bibinfo {year}
  {2011}{\natexlab{c}})}\BibitemShut {NoStop}%
\bibitem [{\citenamefont {Torquato}(2021{\natexlab{a}})}]{Torquato_FVT}%
  \BibitemOpen
  \bibfield  {author} {\bibinfo {author} {\bibfnamefont {S.}~\bibnamefont
  {Torquato}},\ }\bibfield  {title} {\bibinfo {title} {Structural
  characterization of many-particle systems on approach to hyperuniform
  states},\ }\href@noop {} {\bibfield  {journal} {\bibinfo  {journal} {Phys.
  Rev. E}\ }\textbf {\bibinfo {volume} {103}},\ \bibinfo {pages} {052126}
  (\bibinfo {year} {2021}{\natexlab{a}})}\BibitemShut {NoStop}%
\bibitem [{\citenamefont {Rissone}\ \emph {et~al.}(2021)\citenamefont
  {Rissone}, \citenamefont {Corwin},\ and\ \citenamefont
  {Parisi}}]{Rissone_TSC}%
  \BibitemOpen
  \bibfield  {author} {\bibinfo {author} {\bibfnamefont {P.}~\bibnamefont
  {Rissone}}, \bibinfo {author} {\bibfnamefont {E.~I.}\ \bibnamefont
  {Corwin}},\ and\ \bibinfo {author} {\bibfnamefont {G.}~\bibnamefont
  {Parisi}},\ }\bibfield  {title} {\bibinfo {title} {Long-range anomalous decay
  of the correlation in jammed packings},\ }\href@noop {} {\bibfield  {journal}
  {\bibinfo  {journal} {Phys. Rev. Lett.}\ }\textbf {\bibinfo {volume} {127}},\
  \bibinfo {pages} {038001} (\bibinfo {year} {2021})}\BibitemShut {NoStop}%
\bibitem [{\citenamefont {Abreu}\ \emph {et~al.}(2003)\citenamefont {Abreu},
  \citenamefont {Tavares},\ and\ \citenamefont {Castier}}]{ABREU_Spherocyl}%
  \BibitemOpen
  \bibfield  {author} {\bibinfo {author} {\bibfnamefont {C.~R.}\ \bibnamefont
  {Abreu}}, \bibinfo {author} {\bibfnamefont {F.~W.}\ \bibnamefont {Tavares}},\
  and\ \bibinfo {author} {\bibfnamefont {M.}~\bibnamefont {Castier}},\
  }\bibfield  {title} {\bibinfo {title} {Influence of particle shape on the
  packing and on the segregation of spherocylinders via monte carlo
  simulations},\ }\href
  {https://doi.org/https://doi.org/10.1016/S0032-5910(03)00151-7} {\bibfield
  {journal} {\bibinfo  {journal} {Powder Technol.}\ }\textbf {\bibinfo {volume}
  {134}},\ \bibinfo {pages} {167} (\bibinfo {year} {2003})}\BibitemShut
  {NoStop}%
\bibitem [{\citenamefont {Williams}\ and\ \citenamefont
  {Philipse}(2003{\natexlab{a}})}]{Williams_SpheroCyl}%
  \BibitemOpen
  \bibfield  {author} {\bibinfo {author} {\bibfnamefont {S.~R.}\ \bibnamefont
  {Williams}}\ and\ \bibinfo {author} {\bibfnamefont {A.~P.}\ \bibnamefont
  {Philipse}},\ }\bibfield  {title} {\bibinfo {title} {Random packings of
  spheres and spherocylinders simulated by mechanical contraction},\ }\href
  {https://doi.org/10.1103/PhysRevE.67.051301} {\bibfield  {journal} {\bibinfo
  {journal} {Phys. Rev. E}\ }\textbf {\bibinfo {volume} {67}},\ \bibinfo
  {pages} {051301} (\bibinfo {year} {2003}{\natexlab{a}})}\BibitemShut
  {NoStop}%
\bibitem [{\citenamefont {Donev}\ \emph {et~al.}(2006)\citenamefont {Donev},
  \citenamefont {Burton}, \citenamefont {Stillinger},\ and\ \citenamefont
  {Torquato}}]{Donev_Rect}%
  \BibitemOpen
  \bibfield  {author} {\bibinfo {author} {\bibfnamefont {A.}~\bibnamefont
  {Donev}}, \bibinfo {author} {\bibfnamefont {J.}~\bibnamefont {Burton}},
  \bibinfo {author} {\bibfnamefont {F.~H.}\ \bibnamefont {Stillinger}},\ and\
  \bibinfo {author} {\bibfnamefont {S.}~\bibnamefont {Torquato}},\ }\bibfield
  {title} {\bibinfo {title} {Tetratic order in the phase behavior of a
  hard-rectangle system},\ }\href {https://doi.org/10.1103/PhysRevB.73.054109}
  {\bibfield  {journal} {\bibinfo  {journal} {Phys. Rev. B}\ }\textbf {\bibinfo
  {volume} {73}},\ \bibinfo {pages} {054109} (\bibinfo {year}
  {2006})}\BibitemShut {NoStop}%
\bibitem [{\citenamefont {Yatsenko}\ and\ \citenamefont
  {Schweizer}(2008)}]{Yatsenko_Rod}%
  \BibitemOpen
  \bibfield  {author} {\bibinfo {author} {\bibfnamefont {G.}~\bibnamefont
  {Yatsenko}}\ and\ \bibinfo {author} {\bibfnamefont {K.~S.}\ \bibnamefont
  {Schweizer}},\ }\bibfield  {title} {\bibinfo {title} {Glassy dynamics and
  kinetic vitrification of isotropic suspensions of hard rods},\ }\href
  {https://doi.org/10.1021/la8002492} {\bibfield  {journal} {\bibinfo
  {journal} {Langmuir}\ }\textbf {\bibinfo {volume} {24}},\ \bibinfo {pages}
  {7474} (\bibinfo {year} {2008})}\BibitemShut {NoStop}%
\bibitem [{\citenamefont {Torquato}\ and\ \citenamefont
  {Jiao}(2012)}]{Torquato_Org}%
  \BibitemOpen
  \bibfield  {author} {\bibinfo {author} {\bibfnamefont {S.}~\bibnamefont
  {Torquato}}\ and\ \bibinfo {author} {\bibfnamefont {Y.}~\bibnamefont
  {Jiao}},\ }\bibfield  {title} {\bibinfo {title} {Organizing principles for
  dense packings of nonspherical hard particles: Not all shapes are created
  equal},\ }\href {https://doi.org/10.1103/PhysRevE.86.011102} {\bibfield
  {journal} {\bibinfo  {journal} {Phys. Rev. E}\ }\textbf {\bibinfo {volume}
  {86}},\ \bibinfo {pages} {011102} (\bibinfo {year} {2012})}\BibitemShut
  {NoStop}%
\bibitem [{\citenamefont {Kitaigorodsky}(2012)}]{kitaigorodsky_orgcryst}%
  \BibitemOpen
  \bibfield  {author} {\bibinfo {author} {\bibfnamefont {A.}~\bibnamefont
  {Kitaigorodsky}},\ }\href@noop {} {\emph {\bibinfo {title} {Molecular
  crystals and Molecules}}}\ (\bibinfo  {publisher} {Elsevier Science},\
  \bibinfo {year} {2012})\BibitemShut {NoStop}%
\bibitem [{\citenamefont {Jiao}\ \emph {et~al.}(2008)\citenamefont {Jiao},
  \citenamefont {Stillinger},\ and\ \citenamefont {Torquato}}]{Jiao_OptSD}%
  \BibitemOpen
  \bibfield  {author} {\bibinfo {author} {\bibfnamefont {Y.}~\bibnamefont
  {Jiao}}, \bibinfo {author} {\bibfnamefont {F.~H.}\ \bibnamefont
  {Stillinger}},\ and\ \bibinfo {author} {\bibfnamefont {S.}~\bibnamefont
  {Torquato}},\ }\bibfield  {title} {\bibinfo {title} {Optimal packings of
  superdisks and the role of symmetry},\ }\href@noop {} {\bibfield  {journal}
  {\bibinfo  {journal} {Phys. Rev. Lett.}\ }\textbf {\bibinfo {volume} {100}},\
  \bibinfo {pages} {245504} (\bibinfo {year} {2008})}\BibitemShut {NoStop}%
\bibitem [{\citenamefont {Jiao}\ \emph {et~al.}(2009)\citenamefont {Jiao},
  \citenamefont {Stillinger},\ and\ \citenamefont {Torquato}}]{Jaio_OptSB}%
  \BibitemOpen
  \bibfield  {author} {\bibinfo {author} {\bibfnamefont {Y.}~\bibnamefont
  {Jiao}}, \bibinfo {author} {\bibfnamefont {F.~H.}\ \bibnamefont
  {Stillinger}},\ and\ \bibinfo {author} {\bibfnamefont {S.}~\bibnamefont
  {Torquato}},\ }\bibfield  {title} {\bibinfo {title} {Optimal packings of
  superballs},\ }\href {https://doi.org/10.1103/PhysRevE.79.041309} {\bibfield
  {journal} {\bibinfo  {journal} {Phys. Rev. E}\ }\textbf {\bibinfo {volume}
  {79}},\ \bibinfo {pages} {041309} (\bibinfo {year} {2009})}\BibitemShut
  {NoStop}%
\bibitem [{\citenamefont {Batten}\ \emph {et~al.}(2010)\citenamefont {Batten},
  \citenamefont {Stillinger},\ and\ \citenamefont {Torquato}}]{Batten_SBPhase}%
  \BibitemOpen
  \bibfield  {author} {\bibinfo {author} {\bibfnamefont {R.~D.}\ \bibnamefont
  {Batten}}, \bibinfo {author} {\bibfnamefont {F.~H.}\ \bibnamefont
  {Stillinger}},\ and\ \bibinfo {author} {\bibfnamefont {S.}~\bibnamefont
  {Torquato}},\ }\bibfield  {title} {\bibinfo {title} {Phase behavior of
  colloidal superballs: Shape interpolation from spheres to cubes},\ }\href
  {https://doi.org/10.1103/PhysRevE.81.061105} {\bibfield  {journal} {\bibinfo
  {journal} {Phys. Rev. E}\ }\textbf {\bibinfo {volume} {81}},\ \bibinfo
  {pages} {061105} (\bibinfo {year} {2010})}\BibitemShut {NoStop}%
\bibitem [{\citenamefont {Ni}\ \emph {et~al.}(2012)\citenamefont {Ni},
  \citenamefont {Gantapara}, \citenamefont {de~Graaf}, \citenamefont {van
  Roij},\ and\ \citenamefont {Dijkstra}}]{Ni_SBphase}%
  \BibitemOpen
  \bibfield  {author} {\bibinfo {author} {\bibfnamefont {R.}~\bibnamefont
  {Ni}}, \bibinfo {author} {\bibfnamefont {A.~P.}\ \bibnamefont {Gantapara}},
  \bibinfo {author} {\bibfnamefont {J.}~\bibnamefont {de~Graaf}}, \bibinfo
  {author} {\bibfnamefont {R.}~\bibnamefont {van Roij}},\ and\ \bibinfo
  {author} {\bibfnamefont {M.}~\bibnamefont {Dijkstra}},\ }\bibfield  {title}
  {\bibinfo {title} {Phase diagram of colloidal hard superballs: from cubes via
  spheres to octahedra},\ }\href {https://doi.org/10.1039/C2SM25813G}
  {\bibfield  {journal} {\bibinfo  {journal} {Soft Matter}\ }\textbf {\bibinfo
  {volume} {8}},\ \bibinfo {pages} {8826} (\bibinfo {year} {2012})}\BibitemShut
  {NoStop}%
\bibitem [{\citenamefont {Gurin}\ \emph {et~al.}(2020)\citenamefont {Gurin},
  \citenamefont {Varga},\ and\ \citenamefont {Odriozola}}]{Odriozola_Phase}%
  \BibitemOpen
  \bibfield  {author} {\bibinfo {author} {\bibfnamefont {P.}~\bibnamefont
  {Gurin}}, \bibinfo {author} {\bibfnamefont {S.}~\bibnamefont {Varga}},\ and\
  \bibinfo {author} {\bibfnamefont {G.}~\bibnamefont {Odriozola}},\ }\bibfield
  {title} {\bibinfo {title} {Three-step melting of hard superdisks in two
  dimensions},\ }\href@noop {} {\bibfield  {journal} {\bibinfo  {journal}
  {Phys. Rev. E}\ }\textbf {\bibinfo {volume} {102}},\ \bibinfo {pages}
  {062603} (\bibinfo {year} {2020})}\BibitemShut {NoStop}%
\bibitem [{\citenamefont {Rossi}\ \emph {et~al.}(2011)\citenamefont {Rossi}
  \emph {et~al.}}]{Rossi_SBcolloid}%
  \BibitemOpen
  \bibfield  {author} {\bibinfo {author} {\bibfnamefont {L.}~\bibnamefont
  {Rossi}},\ \bibinfo {author} {\bibfnamefont {S.}~\bibnamefont
  {Sacanna}},\ \bibinfo {author} {\bibfnamefont {W.~T.~M.}~\bibnamefont
  {Irvine}},\ \bibinfo {author} {\bibfnamefont {P.~M.}~\bibnamefont
  {Chaikin}},\ \bibinfo {author} {\bibfnamefont {D.~J.}~\bibnamefont
  {Pine}},\ and\ \bibinfo {author} {\bibfnamefont {A.~P.}~\bibnamefont
  {Philipse}},\ }\bibfield  {title} {\bibinfo {title} {Cubic
  crystals from cubic colloids},\ }\href {https://doi.org/10.1039/C0SM01246G}
  {\bibfield  {journal} {\bibinfo  {journal} {Soft Matter}\ }\textbf {\bibinfo
  {volume} {7}},\ \bibinfo {pages} {4139} (\bibinfo {year} {2011})}\BibitemShut
  {NoStop}%
\bibitem [{\citenamefont {Zhang}\ \emph {et~al.}(2011)\citenamefont {Zhang}
  \emph {et~al.}}]{Zhang_SBass}%
  \BibitemOpen
  \bibfield  {author} {\bibinfo {author} {\bibfnamefont {Y.}~\bibnamefont
  {Zhang}},\ \bibinfo {author} {\bibfnamefont {F.}~\bibnamefont
  {Lu}},\ \bibinfo {author} {\bibfnamefont {D.}~\bibnamefont
  {van~der~Lelie}},\ and\ \bibinfo {author} {\bibfnamefont {O.}~\bibnamefont
  {Gang}},\ }\bibfield  {title} {\bibinfo {title} {Continuous
  phase transformation in nanocube assemblies},\ }\href
  {https://doi.org/10.1103/PhysRevLett.107.135701} {\bibfield  {journal}
  {\bibinfo  {journal} {Phys. Rev. Lett.}\ }\textbf {\bibinfo {volume} {107}},\
  \bibinfo {pages} {135701} (\bibinfo {year} {2011})}\BibitemShut {NoStop}%
\bibitem [{\citenamefont {Meijer}\ \emph {et~al.}(2017)\citenamefont {Meijer}
  \emph {et~al.}}]{meijer_SBsolidsolid}%
  \BibitemOpen
  \bibfield  {author} {\bibinfo {author} {\bibfnamefont {J.-M.}~\bibnamefont
  {Meijer}},\ \bibinfo {author} {\bibfnamefont {A.}~\bibnamefont
  {Pal}},\ \bibinfo {author} {\bibfnamefont {S.}~\bibnamefont
  {Ouhajji}},\ \bibinfo {author} {\bibfnamefont {H.~N.~W.}~\bibnamefont
  {Lekkerkerker}},\ \bibinfo {author} {\bibfnamefont {A.~P.}~\bibnamefont
  {Philipse}},\ and\ \bibinfo {author} {\bibfnamefont {A.~V.}~\bibnamefont
  {Petukhov}},\ }\bibfield  {title} {\bibinfo {title} {Observation
  of solid–solid transitions in 3{D} crystals of colloidal superballs},\
  }\bibfield  {journal} {\bibinfo  {journal} {Nat. Commun}\ }\textbf {\bibinfo
  {volume} {8}},\  (\bibinfo {year} {2017})\BibitemShut {NoStop}%
\bibitem [{\citenamefont {Donev}\ \emph
  {et~al.}(2004{\natexlab{b}})\citenamefont {Donev}, \citenamefont
  {Stillinger}, \citenamefont {Chaikin},\ and\ \citenamefont
  {Torquato}}]{Donev_OptEllip}%
  \BibitemOpen
  \bibfield  {author} {\bibinfo {author} {\bibfnamefont {A.}~\bibnamefont
  {Donev}}, \bibinfo {author} {\bibfnamefont {F.~H.}\ \bibnamefont
  {Stillinger}}, \bibinfo {author} {\bibfnamefont {P.~M.}\ \bibnamefont
  {Chaikin}},\ and\ \bibinfo {author} {\bibfnamefont {S.}~\bibnamefont
  {Torquato}},\ }\bibfield  {title} {\bibinfo {title} {Unusually dense crystal
  packings of ellipsoids},\ }\href@noop {} {\bibfield  {journal} {\bibinfo
  {journal} {Phys. Rev. Lett.}\ }\textbf {\bibinfo {volume} {92}},\ \bibinfo
  {pages} {255506} (\bibinfo {year} {2004}{\natexlab{b}})}\BibitemShut
  {NoStop}%
\bibitem [{\citenamefont {Donev}\ \emph
  {et~al.}(2004{\natexlab{c}})\citenamefont {Donev}, \citenamefont {Torquato},
  \citenamefont {Stillinger},\ and\ \citenamefont {Connelly}}]{Donev_MRJDisk}%
  \BibitemOpen
  \bibfield  {author} {\bibinfo {author} {\bibfnamefont {A.}~\bibnamefont
  {Donev}}, \bibinfo {author} {\bibfnamefont {S.}~\bibnamefont {Torquato}},
  \bibinfo {author} {\bibfnamefont {F.~H.}\ \bibnamefont {Stillinger}},\ and\
  \bibinfo {author} {\bibfnamefont {R.}~\bibnamefont {Connelly}},\ }\bibfield
  {title} {\bibinfo {title} {Jamming in hard sphere and disk packings},\ }\href
  {https://doi.org/10.1063/1.1633647} {\bibfield  {journal} {\bibinfo
  {journal} {J. Appl. Phys.}\ }\textbf {\bibinfo {volume} {95}},\ \bibinfo
  {pages} {989} (\bibinfo {year} {2004}{\natexlab{c}})}\BibitemShut {NoStop}%
\bibitem [{\citenamefont {Torquato}\ and\ \citenamefont
  {Jiao}(2010)}]{TJ_Algo}%
  \BibitemOpen
  \bibfield  {author} {\bibinfo {author} {\bibfnamefont {S.}~\bibnamefont
  {Torquato}}\ and\ \bibinfo {author} {\bibfnamefont {Y.}~\bibnamefont
  {Jiao}},\ }\bibfield  {title} {\bibinfo {title} {Robust algorithm to generate
  a diverse class of dense disordered and ordered sphere packings via linear
  programming},\ }\href {https://doi.org/10.1103/PhysRevE.82.061302} {\bibfield
   {journal} {\bibinfo  {journal} {Phys. Rev. E}\ }\textbf {\bibinfo {volume}
  {82}},\ \bibinfo {pages} {061302} (\bibinfo {year} {2010})}\BibitemShut
  {NoStop}%
\bibitem [{\citenamefont {Donev}\ \emph
  {et~al.}(2005{\natexlab{c}})\citenamefont {Donev}, \citenamefont {Torquato},\
  and\ \citenamefont {Stillinger}}]{DTS_AlgI}%
  \BibitemOpen
  \bibfield  {author} {\bibinfo {author} {\bibfnamefont {A.}~\bibnamefont
  {Donev}}, \bibinfo {author} {\bibfnamefont {S.}~\bibnamefont {Torquato}},\
  and\ \bibinfo {author} {\bibfnamefont {F.~H.}\ \bibnamefont {Stillinger}},\
  }\bibfield  {title} {\bibinfo {title} {Neighbor list collision-driven
  molecular dynamics simulation for nonspherical hard particles. i. algorithmic
  details},\ }\href {https://doi.org/https://doi.org/10.1016/j.jcp.2004.08.014}
  {\bibfield  {journal} {\bibinfo  {journal} {J. Computat. Phys.}\ }\textbf
  {\bibinfo {volume} {202}},\ \bibinfo {pages} {737} (\bibinfo {year}
  {2005}{\natexlab{c}})}\BibitemShut {NoStop}%
\bibitem [{\citenamefont {Donev}\ \emph
  {et~al.}(2005{\natexlab{d}})\citenamefont {Donev}, \citenamefont {Torquato},\
  and\ \citenamefont {Stillinger}}]{DTS_AlgII}%
  \BibitemOpen
  \bibfield  {author} {\bibinfo {author} {\bibfnamefont {A.}~\bibnamefont
  {Donev}}, \bibinfo {author} {\bibfnamefont {S.}~\bibnamefont {Torquato}},\
  and\ \bibinfo {author} {\bibfnamefont {F.~H.}\ \bibnamefont {Stillinger}},\
  }\bibfield  {title} {\bibinfo {title} {Neighbor list collision-driven
  molecular dynamics simulation for nonspherical hard particles.: Ii.
  applications to ellipses and ellipsoids},\ }\href
  {https://doi.org/https://doi.org/10.1016/j.jcp.2004.08.025} {\bibfield
  {journal} {\bibinfo  {journal} {J. Comput. Phys.}\ }\textbf {\bibinfo
  {volume} {202}},\ \bibinfo {pages} {765} (\bibinfo {year}
  {2005}{\natexlab{d}})}\BibitemShut {NoStop}%
\bibitem [{\citenamefont {Debye}\ \emph {et~al.}(1957)\citenamefont {Debye},
  \citenamefont {Anderson},\ and\ \citenamefont
  {Brumberger}}]{Debye_Scattering}%
  \BibitemOpen
  \bibfield  {author} {\bibinfo {author} {\bibfnamefont {P.}~\bibnamefont
  {Debye}}, \bibinfo {author} {\bibfnamefont {H.~R.}\ \bibnamefont
  {Anderson}},\ and\ \bibinfo {author} {\bibfnamefont {H.}~\bibnamefont
  {Brumberger}},\ }\bibfield  {title} {\bibinfo {title} {Scattering by an
  inhomogeneous solid. ii. the correlation function and its application},\
  }\href {https://doi.org/10.1063/1.1722830} {\bibfield  {journal} {\bibinfo
  {journal} {J. Appl. Phys.}\ }\textbf {\bibinfo {volume} {28}},\ \bibinfo
  {pages} {679} (\bibinfo {year} {1957})}\BibitemShut {NoStop}%
\bibitem [{\citenamefont {Torquato}(2021{\natexlab{b}})}]{Torqauto_Spread}%
  \BibitemOpen
  \bibfield  {author} {\bibinfo {author} {\bibfnamefont {S.}~\bibnamefont
  {Torquato}},\ }\bibfield  {title} {\bibinfo {title} {Diffusion spreadability
  as a probe of the microstructure of complex media across length scales},\
  }\href@noop {} {\bibfield  {journal} {\bibinfo  {journal} {Phys. Rev. E}\
  }\textbf {\bibinfo {volume} {104}},\ \bibinfo {pages} {054102} (\bibinfo
  {year} {2021}{\natexlab{b}})}\BibitemShut {NoStop}%
\bibitem [{\citenamefont {Stoyan}\ \emph {et~al.}(1995)\citenamefont {Stoyan},
  \citenamefont {Kendall},\ and\ \citenamefont {Mecke}}]{StochaticGeo_Text}%
  \BibitemOpen
  \bibfield  {author} {\bibinfo {author} {\bibfnamefont {D.}~\bibnamefont
  {Stoyan}}, \bibinfo {author} {\bibfnamefont {W.}~\bibnamefont {Kendall}},\
  and\ \bibinfo {author} {\bibfnamefont {J.}~\bibnamefont {Mecke}},\
  }\href@noop {} {\emph {\bibinfo {title} {Stochastic {G}eometry and {I}ts
  {A}pplications}}}\ (\bibinfo  {publisher} {Wiley, New York},\ \bibinfo {year}
  {1995})\BibitemShut {NoStop}%
\bibitem [{\citenamefont {Quintanilla}(2008)}]{Quint_Autoco}%
  \BibitemOpen
  \bibfield  {author} {\bibinfo {author} {\bibfnamefont {J.~A.}\ \bibnamefont
  {Quintanilla}},\ }\bibfield  {title} {\bibinfo {title} {Necessary and
  sufficient conditions for the two-point phase probability function of
  two-phase random media},\ }\href {https://doi.org/10.1098/rspa.2008.0023}
  {\bibfield  {journal} {\bibinfo  {journal} {Proc. R. Soc. A}\ }\textbf
  {\bibinfo {volume} {464}},\ \bibinfo {pages} {1761} (\bibinfo {year}
  {2008})}\BibitemShut {NoStop}%
\bibitem [{\citenamefont {Torquato}\ and\ \citenamefont
  {Stell}(1985)}]{Torquato_SD1}%
  \BibitemOpen
  \bibfield  {author} {\bibinfo {author} {\bibfnamefont {S.}~\bibnamefont
  {Torquato}}\ and\ \bibinfo {author} {\bibfnamefont {G.}~\bibnamefont
  {Stell}},\ }\bibfield  {title} {\bibinfo {title} {Microstructure of
  two‐phase random media. {V}. {T}he n‐point matrix probability functions
  for impenetrable spheres},\ }\href {https://doi.org/10.1063/1.448475}
  {\bibfield  {journal} {\bibinfo  {journal} {J. Chem. Phys.}\ }\textbf
  {\bibinfo {volume} {82}},\ \bibinfo {pages} {980} (\bibinfo {year}
  {1985})}\BibitemShut {NoStop}%
\bibitem [{\citenamefont
  {Torquato}(2016{\natexlab{a}})}]{Torquato_DisorderHUHet}%
  \BibitemOpen
  \bibfield  {author} {\bibinfo {author} {\bibfnamefont {S.}~\bibnamefont
  {Torquato}},\ }\bibfield  {title} {\bibinfo {title} {Disordered hyperuniform
  heterogeneous materials},\ }\href@noop {} {\bibfield  {journal} {\bibinfo
  {journal} {J. Condens. Matter Phys.}\ }\textbf {\bibinfo {volume} {28}},\
  \bibinfo {pages} {414012} (\bibinfo {year} {2016}{\natexlab{a}})}\BibitemShut
  {NoStop}%
\bibitem [{Sup()}]{Supp}%
  \BibitemOpen
  \href@noop {} {}\bibinfo {note} {See Supplemental Information at [LINK] for
  the form factors of the superballs used in this paper.}\BibitemShut {Stop}%
\bibitem [{\citenamefont {Zachary}\ and\ \citenamefont
  {Torquato}(2009)}]{Zachary2009}%
  \BibitemOpen
  \bibfield  {author} {\bibinfo {author} {\bibfnamefont {C.~E.}\ \bibnamefont
  {Zachary}}\ and\ \bibinfo {author} {\bibfnamefont {S.}~\bibnamefont
  {Torquato}},\ }\bibfield  {title} {\bibinfo {title} {Hyperuniformity in point
  patterns and two-phase random heterogeneous media},\ }\href
  {https://doi.org/10.1088/1742-5468/2009/12/p12015} {\bibfield  {journal}
  {\bibinfo  {journal} {J. Stat. Mech. Theory Exp.}\ }\textbf {\bibinfo
  {volume} {2009}},\ \bibinfo {pages} {P12015} (\bibinfo {year}
  {2009})}\BibitemShut {NoStop}%
\bibitem [{\citenamefont {Zachary}\ and\ \citenamefont
  {Torquato}(2011)}]{Zachary_Skclasses}%
  \BibitemOpen
  \bibfield  {author} {\bibinfo {author} {\bibfnamefont {C.~E.}\ \bibnamefont
  {Zachary}}\ and\ \bibinfo {author} {\bibfnamefont {S.}~\bibnamefont
  {Torquato}},\ }\bibfield  {title} {\bibinfo {title} {Anomalous local
  coordination, density fluctuations, and void statistics in disordered
  hyperuniform many-particle ground states},\ }\href@noop {} {\bibfield
  {journal} {\bibinfo  {journal} {Phys. Rev. E}\ }\textbf {\bibinfo {volume}
  {83}},\ \bibinfo {pages} {051133} (\bibinfo {year} {2011})}\BibitemShut
  {NoStop}%
\bibitem [{\citenamefont {Chen}\ \emph {et~al.}(2018)\citenamefont {Chen},
  \citenamefont {Lomba},\ and\ \citenamefont {Torquato}}]{Hchi_cutoff}%
  \BibitemOpen
  \bibfield  {author} {\bibinfo {author} {\bibfnamefont {D.}~\bibnamefont
  {Chen}}, \bibinfo {author} {\bibfnamefont {E.}~\bibnamefont {Lomba}},\ and\
  \bibinfo {author} {\bibfnamefont {S.}~\bibnamefont {Torquato}},\ }\bibfield
  {title} {\bibinfo {title} {Binary mixtures of charged colloids: a potential
  route to synthesize disordered hyperuniform materials},\ }\href@noop {}
  {\bibfield  {journal} {\bibinfo  {journal} {Phys. Chem. Chem. Phys.}\
  }\textbf {\bibinfo {volume} {20}},\ \bibinfo {pages} {17557} (\bibinfo {year}
  {2018})}\BibitemShut {NoStop}%
\bibitem [{\citenamefont {Prager}(1963)}]{PRAGER_PSD}%
  \BibitemOpen
  \bibfield  {author} {\bibinfo {author} {\bibfnamefont {S.}~\bibnamefont
  {Prager}},\ }\bibfield  {title} {\bibinfo {title} {Interphase transfer in
  stationary two-phase media},\ }\href
  {https://doi.org/https://doi.org/10.1016/0009-2509(63)87003-7} {\bibfield
  {journal} {\bibinfo  {journal} {Chem. Eng. Sci.}\ }\textbf {\bibinfo {volume}
  {18}},\ \bibinfo {pages} {227} (\bibinfo {year} {1963})}\BibitemShut
  {NoStop}%
\bibitem [{\citenamefont {Rubinstein}\ and\ \citenamefont
  {Torquato}(1989)}]{rubinstein_torquato_1989}%
  \BibitemOpen
  \bibfield  {author} {\bibinfo {author} {\bibfnamefont {J.}~\bibnamefont
  {Rubinstein}}\ and\ \bibinfo {author} {\bibfnamefont {S.}~\bibnamefont
  {Torquato}},\ }\bibfield  {title} {\bibinfo {title} {Flow in random porous
  media: mathematical formulation, variational principles, and rigorous
  bounds},\ }\href {https://doi.org/10.1017/S0022112089002211} {\bibfield
  {journal} {\bibinfo  {journal} {J. of Fluid Mech.}\ }\textbf {\bibinfo
  {volume} {206}},\ \bibinfo {pages} {25–46} (\bibinfo {year}
  {1989})}\BibitemShut {NoStop}%
\bibitem [{\citenamefont {Avellaneda}\ and\ \citenamefont
  {Torquato}(1991)}]{Avellaneda_fluid}%
  \BibitemOpen
  \bibfield  {author} {\bibinfo {author} {\bibfnamefont {M.}~\bibnamefont
  {Avellaneda}}\ and\ \bibinfo {author} {\bibfnamefont {S.}~\bibnamefont
  {Torquato}},\ }\bibfield  {title} {\bibinfo {title} {Rigorous link between
  fluid permeability, electrical conductivity, and relaxation times for
  transport in porous media},\ }\href {https://doi.org/10.1063/1.858194}
  {\bibfield  {journal} {\bibinfo  {journal} {Phys. Fluids A}\ }\textbf
  {\bibinfo {volume} {3}},\ \bibinfo {pages} {2529} (\bibinfo {year}
  {1991})}\BibitemShut {NoStop}%
\bibitem [{\citenamefont {Torquato}(2020)}]{TORQUATO_water}%
  \BibitemOpen
  \bibfield  {author} {\bibinfo {author} {\bibfnamefont {S.}~\bibnamefont
  {Torquato}},\ }\bibfield  {title} {\bibinfo {title} {Predicting transport
  characteristics of hyperuniform porous media via rigorous
  microstructure-property relations},\ }\href
  {https://doi.org/https://doi.org/10.1016/j.advwatres.2020.103565} {\bibfield
  {journal} {\bibinfo  {journal} {Adv. in Water Resour.}\ }\textbf {\bibinfo
  {volume} {140}},\ \bibinfo {pages} {103565} (\bibinfo {year}
  {2020})}\BibitemShut {NoStop}%
\bibitem [{\citenamefont {Klatt}\ \emph {et~al.}(2021)\citenamefont {Klatt},
  \citenamefont {Ziff},\ and\ \citenamefont {Torquato}}]{Klatt_Pore}%
  \BibitemOpen
  \bibfield  {author} {\bibinfo {author} {\bibfnamefont {M.~A.}\ \bibnamefont
  {Klatt}}, \bibinfo {author} {\bibfnamefont {R.~M.}\ \bibnamefont {Ziff}},\
  and\ \bibinfo {author} {\bibfnamefont {S.}~\bibnamefont {Torquato}},\
  }\bibfield  {title} {\bibinfo {title} {Critical pore radius and transport
  properties of disordered hard- and overlapping-sphere models},\ }\href
  {https://doi.org/10.1103/PhysRevE.104.014127} {\bibfield  {journal} {\bibinfo
   {journal} {Phys. Rev. E}\ }\textbf {\bibinfo {volume} {104}},\ \bibinfo
  {pages} {014127} (\bibinfo {year} {2021})}\BibitemShut {NoStop}%
\bibitem [{\citenamefont {Torquato}(1985)}]{Torquato_F}%
  \BibitemOpen
  \bibfield  {author} {\bibinfo {author} {\bibfnamefont {S.}~\bibnamefont
  {Torquato}},\ }\bibfield  {title} {\bibinfo {title} {Effective electrical
  conductivity of two‐phase disordered composite media},\ }\href
  {https://doi.org/10.1063/1.335593} {\bibfield  {journal} {\bibinfo  {journal}
  {J. Appl. Phys.}\ }\textbf {\bibinfo {volume} {58}},\ \bibinfo {pages} {3790}
  (\bibinfo {year} {1985})}\BibitemShut {NoStop}%
\bibitem [{\citenamefont {Kim}\ and\ \citenamefont {Torquato}(1991)}]{Kim_F}%
  \BibitemOpen
  \bibfield  {author} {\bibinfo {author} {\bibfnamefont {I.~C.}\ \bibnamefont
  {Kim}}\ and\ \bibinfo {author} {\bibfnamefont {S.}~\bibnamefont {Torquato}},\
  }\bibfield  {title} {\bibinfo {title} {Effective conductivity of suspensions
  of hard spheres by brownian motion simulation},\ }\href
  {https://doi.org/10.1063/1.348708} {\bibfield  {journal} {\bibinfo  {journal}
  {J. Appl. Phys.}\ }\textbf {\bibinfo {volume} {69}},\ \bibinfo {pages} {2280}
  (\bibinfo {year} {1991})}\BibitemShut {NoStop}%
\bibitem [{\citenamefont {Robinson}\ and\ \citenamefont
  {Friedman}(2005)}]{ROBINSON_F}%
  \BibitemOpen
  \bibfield  {author} {\bibinfo {author} {\bibfnamefont {D.~A.}\ \bibnamefont
  {Robinson}}\ and\ \bibinfo {author} {\bibfnamefont {S.~P.}\ \bibnamefont
  {Friedman}},\ }\bibfield  {title} {\bibinfo {title} {Electrical conductivity
  and dielectric permittivity of sphere packings: Measurements and modelling of
  cubic lattices, randomly packed monosize spheres and multi-size mixtures},\
  }\href {https://doi.org/https://doi.org/10.1016/j.physa.2005.03.054}
  {\bibfield  {journal} {\bibinfo  {journal} {Physica A}\ }\textbf {\bibinfo
  {volume} {358}},\ \bibinfo {pages} {447} (\bibinfo {year}
  {2005})}\BibitemShut {NoStop}%
\bibitem [{\citenamefont {Gillman}\ and\ \citenamefont
  {Matouš}(2014)}]{GILLMAN_FA}%
  \BibitemOpen
  \bibfield  {author} {\bibinfo {author} {\bibfnamefont {A.}~\bibnamefont
  {Gillman}}\ and\ \bibinfo {author} {\bibfnamefont {K.}~\bibnamefont
  {Matouš}},\ }\bibfield  {title} {\bibinfo {title} {Third-order model of
  thermal conductivity for random polydisperse particulate materials using
  well-resolved statistical descriptions from tomography},\ }\href
  {https://doi.org/https://doi.org/10.1016/j.physleta.2014.08.032} {\bibfield
  {journal} {\bibinfo  {journal} {Phys. Lett. A}\ }\textbf {\bibinfo {volume}
  {378}},\ \bibinfo {pages} {3070} (\bibinfo {year} {2014})}\BibitemShut
  {NoStop}%
\bibitem [{\citenamefont {Gillman}\ \emph {et~al.}(2015)\citenamefont
  {Gillman}, \citenamefont {Amadio}, \citenamefont {Matouš},\ and\
  \citenamefont {Jackson}}]{gillman_FB}%
  \BibitemOpen
  \bibfield  {author} {\bibinfo {author} {\bibfnamefont {A.}~\bibnamefont
  {Gillman}}, \bibinfo {author} {\bibfnamefont {G.}~\bibnamefont {Amadio}},
  \bibinfo {author} {\bibfnamefont {K.}~\bibnamefont {Matouš}},\ and\ \bibinfo
  {author} {\bibfnamefont {T.~L.}\ \bibnamefont {Jackson}},\ }\bibfield
  {title} {\bibinfo {title} {Third-order thermo-mechanical properties for packs
  of platonic solids using statistical micromechanics},\ }\href
  {https://doi.org/10.1098/rspa.2015.0060} {\bibfield  {journal} {\bibinfo
  {journal} {Proc. R. Soc. A}\ }\textbf {\bibinfo {volume} {471}},\ \bibinfo
  {pages} {20150060} (\bibinfo {year} {2015})}\BibitemShut {NoStop}%
\bibitem [{\citenamefont {Nguyen}\ \emph {et~al.}(2016)\citenamefont {Nguyen},
  \citenamefont {Monchiet}, \citenamefont {Bonnet},\ and\ \citenamefont
  {To}}]{Nguyen_F}%
  \BibitemOpen
  \bibfield  {author} {\bibinfo {author} {\bibfnamefont {M.-T.}\ \bibnamefont
  {Nguyen}}, \bibinfo {author} {\bibfnamefont {V.}~\bibnamefont {Monchiet}},
  \bibinfo {author} {\bibfnamefont {G.}~\bibnamefont {Bonnet}},\ and\ \bibinfo
  {author} {\bibfnamefont {Q.-D.}\ \bibnamefont {To}},\ }\bibfield  {title}
  {\bibinfo {title} {Conductivity estimates of spherical-particle suspensions
  based on triplet structure factors},\ }\href
  {https://doi.org/10.1103/PhysRevE.93.022105} {\bibfield  {journal} {\bibinfo
  {journal} {Phys. Rev. E}\ }\textbf {\bibinfo {volume} {93}},\ \bibinfo
  {pages} {022105} (\bibinfo {year} {2016})}\BibitemShut {NoStop}%
\bibitem [{\citenamefont {Hashin}\ and\ \citenamefont
  {Shtrikman}(1963)}]{HASHIN_F}%
  \BibitemOpen
  \bibfield  {author} {\bibinfo {author} {\bibfnamefont {Z.}~\bibnamefont
  {Hashin}}\ and\ \bibinfo {author} {\bibfnamefont {S.}~\bibnamefont
  {Shtrikman}},\ }\bibfield  {title} {\bibinfo {title} {A variational approach
  to the theory of the elastic behaviour of multiphase materials},\ }\href
  {https://doi.org/https://doi.org/10.1016/0022-5096(63)90060-7} {\bibfield
  {journal} {\bibinfo  {journal} {J Mech Phys}\ }\textbf {\bibinfo {volume}
  {11}},\ \bibinfo {pages} {127} (\bibinfo {year} {1963})}\BibitemShut
  {NoStop}%
\bibitem [{\citenamefont {Torquato}\ and\ \citenamefont
  {Avellaneda}(1991)}]{Torquato_T1}%
  \BibitemOpen
  \bibfield  {author} {\bibinfo {author} {\bibfnamefont {S.}~\bibnamefont
  {Torquato}}\ and\ \bibinfo {author} {\bibfnamefont {M.}~\bibnamefont
  {Avellaneda}},\ }\bibfield  {title} {\bibinfo {title} {Diffusion and reaction
  in heterogeneous media: Pore size distribution, relaxation times, and mean
  survival time},\ }\href {https://doi.org/10.1063/1.461519} {\bibfield
  {journal} {\bibinfo  {journal} {J. Chem. Phys.}\ }\textbf {\bibinfo {volume}
  {95}},\ \bibinfo {pages} {6477} (\bibinfo {year} {1991})}\BibitemShut
  {NoStop}%
\bibitem [{\citenamefont {Torquato}\ and\ \citenamefont
  {Jiao}(2013)}]{Torquato_Perc1}%
  \BibitemOpen
  \bibfield  {author} {\bibinfo {author} {\bibfnamefont {S.}~\bibnamefont
  {Torquato}}\ and\ \bibinfo {author} {\bibfnamefont {Y.}~\bibnamefont
  {Jiao}},\ }\bibfield  {title} {\bibinfo {title} {Effect of dimensionality on
  the percolation threshold of overlapping nonspherical hyperparticles},\
  }\href {https://doi.org/10.1103/PhysRevE.87.022111} {\bibfield  {journal}
  {\bibinfo  {journal} {Phys. Rev. E}\ }\textbf {\bibinfo {volume} {87}},\
  \bibinfo {pages} {022111} (\bibinfo {year} {2013})}\BibitemShut {NoStop}%
\bibitem [{\citenamefont {Lubachevsky}\ and\ \citenamefont
  {Stillinger}(1990)}]{LS_ALG}%
  \BibitemOpen
  \bibfield  {author} {\bibinfo {author} {\bibfnamefont {B.~D.}\ \bibnamefont
  {Lubachevsky}}\ and\ \bibinfo {author} {\bibfnamefont {F.~H.}\ \bibnamefont
  {Stillinger}},\ }\bibfield  {title} {\bibinfo {title} {Geometric properties
  of random disk packings},\ }\href {https://doi.org/10.1007/bf01025983}
  {\bibfield  {journal} {\bibinfo  {journal} {J. Stat. Phys.}\ }\textbf
  {\bibinfo {volume} {60}},\ \bibinfo {pages} {561–583} (\bibinfo {year}
  {1990})}\BibitemShut {NoStop}%
\bibitem [{\citenamefont {Donev}(2006)}]{donev_thesis}%
  \BibitemOpen
  \bibfield  {author} {\bibinfo {author} {\bibfnamefont {A.}~\bibnamefont
  {Donev}},\ }\emph {\bibinfo {title} {Jammed Packings of Hard Particles}},\
  \href@noop {} {Ph.D. thesis} (\bibinfo {year} {2006})\BibitemShut {NoStop}%
\bibitem [{\citenamefont {Coker}\ and\ \citenamefont
  {Torquato}(1995)}]{Coker_Digit}%
  \BibitemOpen
  \bibfield  {author} {\bibinfo {author} {\bibfnamefont {D.~A.}\ \bibnamefont
  {Coker}}\ and\ \bibinfo {author} {\bibfnamefont {S.}~\bibnamefont
  {Torquato}},\ }\bibfield  {title} {\bibinfo {title} {Extraction of
  morphological quantities from a digitized medium},\ }\href
  {https://doi.org/10.1063/1.359134} {\bibfield  {journal} {\bibinfo  {journal}
  {J. Appl. Phys.}\ }\textbf {\bibinfo {volume} {77}},\ \bibinfo {pages} {6087}
  (\bibinfo {year} {1995})}\BibitemShut {NoStop}%
\bibitem [{\citenamefont {Song}\ \emph {et~al.}(2019)\citenamefont {Song},
  \citenamefont {Ding},\ and\ \citenamefont {Wei}}]{Pore_Alg}%
  \BibitemOpen
  \bibfield  {author} {\bibinfo {author} {\bibfnamefont {S.}~\bibnamefont
  {Song}}, \bibinfo {author} {\bibfnamefont {Q.}~\bibnamefont {Ding}},\ and\
  \bibinfo {author} {\bibfnamefont {J.}~\bibnamefont {Wei}},\ }\bibfield
  {title} {\bibinfo {title} {Improved algorithm for estimating pore size
  distribution from pore space images of porous media},\ }\href
  {https://doi.org/10.1103/PhysRevE.100.053314} {\bibfield  {journal} {\bibinfo
   {journal} {Phys. Rev. E}\ }\textbf {\bibinfo {volume} {100}},\ \bibinfo
  {pages} {053314} (\bibinfo {year} {2019})}\BibitemShut {NoStop}%
\bibitem [{\citenamefont {A.}(1979)}]{santalo}%
  \BibitemOpen
  \bibfield  {author} {\bibinfo {author} {\bibfnamefont {L.~A.}\ \bibnamefont
  {Santal\'o}},\ }\href@noop {} {\emph {\bibinfo {title} {Integral geometry and
  geometric probability}}}\ (\bibinfo  {publisher} {Addison-Wesley},\ \bibinfo
  {year} {1979})\BibitemShut {NoStop}%
\bibitem [{\citenamefont {Wilken}\ \emph {et~al.}(2021)\citenamefont {Wilken},
  \citenamefont {Guerra}, \citenamefont {Levine},\ and\ \citenamefont
  {Chaikin}}]{Wilken_ClassIII}%
  \BibitemOpen
  \bibfield  {author} {\bibinfo {author} {\bibfnamefont {S.}~\bibnamefont
  {Wilken}}, \bibinfo {author} {\bibfnamefont {R.~E.}\ \bibnamefont {Guerra}},
  \bibinfo {author} {\bibfnamefont {D.}~\bibnamefont {Levine}},\ and\ \bibinfo
  {author} {\bibfnamefont {P.~M.}\ \bibnamefont {Chaikin}},\ }\bibfield
  {title} {\bibinfo {title} {Random close packing as a dynamical phase
  transition},\ }\href@noop {} {\bibfield  {journal} {\bibinfo  {journal}
  {Phys. Rev. Lett.}\ }\textbf {\bibinfo {volume} {127}},\ \bibinfo {pages}
  {038002} (\bibinfo {year} {2021})}\BibitemShut {NoStop}%
\bibitem [{\citenamefont {Yousefi}\ \emph {et~al.}(2019)\citenamefont
  {Yousefi}, \citenamefont {Malmir},\ and\ \citenamefont
  {Sahimi}}]{Concave_SB}%
  \BibitemOpen
  \bibfield  {author} {\bibinfo {author} {\bibfnamefont {P.}~\bibnamefont
  {Yousefi}}, \bibinfo {author} {\bibfnamefont {H.}~\bibnamefont {Malmir}},\
  and\ \bibinfo {author} {\bibfnamefont {M.}~\bibnamefont {Sahimi}},\
  }\bibfield  {title} {\bibinfo {title} {Morphology and kinetics of random
  sequential adsorption of superballs: From hexapods to cubes},\ }\href
  {https://doi.org/10.1103/PhysRevE.100.020602} {\bibfield  {journal} {\bibinfo
   {journal} {Phys. Rev. E}\ }\textbf {\bibinfo {volume} {100}},\ \bibinfo
  {pages} {020602} (\bibinfo {year} {2019})}\BibitemShut {NoStop}%
\bibitem [{\citenamefont {VanderWerf}\ \emph {et~al.}(2018)\citenamefont
  {VanderWerf}, \citenamefont {Jin}, \citenamefont {Shattuck},\ and\
  \citenamefont {O'Hern}}]{O'Hern_hypos}%
  \BibitemOpen
  \bibfield  {author} {\bibinfo {author} {\bibfnamefont {K.}~\bibnamefont
  {VanderWerf}}, \bibinfo {author} {\bibfnamefont {W.}~\bibnamefont {Jin}},
  \bibinfo {author} {\bibfnamefont {M.~D.}\ \bibnamefont {Shattuck}},\ and\
  \bibinfo {author} {\bibfnamefont {C.~S.}\ \bibnamefont {O'Hern}},\ }\bibfield
   {title} {\bibinfo {title} {Hypostatic jammed packings of frictionless
  nonspherical particles},\ }\href@noop {} {\bibfield  {journal} {\bibinfo
  {journal} {Phys. Rev. E}\ }\textbf {\bibinfo {volume} {97}},\ \bibinfo
  {pages} {012909} (\bibinfo {year} {2018})}\BibitemShut {NoStop}%
\end{thebibliography}

\providecommand{\noopsort}[1]{}\providecommand{\singleletter}[1]{#1}%
\end{document}


\title{Supporting Information: Characterization of Void Space, Large-Scale Structure, and Transport Properties of Maximally Random Jammed Packings of Superballs - Supplemental Material}

\author{Charles Emmett Maher}
\affiliation{Department of Chemistry, Princeton University, Princeton, New Jersey 08544, USA}
\author{Frank H. Stillinger}
\affiliation{Department of Chemistry, Princeton University, Princeton, New Jersey 08544, USA}
\author{Salvatore Torquato}%
\email{torquato@princeton.edu}
\homepage{http://chemlabs.princeton.edu}
\affiliation{Department of Chemistry, Princeton University, Princeton, New Jersey 08544, USA}
\affiliation{Department of Physics, Princeton University, Princeton, New Jersey 08544, USA}
\affiliation{Princeton Institute for the Science and Technology of Materials, Princeton University, Princeton, New Jersey 08544, USA}
\affiliation{Program in Applied and Computational Mathematics, Princeton University, Princeton, New Jersey 08544, USA}


\maketitle
\section{Symbol Glossary}
\begin{longtable}{l|l}
\caption{List of symbols and notation used in the text.}
\\ \hline
Symbol                                       & Description                                                       \\ \hline \hline
$\mathbb{R}^d$                               & $d$-dimensional Euclidean space                                   \\ \hline
$\phi$                                       & Packing fraction                                                  \\ \hline
$\bar{Z}$                                    & Average contacts per particle                                     \\ \hline
$f$                                          & Degree of freedom                                                 \\ \hline
$\phi_R$                                     & Rattler fraction                                                  \\ \hline
$p$                                          & Superball deformation parameter                                   \\ \hline
$a$                                          & Aspect ratio                                                      \\ \hline
$a^*$                                        & Ellipsoid critical aspect ratio
    \\ \hline
$S(Q)$                                       & Structure factor                                                  \\ \hline
$\tilde{\chi}_{_V}(Q)$                        & Spectral density                                                  \\ \hline
$\alpha$                                     & Hyperuniformity scaling exponent                                  \\ \hline
$g_2(r)$                                     & Pair correlation function                                         \\ \hline
$\chi_{_V}(r)$                               & Phase autocovariance function                                     \\ \hline
$\mathcal{S}(t)$                             & Spreadability                                                     \\ \hline
$\delta$                                     & Pore radius
    \\ \hline
$F(\delta)$                                  & Pore-size distribution function                                   \\ \hline
$k$                                          & Fluid permeability                                                \\ \hline
$\tau$                                       & Mean survival time                                                \\ \hline
$T_1$                                        & Principal relaxation time                                         \\ \hline
$\rho_n(\mathbf{r}_1,\dots,\mathbf{r}_n)$    & $n$-particle correlation function                                 \\ \hline
$\rho$                                       & Number density                                                    \\ \hline
$h(r)$                                       & Total correlation function ($g_2(r)-1$)                           \\ \hline
$\tilde{h}(Q)$                               & Fourier transform of $h(r)$
    \\ \hline
$\Lambda$                                    & Lattice                                                           \\ \hline
$F$                                          & Fundamental cell                                                 \\ \hline
$V_F$                                        & Fundamental cell volume                                           \\ \hline
$\mathbb{S}(Q)$                              & Scattering intensity                                              \\ \hline
$\delta(Q)$                                  & Dirac delta function                                              \\ \hline
$\mathcal{V}_i$                              & Volume of phase $i$                                               \\ \hline
$S_n^{(i)}(\mathbf{x}_1,\dots,\mathbf{x}_n)$ & $n$-point probability function for phase $i$                      \\ \hline
$\mathcal{I}^{(i)}(x)$                       & Phase indicator function for phase $i$                            \\ \hline
$\mathbf{R}$                                 & Geometrical parameters of particle shape                       \\ \hline
$m(r; \mathbf{R})$                           & Particle indicator function                                       \\ \hline
$\tilde{m}(Q; \mathbf{R})$                   & Particle form factor                                              \\ \hline
$v_1$                                        & Single-particle volume                                            \\ \hline
$H$                                          & Hyperuniformity index                                             \\ \hline
$P(\delta)$                                  & Pore size probability density                                     \\ \hline
$\langle \delta\rangle$                      & Mean pore size                                                    \\ \hline
$\langle \delta^2\rangle$                    & Second moment of $P(\delta)$                                      \\ \hline
$\mathcal{F}$                                & Formation factor                                                  \\ \hline
$\mathcal{L}$                                & Length scale determined by the 
    \\
                                             & eigenvalues of the Stokes operator
    \\ \hline
$\Theta_n$                                   & Viscous relaxation time                                           \\ \hline
$\sigma_i$                                   & Conductivity of phase $i$                                         \\ \hline
$\zeta_2$                                    & 3-point microstructural parameter                                 \\ \hline
$\mathcal{D}$                                & Diffusion coefficient                                             \\ \hline
$\kappa$                                     & Reaction rate                                                     \\ \hline
$s$                                          & Specific surface                                                  \\ \hline
$\omega_d$                                   & Volume of a $d$-dimensional unit sphere                           \\ \hline
$w(\mathbf{n})$                              & Width of a convex body in direction $\mathbf{n}$
    \\ \hline
$\bar{w}$                                    & Mean width                                                        \\ \hline
$\gamma$                                     & Expansion rate                                                    \\ \hline
$P$                                          & Pressure                                                          \\ \hline
$\mathbf{O}^{-1}$                            & Matrix describing the sphere                                      \\ \hline
$\epsilon$                                   & Voxelization scaling parameter                                    \\ \hline
$\tilde{r}$                                  & Scaled particle position                                          \\ \hline
$\mathbb{N}_0$                               & Natural numbers including 0
    \\ \hline
$l_i$                                        & Length of edge $i$
    \\ \hline
$\theta_i$                                   & Angle between the two faces meeting at edge $i$
    \\ \hline
$\hat{a}$                                    & Debye random media length scale
    \\ \hline
$D$                                          & Sphere diameter                                                   \\ \hline
$A$                                          & Superball major axis                                              
\end{longtable}

\section{Form Factors For Superballs}
\begin{figure}[h]
    \centering
    \includegraphics[width = 0.45\textwidth]{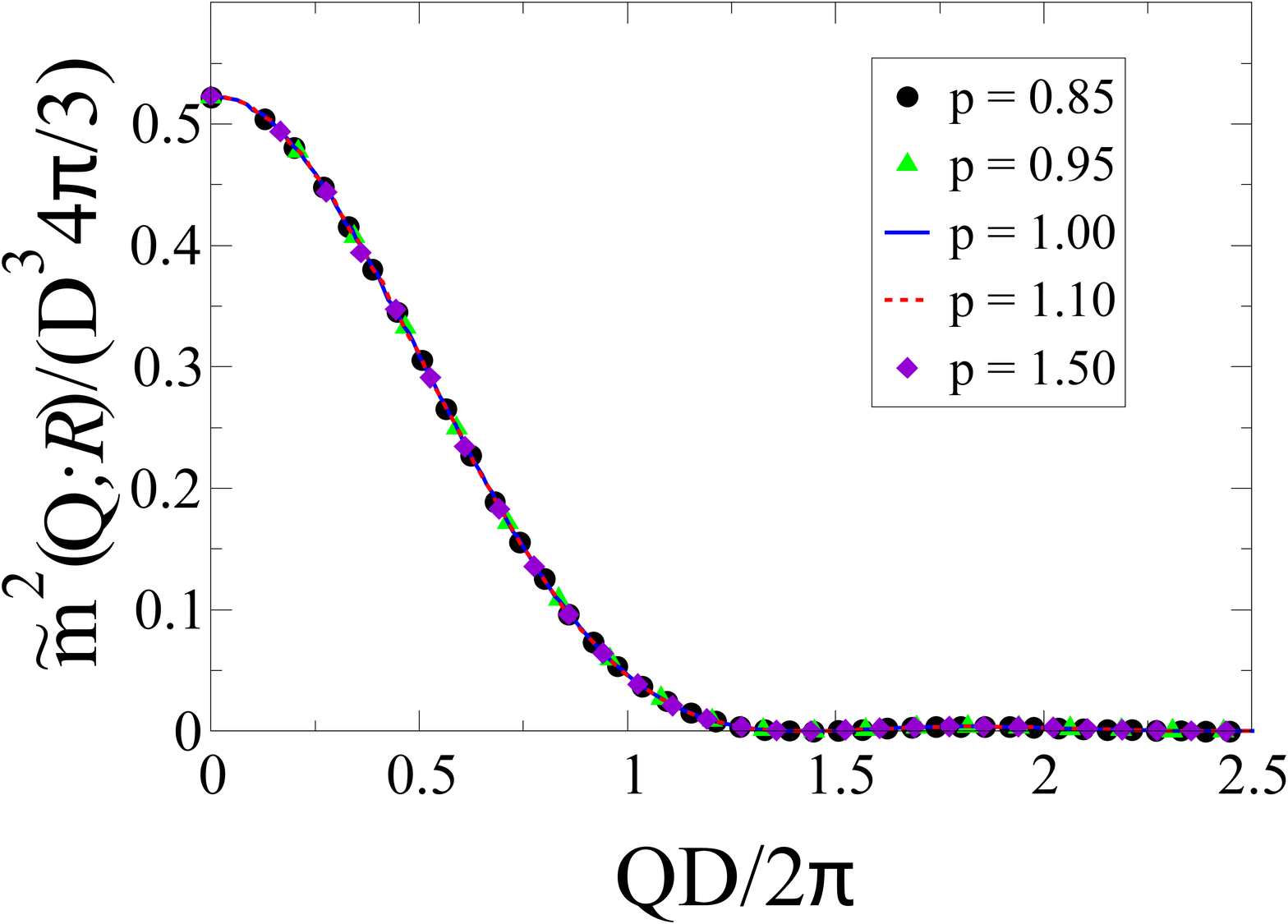}
    \caption{The scaled form factor $\tilde{m}(k;\mathbf{R})/(D^34\pi/3)$ where $D$ is the diameter of the unit sphere as a function of scaled wavenumber $QD/2\pi$.}
    \label{fig:formlin}
\end{figure}

\begin{figure}[h]
    \centering
    \includegraphics[width = 0.45\textwidth]{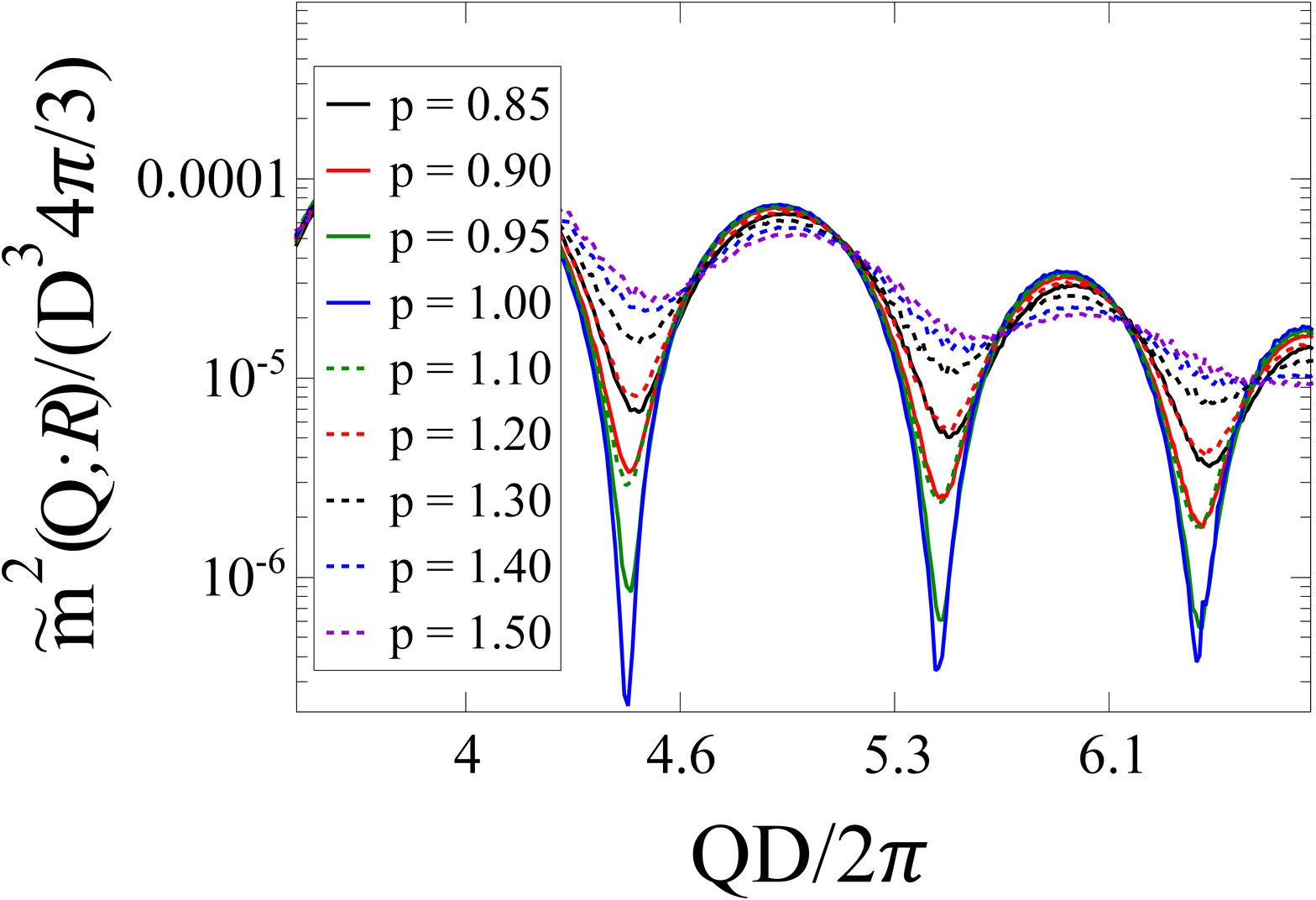}
    \caption{The scaled form factor $\tilde{m}(k;\mathbf{R})/(D^34\pi/3)$ where $D$ is the diameter of the unit sphere as a function of scaled wavenumber $QD/2\pi$ on a log-log scale.}
    \label{fig:formlog}
\end{figure}

Here, we present the angular-averaged form factors $\tilde{m}(k;\mathbf{R})$ (defined in Sec. II A) for a selection of the superballs which have their spectral densities computed in the main text (i.e., $0.85\leq p\leq1.50$).
We apply our novel voxelization procedure, described in Sec. III B, to a a simulation box containing a single superball and apply Eq. (12) to the result.
These voxelizations have a resolution of $500^3$ voxels.
Each of the particles considered has the same volume, chosen here to be equal to that of the unit sphere.
Figure \ref{fig:formlin} shows the scaled form factors of the superballs.
In this range of $p$ values, when scaled by the diameter of a sphere with the same volume as the superball, all of the form factors appear to be nearly identical to that of a sphere, with deviations due to error associated with the voxelization procedure.
Such errors arise because the smooth surface of the superballs in this range of $p$ values cannot be exactly replicated by discrete voxels.
Specifically, there are a small number of voxels near the boundary of the superball erroneously set as empty.
At the resolution used here, these errors result in a very small underestimation of the volume of the superball, but this has a very small impact on the form factor.
Figure \ref{fig:formlog} shows a zoomed-in portion of the scaled form factors on a log-log scale, where we see that as $|1-p|$ increases, peaks at larger wavenumbers start to become less pronounced. 

\section{Mean Width Scaling}
\begin{figure}[h]
    \centering
    \includegraphics[width = 0.48\textwidth]{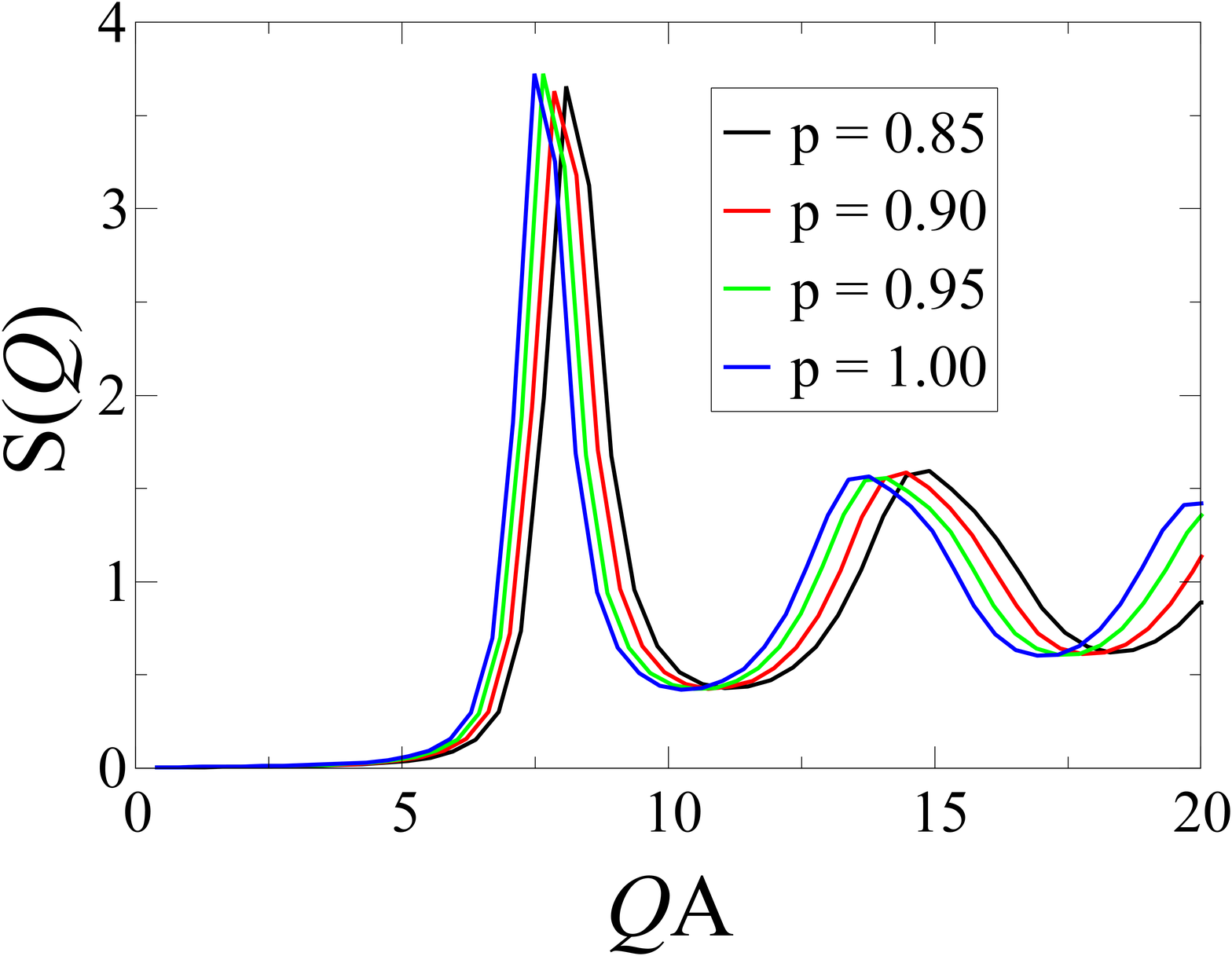}
    \caption{The structure factor $S(Q)$ as a function of wavenumber $Q$ scaled by the superball major axis legth $A$ for values of the deformation parameter $p \leq 1$.}
    \label{fig:Sksm}
\end{figure}

\begin{figure}[h]
    \centering
    \includegraphics[width = 0.48\textwidth]{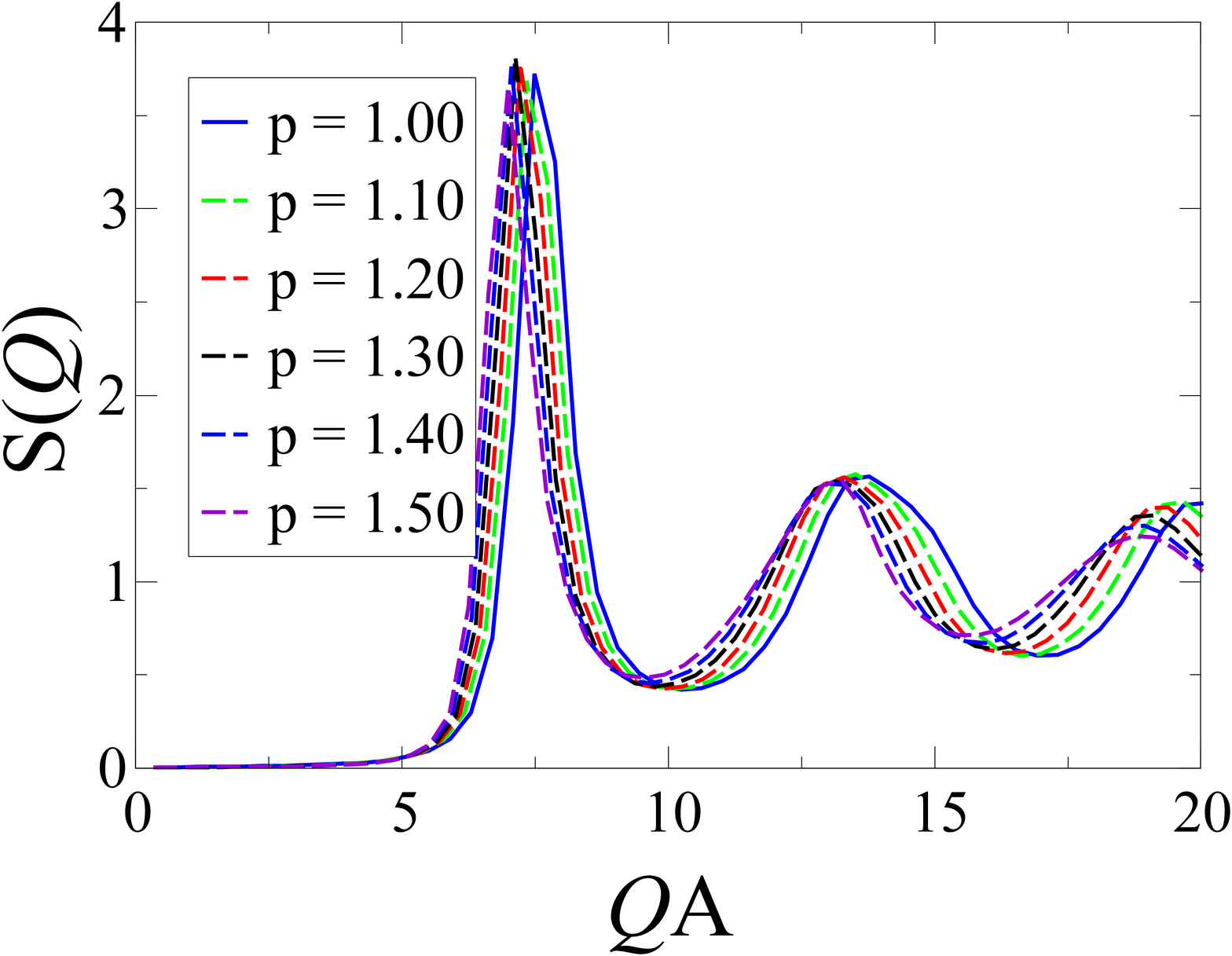}
    \caption{The structure factor $S(Q)$ as a function of wavenumber $Q$ scaled by the superball major axis legth $A$ for values of the deformation parameter $p \geq 1$.}
    \label{fig:Sklg}
\end{figure}

\begin{figure}[h]
    \centering
    \includegraphics[width = 0.48\textwidth]{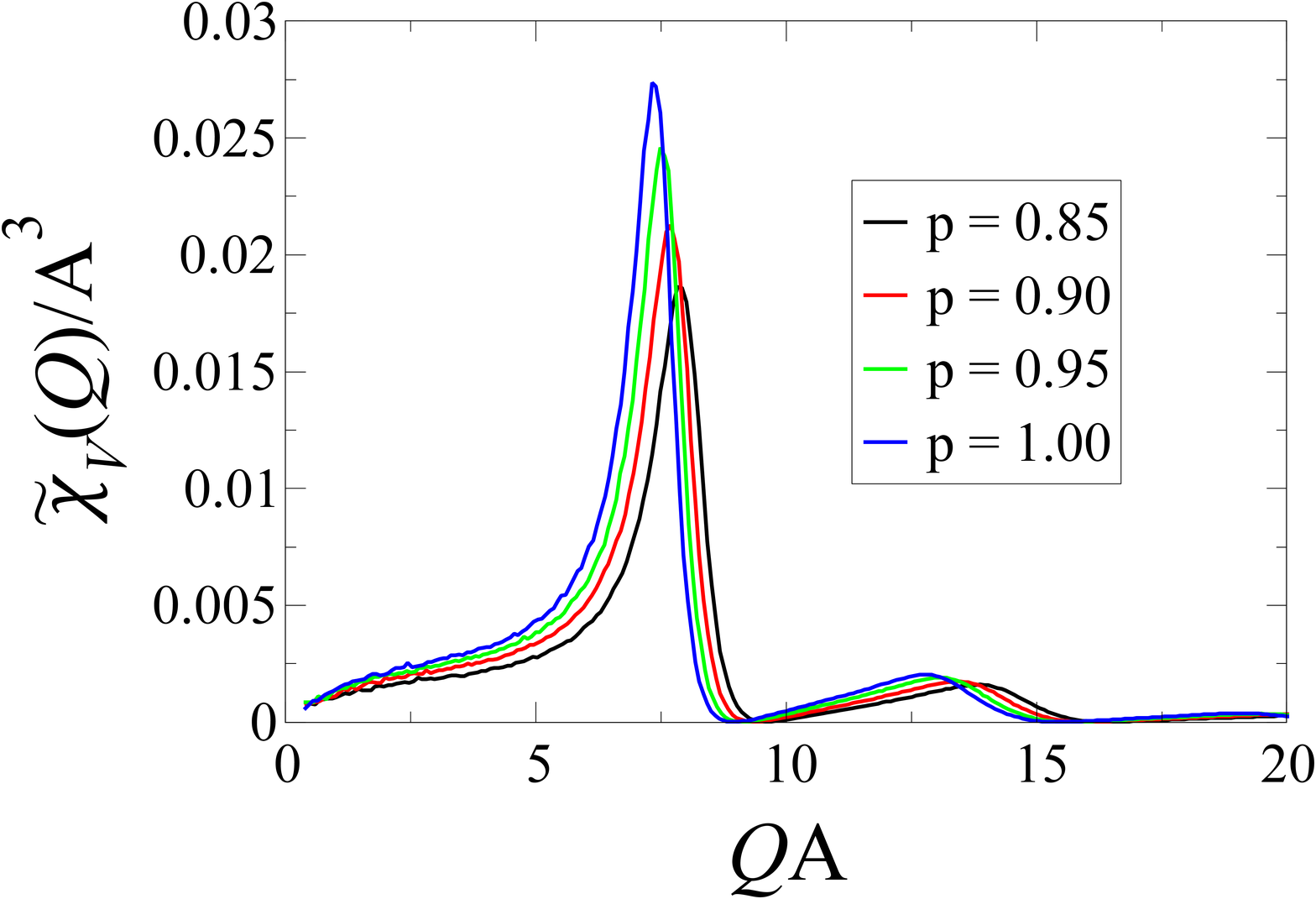}
    \caption{The scaled spectral density $\spD{Q}/A^3$ as a function of wavenumber $Q$ scaled by the superball major axis $A$ for values of the deformation parameter $p \leq 1$.}
    \label{fig:chism}
\end{figure}

\begin{figure}[h]
    \centering
    \includegraphics[width = 0.48\textwidth]{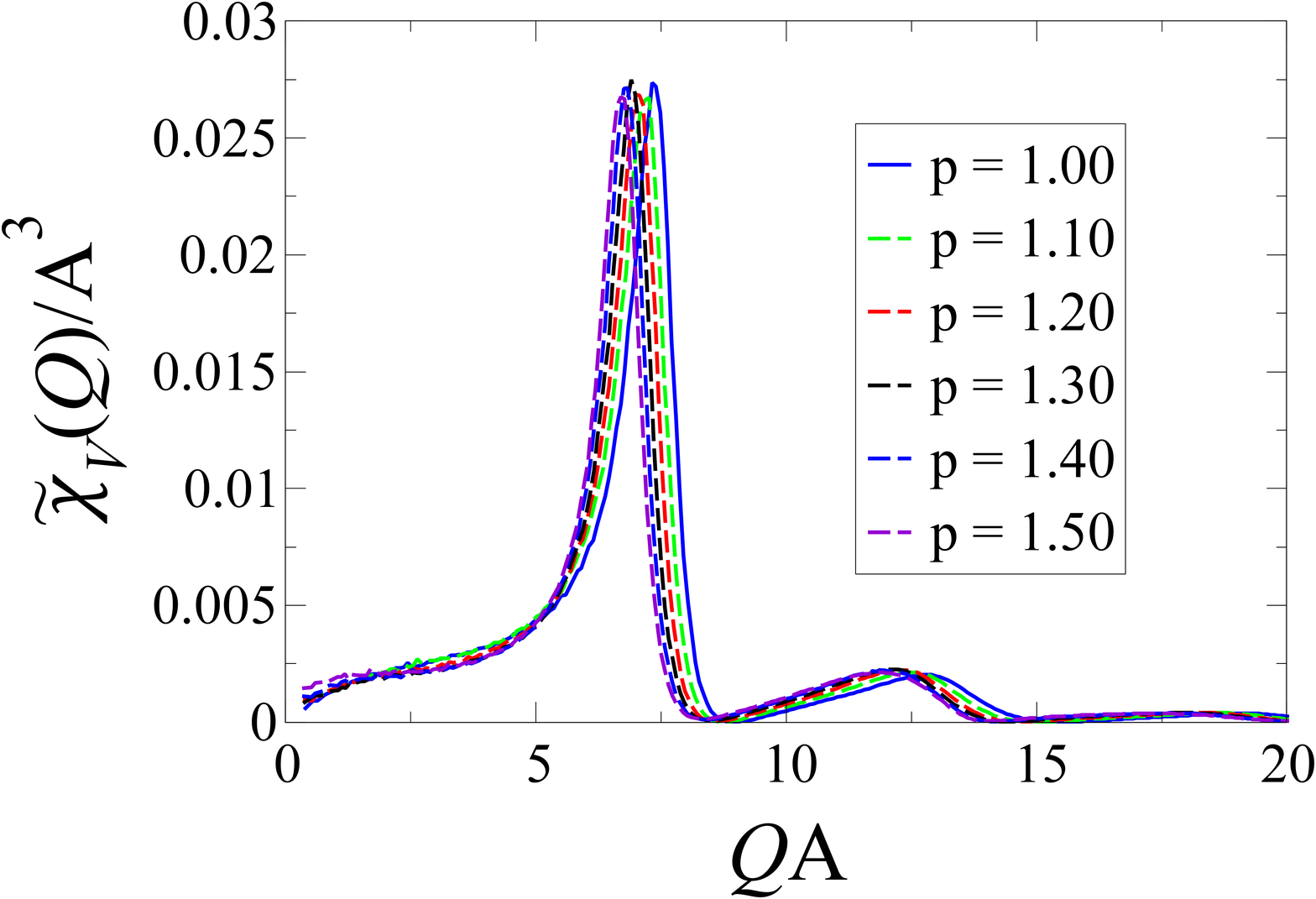}
    \caption{The scaled spectral density $\spD{Q}/A^3$ as a function of wavenumber $Q$ scaled by the superball major axis $A$ for values of the deformation parameter $p \geq 1$.}
    \label{fig:chilg}
\end{figure}

Figures \ref{fig:Sksm}, \ref{fig:Sklg}, \ref{fig:chism}, and \ref{fig:chilg} show the structure factors $S(Q)$ and spectral densities $\spD{Q}$ scaled by the major axis length $A$ of the superballs instead of the mean widths respectively.
Compared to Figs. 8, 9, 10, and 11 in the main text, it is clear that the peaks in these figures are more broadly distributed.
In particular, superballs with $p < 1$ have their peaks pushed to larger wavenumbers, while superballs with $p > 1$ have their peaks brought closer to the origin.
Thus, we claim that the mean width is a more sensible choice of scale than the superball major axis length.




\clearpage

\section{Raw Data for $\phi_R, \phi,$ and $\bar{Z}$}
\begin{longtable}{l|l}
\caption{Raw values used to produce Figure 4, which shows the rattler fraction $\phi_R$ as a function of the deformation parameter $p$, and the associated standard deviation of the data points.}
\\ \hline
$p$                                       & $\phi_R$                                                       \\ \hline \hline
0.85                               & $0.4532 \pm 0.1006$                                 \\ \hline
0.90                               & $0.7656 \pm 0.1226$                                   \\ \hline
0.95                               & $1.246	\pm 0.1821$                                   \\ \hline
0.975                              & $1.996	\pm 0.1429$                                   \\ \hline
1.00                               & $2.42	\pm 0.5223$                                   \\ \hline
1.025                              & $1.966	\pm 0.1890$                                   \\ \hline
1.05                               & $1.126	\pm 0.1333$                                   \\ \hline
1.10                               & $0.6944 \pm 0.1248$                                   \\ \hline
1.20                               & $0.2516 \pm 0.0645$                                   \\ \hline
1.30                               & $0.1384 \pm 0.0504$                                   \\ \hline
1.40                               & $0.0948 \pm 0.0472$                                   \\ \hline
1.50                               & $0.0852 \pm 0.0337$                                   \\ \hline
\end{longtable}

\begin{table}[H]
\centering
\caption{Raw values used to produce Figure 5, which shows the packing fraction $\phi$ as a function of the deformation parameter $p$, and the associated standard deviation of the data points.}
\begin{tabular}{l|l}

\hline
$p$                                       & $\phi$                                                       \\ \hline \hline
0.85                               & $0.67763 \pm 0.0004398$                                 \\ \hline
0.90                               & $0.66522 \pm 0.0004810$                                   \\ \hline
0.95                               & $0.65369 \pm 0.000408$                                   \\ \hline
0.975                              & $0.64812 \pm 0.0003156$                                   \\ \hline
1.00                               & $0.64327 \pm 0.0003754$                                   \\ \hline
1.025                              & $0.64841 \pm 0.0004104$                                   \\ \hline
1.05                               & $0.65362 \pm 0.0004911$                                   \\ \hline
1.10                               & $0.66351 \pm 0.0004847$                                   \\ \hline
1.20                               & $0.68069 \pm 0.0004036$                                   \\ \hline
1.30                               & $0.69406 \pm 0.0004449$                                   \\ \hline
1.40                               & $0.70485 \pm 0.0004811$                                   \\ \hline
1.50                               & $0.71448 \pm 0.0006174$                                   \\ \hline
\end{tabular}
\end{table}

\begin{longtable}{l|l}
\caption{Raw values used to produce Figure 7, which shows the average contact number $\bar{Z}$ as a function of the deformation parameter $p$, and the associated standard deviation of the data points.}
\\ \hline
$p$                                       & $\bar{Z}$                                                       \\ \hline \hline
0.85                               & $7.325278 \pm 0.01710$                               \\ \hline
0.90                               & $6.969658 \pm 0.01404$                               \\ \hline
0.95                               & $6.543314 \pm 0.01800$                                 \\ \hline
0.975                              & $6.20387 \pm 0.01125$                                \\ \hline
1.00                               & $6 \pm 0$                                 \\ \hline
1.025                              & $6.26906 \pm 0.00782$                                 \\ \hline
1.05                               & $6.60804 \pm 0.00865$                                 \\ \hline
1.10                               & $7.013264 \pm 0.01627$                                  \\ \hline
1.20                               & $7.513714 \pm 0.01415$                                 \\ \hline
1.30                               & $7.765312 \pm 0.01395$                                 \\ \hline
1.40                               & $7.880766 \pm 0.02107$                                \\ \hline
1.50                               & $7.925428 \pm 0.02245$                                  \\ \hline
\end{longtable}

